\documentclass[12pt]{extarticle}
\usepackage[utf8]{inputenc}
\usepackage[pdfencoding=auto, psdextra]{hyperref}
\usepackage{comment}

\usepackage[T1]{fontenc}

\usepackage{amsthm}
\usepackage{tikz}
\usetikzlibrary{fadings}
\usetikzlibrary{patterns}
\usetikzlibrary{shadows.blur}
\usetikzlibrary{shapes}

\newcommand{\qw}{q^{w}}
\newcommand{\tw}{\theta^{w}}

\newcommand{\hw}{h^{w}}
\newcommand{\rw}{\rho^{w}}

\newcommand{\eps}{\varepsilon}
\renewcommand{\epsilon}{\varepsilon}

\newcommand{\It}{I_{tan}}
\newcommand{\bIt}{\bar{I}_{tan}}

\newcommand{\KMIN}{K_{\min}}
\newcommand{\HOV}{\mathrm{HOV}}

\theoremstyle{plain}
\newtheorem{theorem}{Theorem}[section]
\theoremstyle{plain}
\newtheorem{lemma}[theorem]{Lemma}
\theoremstyle{definition}
\newtheorem{definition}{Definition}[section]
\theoremstyle{remark}
\newtheorem{remark}{Remark}[section]
\counterwithin{figure}{section}

\usepackage{enumitem}
\usepackage{mathtools}
\usepackage{amsmath,amsfonts, amssymb}
\usepackage{mathabx}
\usepackage{graphics}
\usepackage{graphicx}
\usepackage{float}
\usepackage[giveninits=true]{biblatex}
\usepackage{geometry}
 \bibliography{vered}

\numberwithin{equation}{section}

\begin{document}
\title{Hovering Sets in Near Pseudo-Integrable Hamiltonian Impact Systems}

\author{ Idan Pazi\thanks{Weizmann Institute of Science, Rehovot, Israel. \texttt{idan.pazi@weizmann.ac.il}}
  \and
   Alexandra Zobova\thanks{Weizmann Institute of Science, Rehovot, Israel, current address: Tel Aviv University, Ramat Aviv, Israel. \texttt{alexzobova@tauex.tau.ac.il}}
  \and
  Vered Rom-Kedar\thanks{The Estrin Family Chair of Computer Science and Applied Mathematics, Weizmann Institute of Science, Rehovot, Israel. Corresponding author: \texttt{vered.rom-kedar@weizmann.ac.il}}
}

\date{\today} 
\maketitle

\newcommand{\keywords}[1]{\begingroup\small\textbf{Keywords: }#1\par\endgroup}
\newcommand{\MSC}[1]{\begingroup\small\textbf{MSC (2020): }#1\par\endgroup}

\begin{abstract}
 The transition from rotational to discontinuous behavior of the return map of the perturbed oscillators-step system, a paradigm model for a perturbation of a pseudo-integrable Hamiltonian impact system, is studied.  The form of the return map is derived, and a truncated form of this map is simulated and analyzed.  For a set of parameters the existence of a hovering set, a set of non-resonant orbits that pass sometimes above the step and sometimes to its side, without ever impacting it, is established and quantified. Its destruction as the sign of the perturbation term is reversed is established.
\end{abstract}

 \keywords{Hamiltonian impact systems; piecewise-smooth dynamics; grazing bifurcation; pseudo-integrable systems}
 \MSC{37J40, 37c83, 70H08, 70H06}



\tableofcontents

\section{Introduction}

The motion of a particle in a plane within a given generic smooth potential field is generally chaotic and challenging to analyze. The special case in which the potential is separable leads to integrable motion, which, by the Arnold-Liouville theorem, is conjugated to directional motion on invariant tori for open dense sets of initial conditions. The behavior of such recurrent motion under smooth perturbations can be examined by constructing iso-energy two-dimensional return maps of the flow, reducing the four-dimensional space to two-dimensional smooth symplectic maps. In the integrable setting, this map represents an action-dependent family of rotations. In the KAM non-degenerate case (the focus of our discussion), the twist condition is satisfied on open intervals of the actions. Thus, under perturbations, as long as the perturbed return map is well defined, the study of local dynamics near these families of tori can be reduced to the analysis of two-dimensional symplectic twist maps. A prime example of such a map is the standard map, which can be derived as a Poincaré return map of the kicked rotor or the bouncing ball system. Alternatively, it can be viewed as a leading-order, first Fourier mode expansion, of a general near-integrable two-degree-of-freedom Hamiltonian system. Indeed, it was recently established that compositions of horizontal and vertical shears are dense in the group of smooth (and even analytic) Hamiltonian diffeomorphisms, including return maps of such systems, thereby strengthening this point of view \cite{berger2024generators}.

The above methodology works for smooth systems, namely when both the potentials and the perturbation terms have sufficiently many bounded derivatives. Yet, in some applications the dynamics may also include localized non-smooth or near-discontinuous components, e.g., when the particle impacts a boundary in the configuration space or when the potential has localized steep fronts traditionally modeled by impulsive forces. The study of such a combination of Hamiltonian smooth and localized non-smooth dynamics falls under the category of Hamiltonian Impact Systems (HIS). These systems obey the same reflection law as classical mathematical billiards, yet, in between impacts, they allow the particles to change their momenta according to the non-trivial smooth potential gradient in the domain's interior \cite{kozlov1991billiards}. Such systems appear, usually with additional realistic elements such as dissipation and friction, in numerous engineering applications, see \cite{babitsky2013theory,THOTA2006163,belykh2023beyond,sushko2025organising} and references therein. 

Two classes of globally analyzable HIS, in which the energy level set is foliated by an additional constant of motion, are the integrable HIS (IHIS) \cite{PnueliRomKedarIntegrability} and the pseudo-integrable HIS (PIHIS) \cite{BeckerStep,frkaczek2023non}. 

The IHIS consist of HIS with energy level sets that are, similar to the integrable smooth case, foliated by families of invariant tori on which the motion is rotational, and these families of tori connect at singular level sets. The iso-energy  Poincaré return map near such families of tori results, as in the smooth case, in an action-dependent family of rotations, where the rotation dependence on the action is smooth away from tangencies and is piecewise continuous, with a square-root singularity near a tangent (grazing) torus. Under perturbations, away from tangencies, the map is a smooth near integrable twist map \cite{PnueliRomKedarIntegrability}. Near tangencies, the piecewise smooth rotations induces, under perturbations, intricate chaotic dynamics  \cite{PnueliRomKedarTangency,THOTA2006163}.

A class of pseudo-integrable HIS (PIHIS) systems was introduced in \cite{BeckerStep}. It corresponds to a particle moving in the interior of a rectilinear domain with at least one corner angle larger than $\pi/2$, where the motion in the interior of the domain is governed by a separable potential, with the configuration space axes aligned with the domain's rectilinear edges. For such systems, the energy level sets are also foliated by families of invariant surfaces, the fixed partial energies surfaces. Yet, beyond a certain energy, there are intervals of partial energies for which the surfaces are of genus two and higher. As explained in  \cite{BeckerStep,frkaczek2023non}, the motion on each such surface is conjugated to a directed motion on a pseudo-integrable rectilinear  billiard and thus to a directed motion on a translation surface  \cite{richens1981pseudointegrable,zorich2006flat}. Figure \ref{fig:two-tang} presents such a system, the oscillator-step system, where the potential is a sum of one dimensional  horizontal and vertical potentials and the impacts occur at a step aligned with the axes. A return map to a circle on each of the partial energies surfaces is an interval exchange map  (IEM) on that circle. The intricate ergodic properties of the motion on such surfaces for some classes of PIHIS were studied in \cite{frkaczek2023non,frkaczek2024resonant}. It follows that the iso-energy return map for such systems results in an action-dependent family of IEMs. 

Here, we study how the return map and its dynamics are deformed under small perturbations near the onset of impacts, when the genus of the partial energies surfaces changes from one to two. The outcome is a construction of a piecewise smooth invertible area-preserving map.

The paper is ordered as follows: In Section \ref{sec:setup} we introduce the necessary notations and recall the relevant background from \cite{PnueliRomKedarTangency}. In Section \ref{sec:mainresults} we formulate the main results of this paper: Theorems \ref{thm:cornersing}, \ref{thm:regionsJdyn}, and \ref{thm:returnmap}, in which the return map is derived, and Theorem \ref{thm:hovering}, in which the existence of hovering dynamics in the truncated model is established. Section \ref{sec:truncatedmapnum} includes an analytical and numerical investigation of the truncated map model for the return map, where, importantly, this model retains the same time-reversal symmetry as the return map of the perturbed system. The hovering dynamics is established for this map, leading to a proof of Theorem \ref{thm:hovering}. Finally, Section \ref{sec:discussion} summarizes the results and outlines directions for future studies.
The extended Appendix \ref{sec:appendixA} includes the needed constructions and the proofs of Theorems \ref{thm:cornersing}, \ref{thm:regionsJdyn} and \ref{thm:returnmap}. Appendix \ref{sec:calculationofparam} includes the calculation of the parameters appearing in the leading order terms of the return map for some specific potentials.

 \paragraph{Acronyms:}
    HIS    - Hamiltonian Impact System;
    IHIS   - Integrable Hamiltonian Impact System;
    PIHIS - Pseudo Integrable Hamiltonian Impact System;
    IEM    - Interval Exchange Map (here, on the circle);

\section{Setup\label{sec:setup}}
Consider a two-degree-of-freedom smooth\footnote{Hereafter, $C^r$-smooth systems with $r>4$ so that, with additional  $C^r$-smooth perturbations, KAM theory applies.} integrable Hamiltonian of the form: 
\begin{equation}
    H_{int}(z)=H_{int}(p_1,p_2,q_1,q_2) = 
    \frac{p_1^2}2 + V_1(q_1) + \frac{p_2^2}2 + V_2(q_2) \label{eq:Hint}
\end{equation}
which satisfies the following conditions \cite{PnueliRomKedarTangency}:
\begin{enumerate}
    \item Each potential $V_i(q_i)$ depends on only one coordinate.
    \item Each potential $V_i(q_i)$ has a single minimum, located at $q_{i}^0 = 0$, $V_i(q_{i}^0) = 0$, and both potentials are convex $q_i\cdot V_i'(q_i)>0$ for $q_i\neq 0$.
\end{enumerate}
For this unperturbed smooth Hamiltonian, corresponding to two uncoupled oscillators, the partial energies: 
$$
h_i =\frac{p_i^2}2 + V_i(q_i),\ 
$$
 are preserved by the flow.

Let $H(z;\epsilon)$ denote the smooth perturbed Hamiltonian:  
\begin{equation}\label{eq:perturbedhamil}
   H(z;\epsilon) = H_{int}(p_1,p_2,q_1,q_2) +\epsilon V_c(q_1,q_2),
\end{equation}
where $\epsilon$ is a small parameter; $V_c(q_1,q_2)$ represents a smooth coupling potential between the oscillators, and we assume it is bounded in $C^r$ and that the coupling is non-trivial ($\frac{\partial^2 V_c}{\partial q_1 \partial q_2} \not\equiv 0$).

\begin{figure}
    \centering
    \resizebox{0.8\columnwidth}{!}{%
        \input{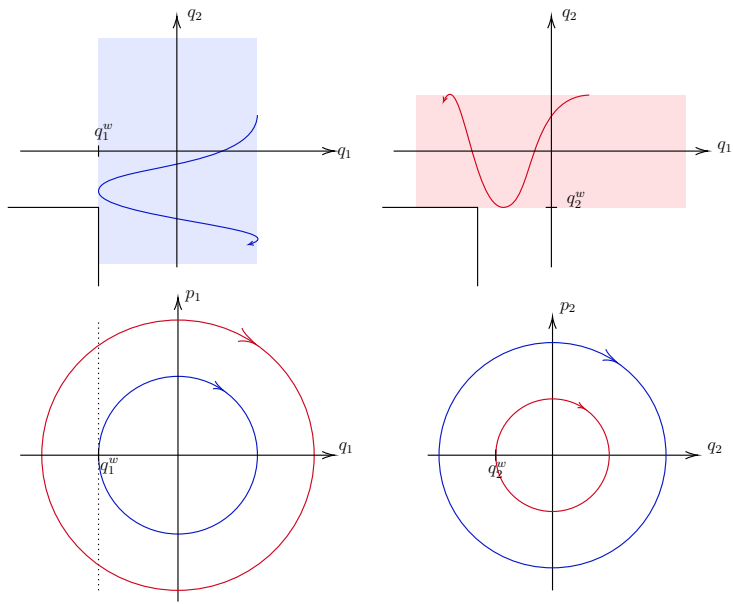}
    }
    \caption{Projections of the two iso-energy tangent tori of the unperturbed system to 3 subspaces. Trajectories belonging to the blue torus are tangent to the line corresponding to the right wall of the step and those belonging to the red torus are tangent to the line corresponding to the upper wall of the step. Upper row: projection to the configuration space with a curve indicating a segment of a trajectory on the corresponding torus. Lower row: projections to the $(q_1,p_1)$ and $(q_2,p_2)$ spaces.  Hereafter we consider the local dynamics near the first, blue, torus.}
    \label{fig:two-tang}
\end{figure}

Introduce a step in the configuration space $(q_1, q_2)$, see Figure \ref{fig:two-tang}:
\begin{equation}\label{eq:stepdef}
    \mathrm{St}(q_1,q_2) = \{(q_1,\,q_2): q_1< \qw_1,\ q_2<\qw_2\}.
\end{equation}
We assume the following \cite{BeckerStep, PnueliRomKedarTangency}:
\begin{enumerate}
\item Impacts at the right/upper wall of the step are purely elastic and, therefore, are identified with reflections of the corresponding momentum (hereafter denoted by $\mathcal{R}_1 ,\mathcal{R}_2 $):
\begin{equation}
    \label{eq:R1}
    \mathcal{R}_1 z^{w_1} = \mathcal{R}_1(\qw_1,p_1,q_2,p_2) = 
    (\qw_1,-p_1,q_2,p_2),\ p_1<0,\  q_2<\qw_2,
\end{equation}
\begin{equation}
    \label{eq:R2}
    \mathcal{R}_2 z^{w_2} = \mathcal{R}_2(q_1,p_1,\qw_2,p_2) = 
    (q_1,p_1,\qw_2,-p_2),\ p_2<0,\  q_1<\qw_1;
\end{equation}
\item The critical points $q_1^0 = q_2^0 = 0$ of both potentials are outside the step:
$\qw_1 < 0$, $\qw_2 < 0$;
\item When a trajectory hits the corner: $q_1 = \qw_1$ and $q_2 = \qw_2$ simultaneously the trajectory stops.
\end{enumerate}

Define the step energies:
\begin{equation}\label{def:heps}
    \hw_i \coloneqq  V_i(\qw_i),\quad \hw_\eps \coloneqq \hw_1 + \hw_2 +\eps V_c(\qw_1,\qw_2)
\end{equation}
On a fixed energy level set $h$, $\{z|H_{int}(z)=h\}$, of the unperturbed Hamiltonian, for $h>\hw_0=\hw_1 + \hw_2 $ there exist two classes of tangent trajectories (Figure \ref{fig:two-tang}):
one class corresponds to the tangency to the right wall of the step
(blue lines, where $h_1=h_1^w$), while the other to the upper side of the step (red lines, where  $h_2=h_2^w$).

Without loss of generality, we will consider the dynamics at the onset of hitting the right wall of the step. Namely, we focus on the neighborhood of the trajectories that are tangent to the right wall. For $\eps=0$ these are the trajectories that belong to the tangent torus that corresponds to the product of the blue circles in the $(q_1,p_1)$ and the $(q_2,p_2)$ planes, as shown in Figure \ref{fig:two-tang}, namely for initial conditions with $H_1(q_1,p_1) \approx h_1^w$. The same analysis applies to the region near the tangency to the upper wall (for $\eps=0$ the product of the red circles of Figure \ref{fig:two-tang}) by reversing the roles of horizontal and vertical directions in the below constructions.

Following \cite{PnueliRomKedarTangency}, we introduce a two dimensional cross-section $\Sigma_h$ for a fixed total energy $h$:
\begin{eqnarray}
\Sigma_h &=& \{(q_1,p_1,q_2,p_2): p_1 = 0, \frac{d p_1}{dt}<0,\, H(q_1,p_1,q_2,p_2;\eps) = h\}.
\end{eqnarray}
Due to the properties of the unperturbed Hamiltonian \cite{PnueliRomKedarTangency}, for sufficiently small $\eps$, 
this cross-section is parameterized by the variables $q_2,\ p_2$, or, equivalently, by the action-angle variables $(I,\theta)=S_2(q_2,p_2)$ of the unperturbed Hamiltonian $H_2=H_2(I)$, where for convenience, as in  \cite{PnueliRomKedarTangency,BeckerStep},  on $\Sigma_h$, we set $\theta=0$ at $p_2=0,\frac{dp_2}{dt} < 0$, so, on  $\Sigma_h$, taking  $\theta \in [-\pi,\pi)$ implies that $\mathrm{sign}(\theta)=-\mathrm{sign}(p_2)$. 

The main challenge is to describe the dynamics of the return map to $\Sigma_h$:
$$
\mathcal{F}_\epsilon(\theta,I)_{\Sigma_h \rightarrow \Sigma_h}: \  (\theta,I) \mapsto ( \bar \theta,\bar I).
$$

For $\eps=0$, since $H_2(q_2,p_2)=H_2(I)$ is constant, the return map $\mathcal{F}_0(\theta,I)$ keeps the circles $I=\mathrm{const}$ invariant. 
As established in \cite{BeckerStep}, the unperturbed dynamics on these invariant circles  becomes non trivial when   $h>h^w_0$, where some of the circles include trajectories that hit the step. For these circles, the return map is a discontinuous interval exchange transformation on the circle. The aim of this work is to study the Poincaré first return map $\mathcal{F}_\epsilon$ near this transition, namely close to tangencies of the flow to the step. 

The strategy is to use an auxiliary section near the step (Section \ref{sec:constructret}) at which the division to different intervals is easy to deduce (Section \ref{sec:proofofcornersing}) and then use the smooth flow to carry the initial conditions on this section backward and forward to $\Sigma_h$ (Section \ref{sec:proofreturnmap}). A truncated model of the return map is then studied analytically and numerically (Section \ref{sec:truncatedmapnum}). 

\subsection{The tangential curve}
Next we set up the notation and review some of the results and methods developed in  \cite{PnueliRomKedarTangency}   who considered the near-tangent behavior of an HIS system of the form \eqref{eq:Hint} with impacts from an infinite wall which is parallel to one of the axis.

First, note an important symmetry of the first return maps of a mechanical HIS to $\Sigma_h$:
\begin{lemma}\label{lem:timerevgen} For any initial condition for which $\mathcal{F}_\eps (\theta, I) $ is defined, the first return map to  $\Sigma_h$ obeys the time reversal symmetry with respect to the reflection symmetry \eqref{eq:R2}:
\begin{equation} \label{eq:timereversal}
    \mathcal{R}_2 \mathcal{F}_\eps  =\mathcal{F}^{-1}_\eps \mathcal{R}_2 .
\end{equation}   
\end{lemma}
\begin{proof}
With this choice of $\theta=0$ on $\Sigma_h$, the restriction of the reflection $\mathcal{R}_2 z $  to the coordinates $(\theta,I)$ on $\Sigma_h$ becomes  $\mathcal{R}_2 (\theta,I)=(-\theta,I) $.  As  $\mathcal{R}_2 z |_{\Sigma_h}$ sends $p_2$ to $-p_2$ and $p_1=0$ there, this reflection also  reverses the direction of motion; for any $(\theta,I)\in \Sigma_h$ for which $(\bar \theta,\bar I)=\mathcal{F}_\eps (\theta, I) \in \Sigma_h $ is defined, the mechanical form of the potential and of the impacts implies that reversing the direction of motion at the initial point, namely setting $ (\theta^*, I^*)=(-\bar \theta,\bar I)=\mathcal{R}_2 \mathcal{F}_\eps (\theta, I) $, leads to a motion in the reverse direction along the same trajectory. Hence,  the map of $(\theta^*, I^*)$ leads back to the same configuration point on $\Sigma_h$ with the opposite vertical momenta, namely to the reflection of the original point: $ \mathcal{F}_\eps (\theta^*, I^*)= \mathcal{R}_2 (\theta, I)$. Thus, indeed,  $ \mathcal{F}_\eps \mathcal{R}_2  \mathcal{F}_\eps = \mathcal{R}_2 $.
\end{proof}
Notice that the proof applies to any HIS for which the return time to $\Sigma_h$ is bounded, see Section \ref{sec:timereversal} for more details and remarks regarding this property.

Next we define the tangential curve and its singular part:
\begin{definition}
\label{Def:sing_curve}
For each $h>h_0^w$,  the \emph{tangential curve}, $\sigma_{tan}^\eps$, is the set of all initial conditions in $\Sigma_h$, which, under the step system flow,  touch the line $q_1=\qw_1$ tangentially before their first return to $\Sigma_h$. A part of the curve $\sigma_{tan}^\eps$, the tangential-singular part, $\sigma^\eps_{tan-R}$, touches the right wall of the step, and thus corresponds to singularities of $\mathcal{F}_\eps$, while the other part, $\sigma_{tan}^\eps \setminus \sigma^\eps_{tan-R}$,  does not,  corresponding to regular trajectories.
\end{definition}
The tangential curve $\sigma_{tan}^\eps$ is identical to the tangential curve defined in \cite{PnueliRomKedarTangency} whereas its division to singular and regular parts arises here due to the step. Recall that for all $h>h_\eps^w$ and sufficiently small  $\eps$: 
\begin{itemize}
    \item   $\sigma_{tan}^\eps$  is a dividing circle on the $\Sigma_h$ cylinder $(\theta, I)$.
    \item At $\eps=0$ the tangential curve and its image are identical and are given by the tangential circle:
    \begin{equation}\label{eq:Itan}
 \It(h) \coloneqq H_2^{-1}(h-\hw_1).
\end{equation}
    \item The curve $\sigma_{tan}^\eps$   and its first image under first return map $\mathcal{F}_\eps$ are graphs over $\theta$ of smooth functions  $\It^\eps(\theta;h)$ and $\bIt^\eps(\theta;h)$ that are $\eps$-close in the $C^r$ norm to the constant function $\It(h)$ of \eqref{eq:Itan}: 
    $$
    \sigma^\eps_{tan} = \{(\theta, \It^\eps(\theta),\ \theta\in[-\pi,\pi]\},
    $$
    $$
    \bar{\sigma}_{tan}^\eps = \mathcal{F}_\eps\sigma^\eps = \{(\theta, \bIt^\eps(\theta),\ \theta\in[-\pi,\pi]\}.
    $$
    
\item The time reversal symmetry implies that the tangential curve and its image are related by the reflection symmetry\footnote{notice that this symmetry does not imply that  the image of an initial condition on this curve coincides with is its reflection - in general it does not! see \cite{PnueliRomKedarTangency} and Section \ref{sec:timereversal}} :  $\bIt^\eps(\theta) = \It^\eps(-\theta)$ (see also Theorem \ref{thm:timereversal}). 
\end{itemize}
It is convenient to introduce the smooth, near identity symplectic change of coordinates $S^\eps:(\theta,I) \rightarrow (\phi,K)$ to the \textit{normal coordinates}:
\begin{equation}\label{eq:kphidef}
(\phi,K)=S^\eps(\theta,I) =(\theta, I - \It^\eps(\theta)),
\end{equation}
where for shorthand notation we drop the superscript $\eps$ on $S$ when there is no need to emphasize its dependence on $\eps$. Trivially, the inverse map is: 
\begin{equation}\label{eq:kphiinvdef}
(\theta,I) =S^{-1}(\phi,K)=(\phi, K+ \It^\eps(\phi)).
\end{equation}

In these normal coordinates $
    \sigma^\eps_{tan} = \{(\phi, K=0),\ \phi \in[-\pi,\pi]\}
    $ and  $
    \bar{\sigma}_{tan}^\eps = \mathcal{F}_\eps\sigma_{tan}^\eps = \{(\bar \phi, \bar K)|\bar K= \bIt^\eps(\bar \phi)- \It^\eps(\bar \phi),\ \bar \phi\in[-\pi,\pi]\}
    $ so \begin{equation}\label{eq:barsigmanormal}
        \bar{\sigma}_{tan}^\eps = \mathcal{F}_\eps\sigma_{tan}^\eps =\{(\phi,K)|K= \It^\eps(-\phi)- \It^\eps(\phi),\ \phi\in[-\pi,\pi]\}.
    \end{equation}
Namely the tangential curve image is the graph of an odd, bounded and smooth function $\eps f(\phi;\eps)$:
\begin{equation}\label{eq:defoff}
  \eps f(\phi;\eps)\coloneqq  \It^\eps(-\phi)- \It^\eps(\phi), \qquad f(\phi;\eps)=-f(-\phi;\eps).
\end{equation}
In the normal coordinates, $K>0$ (respectively $K<0$) corresponds to initial conditions that do not (respectively, do) cross the line $q_1=q_1^w$. 

Summarizing, on the two-dimensional Poincaré section\footnote{The restriction of the map $\mathcal{F}_\eps$ to the iso-energy level  $\Sigma_h$ is achieved by adjusting the corresponding initial $q_1$ value so that $H(q_1,p_1=0,q_2,p_2;\eps)=h$ and $q_1$ is close to its maximal value (more details are included in the proofs below). } $\Sigma_h$, we alternately use, as needed, the following three sets of symplectic coordinates which are, for all $|\eps|<\eps_c$ for which $\It^\eps(\phi)$ is a smooth graph, smoothly conjugated: \begin{equation} (\phi,K)= S^\eps(\theta,I)=S^\eps\circ S_2 (q_2,p_2). \end{equation}

\subsection{The parameters at onset \label{sec:parametersdef}}
To establish the asymptotic form of the  return map dependence on parameters we introduce the following notations:
\begin{itemize}
    \item The unperturbed periods and frequencies of the horizontal and vertical oscillators:
\begin{align*}
T_i(h_i) &\coloneqq 2\int\limits_{q_{i,min}}^{q_{i,max}}
\frac{dq_i}{\sqrt{2(h_i - V_i(q_i))}},\quad  q_{i,min/max} \coloneqq V_i^{-1}(h_i) \\ 
\omega_i(h_i) &\coloneqq \frac{2\pi}{T_i},
\end{align*}
\item The rotation in $\theta$ of the return map to $\Sigma_h$ for unperturbed non-impacting trajectories:
\begin{equation}\label{eq:omega0ofI}
    \Omega_0 (I)\coloneqq\omega_2(I)\cdot T_1(h-H_2(I)), \ \Omega_0\coloneqq\Omega_0 (\It(h)).
\end{equation}
\item The local twist at $\It(h)$: 
\begin{equation}\label{eq:deftau0}
   \tau_0\coloneqq \frac{d  \Omega_0 (I)}{d I}\Big|_{\It(h)}.
\end{equation}
\item For actions $I>\It(h)$,  $\tw(I) $ is the angle variable that corresponds to  $q_2^w$  and $p_2^w(I)=-\sqrt{2(H_2(I) - V_2(\qw_2))}$, so that $(\tw(I),I)=S_2(\qw_2,p_2^w(I))$:
\begin{equation}\label{eq:thw}
\tw(I) = \frac{2 \pi}{{T}_2(H_2(I))}\int_{q_{2}^w}^{q_{i,max}}
\frac{dq_i}{\sqrt{2(H_2(I) - V_2(q_2))}}, \quad \tw \coloneqq \tw(\It(h)).  
\end{equation}
\item The factor of the leading order increment in $\theta$ when traveling above the step (see \eqref{eq:cornersingbordrho} below):
\begin{equation}\label{eq:deflambdah}
    \lambda \coloneqq \frac{\sqrt{2\omega_2(\It(h))^3}}{|V_1'(q_1^w)|}
\end{equation}
\item The twist associated with the reflection from the upper side of the step:
\begin{equation}\label{eq:deftau1}
    \tau_1 \coloneqq -2 \frac{d \tw(I)}{dI}\Big|_{\It(h)}.
\end{equation}Notice that $\tau_1$ is bounded for energies which are larger and bounded away from the corner energy, namely for  $h>h_0^w $. 

\end{itemize}
 
Hereafter we always consider $h>h_0^w $ and the dependence on $h$ is usually omitted. Nevertheless, notice that all the parameters may depend on $h$ and the limit $h\searrow h_0^w$ is singular (e.g. $\tau_1$ is unbounded in this limit). Explicit calculations of these parameters for combinations of the quadratic potential, $V_i(q_i)=\frac{\omega_i^2q_i^2}{2}$,  and the Tan potential, $V_i(q_i)=\frac{\omega_i^2}{2\alpha_i^2}\tan^2(\alpha_i q_i)$,  are listed in Appendix \ref{sec:calculationofparam}.  We find that only in the case at which both potentials are non-linear in the action (i.e., both potentials are not the quadratic potential) all the parameters depend on $h$.  Our theory regarding the hovering set applies to the case at which at least one of the potentials is not quadratic (otherwise the return map at the non-impacting regime has zero twist and KAM theory does not apply there).

\section{Main results\label{sec:mainresults}}

We introduce below the region above the tangential curve, $J_{0u}^\eps$, the corner-singularity curves that lie below the tangential curve,  $\sigma_{ab}^\eps, ab \in \{R0,01,1R\}$, and the regions enclosed in between these curves, denoted by  $J^\epsilon_R,J^\epsilon_0, J^\epsilon_1  $. 
In Theorem \ref{thm:cornersing} we establish that for $h>h_0^w$ and for sufficiently small $\epsilon$ the corner-singularity curves, expressed by the normal coordinates, are graphs over $\sqrt{-K}$ that depend continuously on $\epsilon$, extending vertically across a band of $K$ values of height $\Delta = \Delta (h)$. Namely, in the normal coordinates,  $\sigma_{ab}^\eps= \{(\phi_{ab}^\eps(K),K))|     K \in [-\Delta,0]  \}$, and the asymptotic form of $\phi_{ab}^\eps(K)$ for small $K$ is established.  
In Theorem \ref{thm:regionsJdyn} we prove that these regions correspond to different dynamical regions:  $J^\epsilon_R$ corresponds to trajectories segments that hit the right boundary of the step before their first return to $\Sigma_h$,  $J^\epsilon_1$  corresponds to segments that hit the upper boundary of the step exactly once,   $J^\epsilon_0$ corresponds to segments that pass above the step without hitting it, and $J_{0u}^\eps$ corresponds to segments that pass to the side of the step without hitting it (i.e. do not cross the line $q_1=\qw_1$). Theorem \ref{thm:reverseJs} establishes that the boundaries of these regions images, in the $(\theta,I)$ coordinates, are reflections of the corner-singularity curves. Theorem \ref{thm:reversalinkphi} establishes the corresponding symmetries in the normal coordinates,  $(\phi,K)$. It follows that the return map is area preserving: the return map in the interior of each of the regions is symplectic as it corresponds to the symplectic return map defined by the Hamiltonian impact flow, and the region images do not overlap and do not leave any gaps. Namely, the return map corresponds to a piecewise smooth, discontinuous area preserving map. 
Theorem \ref{thm:returnmap} establishes the form of the return map in each of these regions to leading order in $\sqrt{-K},\eps$. Lemma \ref{lem:truncreturnmap} establishes that a family of truncated maps that imitate the full dynamics preserves the same time reversal symmetries as the perturbed return map (see also \cite{PaziRK}).  
The proofs of these theorems require additional  constructions and detailed computations which appear in Appendix \ref{sec:appendixA}. Section \ref{sec:truncatedmapnum} presents an analysis and numerical simulations of this model of the truncated map. Theorem \ref{thm:hovering} implies that for open sets of parameters of this model there is an open set of initial conditions, of measure  $C \eps + O(\eps^2), C>0$ of hovering orbits. Moreover, for the same parameter values, flipping the sign of $\eps$ destroys this set. Finally,  the existence of resonant islands that visit different dynamical regions is demonstrated numerically. Interestingly, invariant circles that cross the corner-singularity curves are not observed, leading to the conjecture that generically such curves do not exist. 

\subsection{The corner-singularity curves}

By the definition of the tangential curve, there exist angles $\theta_{01R},\theta_{R0}$ (see below for the subscripts notation) that correspond to the left and right boundaries of the tangent-singularity segment on the tangential curve. Introduce the notation $[a,b]_c$ for the interval on the circle so that for all $a,b \in[-\pi,\pi]$: \begin{equation}
    [a,b]_c \coloneqq \begin{cases}
[a,b]         & a \le b  \\
     [a,b+2\pi]    & b<a 
    \end{cases}
\end{equation}
Then, the tangent singularity segment is
\[
    \sigma^\eps_{tan-R} = \{(\theta, \It^\eps(\theta)),\ \theta\in[\theta_{01R}^\eps,\theta_{R0}^\eps]_c\eqqcolon\Theta_R^\eps\}    \]
and the flow emanating from either $(\theta_{01R}^\eps,\It^\eps(\theta_{01R}^\eps))$ or $(\theta_{R0}^\eps, \It^\eps(\theta_{R0}^\eps))$   is tangent to the  line $q_1=\qw_1$ exactly at the corner point. For $h>\hw_\eps$ the segment $\Theta_R^\eps$ has positive length whereas at  $h=\hw_\eps$ the tangential curve only touches the corner, so $\theta_{R0}^\eps=\theta_{01R}^\eps$.  Since $\qw_1<0$, it follows that $|\Theta_R^0|<2\pi$. For $h>\hw_0$   we get that $|\Theta_R^0|>0$ and that the dependence of this interval boundary  on $\eps$ is smooth. Hence, for sufficiently small $\eps$,  $|\Theta_R^\eps|$ is strictly  inside the interval $ (0,2 \pi)$. 
Hereafter, all intervals are considered on the circle and the subscript $c$ is omitted. 

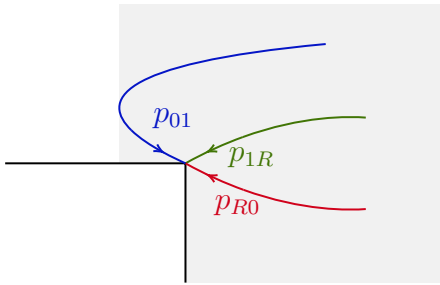
\begin{figure}[H]
\centering
 
\tikzset{
pattern size/.store in=\mcSize, 
pattern size = 5pt,
pattern thickness/.store in=\mcThickness, 
pattern thickness = 0.3pt,
pattern radius/.store in=\mcRadius, 
pattern radius = 1pt}
\makeatletter
\pgfutil@ifundefined{pgf@pattern@name@_phfdnif3b}{
\pgfdeclarepatternformonly[\mcThickness,\mcSize]{_phfdnif3b}
{\pgfqpoint{0pt}{0pt}}
{\pgfpoint{\mcSize+\mcThickness}{\mcSize+\mcThickness}}
{\pgfpoint{\mcSize}{\mcSize}}
{
\pgfsetcolor{\tikz@pattern@color}
\pgfsetlinewidth{\mcThickness}
\pgfpathmoveto{\pgfqpoint{0pt}{0pt}}
\pgfpathlineto{\pgfpoint{\mcSize+\mcThickness}{\mcSize+\mcThickness}}
\pgfusepath{stroke}
}}
\makeatother

  
\tikzset {_z3n022qy6/.code = {\pgfsetadditionalshadetransform{ \pgftransformshift{\pgfpoint{0 bp } { 0 bp }  }  \pgftransformrotate{-90 }  \pgftransformscale{2 }  }}}
\pgfdeclarehorizontalshading{_u1mzkxgia}{150bp}{rgb(0bp)=(1,1,1);
rgb(37.5bp)=(1,1,1);
rgb(55.535714285714285bp)=(1,1,1);
rgb(100bp)=(1,1,1)}
\tikzset{_t8dx1kwl7/.code = {\pgfsetadditionalshadetransform{\pgftransformshift{\pgfpoint{0 bp } { 0 bp }  }  \pgftransformrotate{-90 }  \pgftransformscale{2 } }}}
\pgfdeclarehorizontalshading{_mgk4mts30} {150bp} {color(0bp)=(transparent!100);
color(37.5bp)=(transparent!100);
color(55.535714285714285bp)=(transparent!0);
color(100bp)=(transparent!0) } 
\pgfdeclarefading{_vzwzx6irw}{\tikz \fill[shading=_mgk4mts30,_t8dx1kwl7] (0,0) rectangle (50bp,50bp); } 

  
\tikzset {_fcvae8oac/.code = {\pgfsetadditionalshadetransform{ \pgftransformshift{\pgfpoint{0 bp } { 0 bp }  }  \pgftransformrotate{-90 }  \pgftransformscale{2 }  }}}
\pgfdeclarehorizontalshading{_gsv94bdi1}{150bp}{rgb(0bp)=(1,1,1);
rgb(40.357142857142854bp)=(1,1,1);
rgb(62.5bp)=(1,1,1);
rgb(100bp)=(1,1,1)}
\tikzset{_cxqxxhziq/.code = {\pgfsetadditionalshadetransform{\pgftransformshift{\pgfpoint{0 bp } { 0 bp }  }  \pgftransformrotate{-90 }  \pgftransformscale{2 } }}}
\pgfdeclarehorizontalshading{_70t9z84jb} {150bp} {color(0bp)=(transparent!0);
color(40.357142857142854bp)=(transparent!0);
color(62.5bp)=(transparent!100);
color(100bp)=(transparent!100) } 
\pgfdeclarefading{_r0ipgq9pm}{\tikz \fill[shading=_70t9z84jb,_cxqxxhziq] (0,0) rectangle (50bp,50bp); } 

  
\tikzset {_wsgrer1dp/.code = {\pgfsetadditionalshadetransform{ \pgftransformshift{\pgfpoint{0 bp } { 0 bp }  }  \pgftransformrotate{-180 }  \pgftransformscale{2 }  }}}
\pgfdeclarehorizontalshading{_zdhgjo95f}{150bp}{rgb(0bp)=(1,1,1);
rgb(40.357142857142854bp)=(1,1,1);
rgb(62.5bp)=(1,1,1);
rgb(100bp)=(1,1,1)}
\tikzset{_g0agvpu2j/.code = {\pgfsetadditionalshadetransform{\pgftransformshift{\pgfpoint{0 bp } { 0 bp }  }  \pgftransformrotate{-180 }  \pgftransformscale{2 } }}}
\pgfdeclarehorizontalshading{_cc8om401u} {150bp} {color(0bp)=(transparent!0);
color(40.357142857142854bp)=(transparent!0);
color(62.5bp)=(transparent!100);
color(100bp)=(transparent!100) } 
\pgfdeclarefading{_nf9ifjjtq}{\tikz \fill[shading=_cc8om401u,_g0agvpu2j] (0,0) rectangle (50bp,50bp); } 

  
\tikzset {_dkeh95v9q/.code = {\pgfsetadditionalshadetransform{ \pgftransformshift{\pgfpoint{0 bp } { 0 bp }  }  \pgftransformrotate{0 }  \pgftransformscale{2 }  }}}
\pgfdeclarehorizontalshading{_m5i4n5hcw}{150bp}{rgb(0bp)=(1,1,1);
rgb(40.357142857142854bp)=(1,1,1);
rgb(62.5bp)=(1,1,1);
rgb(100bp)=(1,1,1)}
\tikzset{_7boa0nj1z/.code = {\pgfsetadditionalshadetransform{\pgftransformshift{\pgfpoint{0 bp } { 0 bp }  }  \pgftransformrotate{0 }  \pgftransformscale{2 } }}}
\pgfdeclarehorizontalshading{_zmt7q0bg5} {150bp} {color(0bp)=(transparent!0);
color(40.357142857142854bp)=(transparent!0);
color(62.5bp)=(transparent!100);
color(100bp)=(transparent!100) } 
\pgfdeclarefading{_7d5b8mwbn}{\tikz \fill[shading=_zmt7q0bg5,_7boa0nj1z] (0,0) rectangle (50bp,50bp); } 
\tikzset{every picture/.style={line width=0.75pt}} 

\begin{tikzpicture}[x=0.75pt,y=0.75pt,yscale=-1,xscale=1]

\draw  [draw opacity=0][fill={rgb, 255:red, 241; green, 241; blue, 241 }  ,fill opacity=1 ] (67.01,30) -- (230,30) -- (230,170) -- (67.01,170) -- cycle ;
\draw  [draw opacity=0][fill={rgb, 255:red, 255; green, 255; blue, 255 }  ,fill opacity=1 ] (10,110) -- (100,110) -- (100,170) -- (10,170) -- cycle ;
\draw    (10,110) -- (100,110) -- (100,170) ;
\draw  [draw opacity=0][pattern=_phfdnif3b,pattern size=22.5pt,pattern thickness=0.75pt,pattern radius=0pt, pattern color={rgb, 255:red, 0; green, 0; blue, 0}] (10,110) -- (100,110) -- (100,170) -- (10,170) -- cycle ;
\draw [color={rgb, 255:red, 65; green, 117; blue, 5 }  ,draw opacity=1 ]   (100,110) .. controls (110.02,104.37) and (119.62,100.07) .. (128.6,96.8) ;
\draw [shift={(111.08,104.21)}, rotate = 335.07] [color={rgb, 255:red, 65; green, 117; blue, 5 }  ,draw opacity=1 ][line width=0.75]    (4.37,-1.96) .. controls (2.78,-0.92) and (1.32,-0.27) .. (0,0) .. controls (1.32,0.27) and (2.78,0.92) .. (4.37,1.96)   ;
\draw  [draw opacity=0][shading=_u1mzkxgia,_z3n022qy6,path fading= _vzwzx6irw ,fading transform={xshift=2}] (230,161) -- (230,171) -- (10,171) -- (10,161) -- cycle ;
\draw  [draw opacity=0][shading=_gsv94bdi1,_fcvae8oac,path fading= _r0ipgq9pm ,fading transform={xshift=2}] (230,39) -- (230,29) -- (10,29) -- (10,39) -- cycle ;
\draw  [draw opacity=0][shading=_zdhgjo95f,_wsgrer1dp,path fading= _nf9ifjjtq ,fading transform={xshift=2}] (230,170) -- (230,30) -- (220,30) -- (220,170) -- cycle ;
\draw  [draw opacity=0][shading=_m5i4n5hcw,_dkeh95v9q,path fading= _7d5b8mwbn ,fading transform={xshift=2}] (19,170) -- (19,30) -- (9,30) -- (9,170) -- cycle ;
\draw [color={rgb, 255:red, 208; green, 2; blue, 27 }  ,draw opacity=1 ]   (100,110) .. controls (110.15,115.71) and (119.87,120.04) .. (128.95,123.33) ;
\draw [shift={(111.23,115.86)}, rotate = 24.87] [color={rgb, 255:red, 208; green, 2; blue, 27 }  ,draw opacity=1 ][line width=0.75]    (4.37,-1.96) .. controls (2.78,-0.92) and (1.32,-0.27) .. (0,0) .. controls (1.32,0.27) and (2.78,0.92) .. (4.37,1.96)   ;
\draw [color={rgb, 255:red, 65; green, 117; blue, 5 }  ,draw opacity=1 ]   (128.6,96.8) .. controls (158.52,85.89) and (181.54,86.38) .. (190,87) ;
\draw [color={rgb, 255:red, 208; green, 2; blue, 27 }  ,draw opacity=1 ]   (128.95,123.33) .. controls (158.7,134.11) and (181.57,133.61) .. (190,133) ;
\draw [color={rgb, 255:red, 5; green, 21; blue, 198 }  ,draw opacity=1 ]   (72.96,93.34) .. controls (58.6,79.18) and (62.73,59.97) .. (170,50) ;
\draw [color={rgb, 255:red, 5; green, 21; blue, 198 }  ,draw opacity=1 ]   (100,110) .. controls (95.62,107.94) and (81.23,101.48) .. (72.96,93.34) ;
\draw [shift={(88.71,104.38)}, rotate = 210.25] [color={rgb, 255:red, 5; green, 21; blue, 198 }  ,draw opacity=1 ][line width=0.75]    (4.37,-1.96) .. controls (2.78,-0.92) and (1.32,-0.27) .. (0,0) .. controls (1.32,0.27) and (2.78,0.92) .. (4.37,1.96)   ;

\draw (126,131) node  [color={rgb, 255:red, 208; green, 2; blue, 27 }  ,opacity=1 ]  {$p_{R0}$};
\draw (133,107) node  [color={rgb, 255:red, 65; green, 117; blue, 5 }  ,opacity=1 ]  {$p_{1R}$};
\draw (93.94,87.2) node  [color={rgb, 255:red, 5; green, 21; blue, 198 }  ,opacity=1 ]  {$p_{01}$};

\end{tikzpicture}
    \caption{Incoming trajectories impacting the corner.}
    \label{fig:corner_directions}
\end{figure}    
For $h>h_\eps^w$, the mechanical form of $H$ determines the 4  momenta pairs  $(\pm p_1^w,\pm  P_2^\eps(p_1^w;h))$ associated with the corner point  $(q_1^w,q_2^w)$: \begin{equation}\label{eq:p2corner}
    P_2^\eps(p_1^w;h)=\sqrt{2(h-h_\eps^w)-(p_1^w)^2}, \quad p_1^w \in [0,\sqrt{2(h-h_\eps^w)} \eqqcolon p_{1,max}^{w}(h;\eps)],
\end{equation}
where  $h_\eps^w$ is defined by \eqref{def:heps}. Due to the step, on a given energy surface $h$, the corner can be reached by only 3 different incoming directions:  $p_{R0}=(-p_1^w,  P_2^\eps(p_1^w;h))$, $p_{01}=(p_1^w,-  P_2^\eps(p_1^w;h))$, $p_{1R}=(-p_1^w,-  P_2^\eps(p_1^w;h))$ illustrated in Figure \ref{fig:corner_directions}. These directions determine the initial conditions on $\Sigma_h$ that hit the corner point, namely, the three corner-singularity curves that are parametrized by $p_1^w$:
\begin{definition}\label{Def:corner_curve}The \emph{corner-singularity curves},  $\sigma_{ab}^\eps,\ ab \in \{R0,01,1R\}$ are the set of all initial conditions in $\Sigma_h$, which, under the step system flow,  hit the corner point in the direction $p_{ab}$  (and thus do not return to $\Sigma_h$).   
\end{definition}

\begin{figure}
    \centering
    \resizebox{0.9\columnwidth}{!}{\input{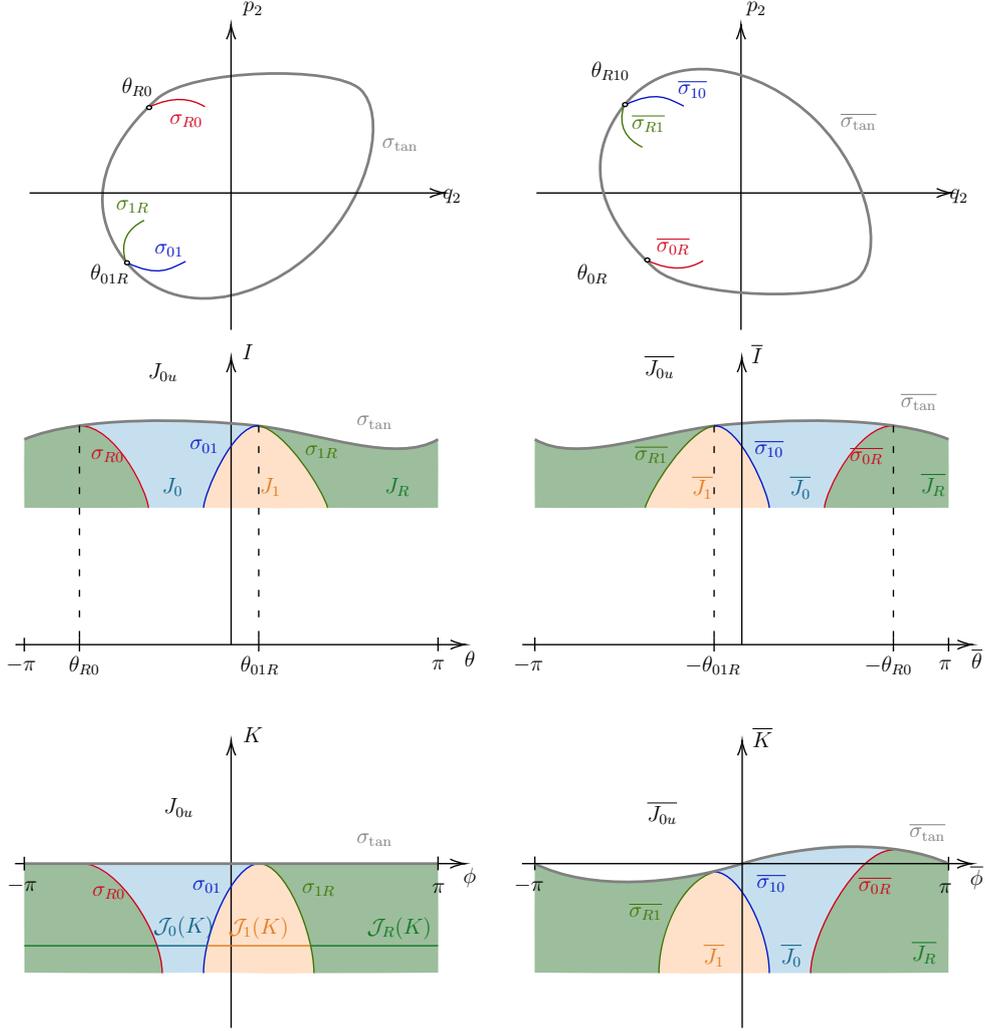}}
        \caption{The schematic structure of the regions  $J_a^\eps, \ a\in \{R,0,1\}$ and the singularity set $\sigma^\eps_{tan},\sigma^\eps_{ab}, \ ab\in \{R0,01,1R\}$ (left column) and their images (right column).  For clarity of presentation the superscript $\eps$ was removed from all labels. First row: the singularity set in the $(q_2,p_2)$ plane. Second row: the singularity set and the regions in the $(\theta,I)$  coordinates. The image of the regions are their symmetric reflections with respect to $\theta$: $ \bar J_a^\eps=\mathcal{R}_2 J_a^\eps, \ a\in \{R,0,1\}$, and the corresponding  corner-singularity curves also obey this symmetry.  Third row: the regions in the $(\phi,K)$  coordinates. The tangency circle $K=0$ maps to $\bar K= \eps f(\bar \phi)$.  The intervals $\mathcal{J}_a^\eps(k)$ of \eqref{eq:mathcalJdef} are the intersection of the circle $K=k$ with the region $J_a^\eps$.  Here, the image of the regions are their symmetric reflections with respect to $\phi$ with the additional shift by $-\eps f(\phi)$: $ \bar J_a^\eps=\mathcal{R} J_a, \ a\in \{R,0,1\}$, with $\mathcal{R}$ defined in Theorem \ref{thm:reversalinkphi}.  }
    \label{fig:structure_Sigma}
\end{figure}

\begin{definition}\label{Def:singularityset}  The \emph{corner singularity set} $\sigma_{cor}^\eps= \sigma_{R0}^\eps\cup \sigma_{01}^\eps\cup \sigma_{1R}^\eps$ includes all initial conditions on $\Sigma_h$ at which the return map is not defined.  The \emph{singularity set}  $\sigma^\eps= \sigma_{cor}^\eps\cup \sigma^\eps_{tan-R}$  includes all initial conditions in $\Sigma_h$ at which the dynamics of the return map is non-smooth. 
\end{definition}
The corner-singularity curves of  $\sigma_{cor}^\eps$, emanate from the first tangential curve, exactly at the borders of the tangent singularity segment, namely at  $(\theta_{R0}^\eps, \It^\eps(\theta_{R0}^\eps))$ and $(\theta_{01R}^\eps,\It^\eps(\theta_{01R}^\eps))$, see Figure \ref{fig:structure_Sigma}.

The singularity curves are utilized to subdivide a band around the tangential curve into the different dynamical regions, as shown schematically in Figure \ref{fig:structure_Sigma}: 
\begin{definition}
\label{Def:impactregions}
The \emph{tangential band}  $\mathcal{B}^\eps\coloneqq\{(\phi,K)|\phi \in [-\pi,\pi], K \in [-\Delta,\Delta]\}$ is composed of two layers\footnote{Notice that for $h>h^w_\eps$ there exists $\Delta_0>0$ such that $H_2(\It(h) \pm \Delta_0) +h_1^w < h $, so for any $(\phi,K) \in \mathcal{B}^\eps $,  for sufficiently small $\eps$,  there exists a real $p_1$ such that $H(\qw_1,p_1,q_2(\phi,K),p_2(\phi,K))=h$, namely,  $\mathcal{B}^\eps\subset \Sigma_h$.
 } , the  \emph{potentially impacting band}  $\mathcal{B}_{im}\coloneqq\{(\phi,K)|\phi \in [-\pi,\pi], K \in [-\Delta,0]\} $ and the \emph{non-impacting region} $J^\eps_{0u}\coloneqq\{(\phi,K)|\phi \in [-\pi,\pi], K \in [0,\Delta]\} $. The interior of the potentially impacting band, $\mathcal{B}_{im}$, is further divided  to  three  open sub-regions $J^\epsilon_R,J^\epsilon_0, J^\epsilon_1  $  that lie, correspondingly, in between the corner singularity curves $(\sigma_{1R}^\eps,\sigma_{R0}^\eps),(\sigma_{R0}^\eps,\sigma_{01}^\eps),(\sigma_{01}^\eps,\sigma_{1R}^\eps)$, namely, the vertical borders of the open region $J^\eps_b$ are $(\sigma_{ab}^\eps,\sigma_{bc}^\eps)$, where $abc \in \{01R, 1R0, R01\}$.  
\end{definition}
Notice that the regions are, in general, not invariant - their dynamical properties refer only to the trajectory segment until their first return to $\Sigma_h$.

\begin{theorem}\label{thm:cornersing}
For any $h>h_0^w$, there exists a width of the potentially impacting layer   $\Delta>0$, such that,  for sufficiently small $\eps $, the corner-singularity curves  $\sigma_{ab}^\eps,\ ab \in \{R0,01,1R\}$  are non-intersecting graphs of the normal coordinates  $\sigma_{ab}^\eps= \{(\phi_{ab}^\eps(K),K))|     K \in [-\Delta,0]  \}$. Near $K=0$ the dependence of $\phi_{ab}^\eps(K)$  is smooth  in  $\eps,\sqrt{-K}$  and is of the form: 
\begin{equation}\label{eq:cornersingbordrho}
        \phi^{\eps}_{ab}(\sqrt{-K})=  \phi^{\eps}_{ab}(0)+ A_{ab}\lambda \sqrt{-K} + \tau_{ab} K + O(\eps \sqrt{-K},(-K)^{3/2})  
\end{equation}
where  $ A_{R0}=A_{1R}=1,\ A_{01}=-1$, and   $\tau_{1R}=\tau_{01}=\frac{1}{2}(-\tau_0+\tau_1),\tau_{R0}=-\frac{1}{2}(\tau_0+\tau_1)$.
The curves $\sigma_{01}^\eps$ and $\sigma_{1R}^\eps$  emanate from the same angle on the tangency line:   $\phi^{\eps}_{01}(0)=\phi^{\eps}_{1R}(0)$ and, for all the curves the emanating corner-singularity angles depend smoothly on $\eps$: $\phi^{\eps}_{ab}(0)=  \phi^{0}_{ab}(0)+O(\eps)$. To leading order in $\eps$, the emanating angles are 
\begin{equation*}
    \phi_{01}^0(0)=  \tw -\frac{\Omega_0}{2},\quad \phi^0_{R0} (0)=-\tw -\frac{\Omega_0}{2}.
\end{equation*}
where $\Omega_0,\tau_0,\tw,\tau_1,\lambda$ are defined by  \eqref{eq:omega0ofI},\eqref{eq:deftau0},\eqref{eq:thw},\eqref{eq:deftau1},\eqref{eq:deflambdah}.
\end{theorem}
The regions in between the corner singularity curves correspond to different dynamics, see Figure~\ref{fig:structure_Sigma}:
\begin{theorem}\label{thm:regionsJdyn}
The open regions $(J^\epsilon_R ,J^\epsilon_0, J^\epsilon_1)$ divide, in this order on the $(\theta,I)$ cylinder, the  potentially impacting band $\mathcal{B}_{im}\setminus \{\sigma_{01}^{\eps}\cup \sigma_{1R}^{\eps}\cup \sigma_{R0}^{\eps}\}$. Trajectories segments emanating from the region $J_R^\eps$ hit the right wall of the step once before returning to  $\Sigma_h$, those  emanating from  $J_1^\eps$  hit the upper wall of the step once before returning, and those belonging to $J_0^\eps$  hover above the step  without hitting it before returning to  $\Sigma_h$. 
\end{theorem}

\subsection{Derivation of the return map}

Denote the reversing and cyclic permutations of the symbols $a,b,c\in\{0,1,R\}$  by $p_{rev}(abc)=(cba)$ and  $p_{cyc}(abc)=(cab)$, where, for example $(abc)=(R01)$.  

The time reversal symmetry of $\mathcal{F}_\eps$ in the $(\theta,I)$ coordinates is simply the reflection $\mathcal{R}_2$, whereas, in the normal coordinates,  $\mathcal{F}_\eps(\phi,K) \coloneqq S  \mathcal{F}_\eps {S}^{-1}(\phi,K) =S  \mathcal{F}_\eps (\theta,I) $, where, with a slight abuse of notation, we use the same symbol for the map in the $(\theta,I)$ and  in the $(\phi,K)$ coordinates) we have an $\eps$-dependent time reversal symmetry (see more details and proofs in Section \ref{sec:timereversal}):
\begin{theorem}\label{thm:reversalinkphi}
The time reversal symmetry in the normal coordinates is   $\mathcal{R}_\eps\coloneqq\mathcal{R}(\phi,K)=(-\phi,K-\eps f(\phi)) $ where  $f$ is the odd function defined by \eqref{eq:defoff} namely,   $\mathcal{R}_\eps    \mathcal{ F}_\eps (\phi,K)  = {\mathcal{ F}_\eps}^{-1} \mathcal{R}_\eps(\phi,K)$.
\end{theorem}
As demonstrated in Figure~\ref{fig:structure_Sigma},  the time reversal symmetry implies: 
\begin{theorem} \label{thm:reverseJs} The return map reverses the order of the regions on the cylinder from  $(J^\epsilon_R,J^\epsilon_0, J^\epsilon_1)$  to  $( \bar J^\epsilon_1,\bar J^\epsilon_0 ,\bar J^\epsilon_R) $.   Moreover, the left and right boundaries of  $\bar J^\epsilon_b$, given by $\bar \sigma_{ab}^\eps$ and  $\bar \sigma_{bc}^\eps$ respectively, with $(abc) \in p_{cyc}(10R)$  are given by the time reversal symmetry of the corresponding boundaries of $ J^\epsilon_b$, namely, the symmetric pairs of  $ \sigma_{ba}^\eps$  and  $\sigma_{cb}^\eps$, respectively (where the symmetric pair is defined by  $\mathcal{R}_2$ for $\mathcal{F}_\eps (\theta,I) $  and by $\mathcal{R}_\eps$ for $\mathcal{F}_\eps(\phi,K)$).
\end{theorem}

Next we establish that in the tangential band  $ \mathcal{B}^\eps $,   namely, for $|K|<\Delta(h)$,  the return map to $\Sigma_h$  becomes a family of perturbed interval exchange transformations. 

For any $k\ge 0$ denote by $\mathcal{J}_{0}(k)$ the circle $\{\phi \in [-\pi,\pi]\}$  whereas for any $k< 0$  denote by $\mathcal{J}^\eps_{b}(k)$  the open interval in $\phi$ that corresponds to the intersection of the circle $K=k $ with the region $J_b^\eps$, see last row in Figure \ref{fig:structure_Sigma}:
\begin{equation}\label{eq:mathcalJdef}
\mathcal{J}^\eps_{b}(K)\coloneqq  \begin{cases}
     [-\pi,\pi] & b=0, K \ge 0 \\
     \emptyset & b\neq0, K \ge 0 \\
     (\phi_{ab}^\eps(K),\phi_{bc}^\eps(K)) & abc \in \{01R,1R0,R01\} ,K<0.
\end{cases} 
\end{equation}
Thus, the open regions $ J_b^\eps \subset \mathcal{B}^\eps, \ b \in \{0,1,R,0u \}$ are parametrized by the $K$ dependent $\phi$ intervals; Any $(\phi,K)=S^\eps(\theta,I) \in\mathcal{B^\eps} \setminus \sigma_{cor}^{\eps}$  belongs to a unique region $J_b^\eps, b \in \{0u,0,1,R\}$ and, for $K\ne 0$,  $(\phi,K) \in J_b^\eps$  if and only if $\phi \in \mathcal{J}_b^\eps(K) $ (where, for $K \ge 0$, to simplify notation, we identify $ \mathcal{J}_{0u}^\eps(K)$ and  $ \mathcal{J}_{0}^\eps(K) $).  The tangency circle at $K=0$ is  $\mathcal{J}_{0}(0)$ (so $\mathcal{J}_{0}(0)= \lim_{K \searrow 0} \mathcal{J}_{0}(K) =\lim_{K \nearrow 0} \left(\mathcal{J}_{0}(K) \cup\mathcal{J}_{R}(K)\right)\cup  \{ \phi_{0R}^\eps(0)\}\cup  \{ \phi_{R0}^\eps(0)\}$). The proof of  Theorem \ref{thm:cornersing} implies that $|\mathcal{J}^\eps_{R}(K)|$ is smooth in $\eps,K$ for $K \in [-\Delta,0]$ whereas $|\mathcal{J}^\eps_{0}(K)|$ and  $|\mathcal{J}^\eps_{1}(K)|$ are smooth in $\eps,\sqrt{-K}$ for $K \in [-\Delta,0]$.

For $K \le 0$ denote the rotation, mod $2\pi$,  of the left boundary of the interval $\mathcal{J}_{0}(K) $ by  $ \Omega_\eps(K)$:
\begin{equation}\label{eq:Omegaepsdef}
    \Omega_\eps(K) \coloneqq|[\phi_{R0}^\eps(K),\bar \phi_{10}^\eps(K)]|_c=|- \phi^{\eps}_{01}(\sqrt{-K})-\phi^{\eps}_{R0}(\sqrt{-K})|_c. 
\end{equation}
Since the dynamics in $\mathcal{J}_{0}(K) $ is smooth, for sufficiently small $\eps$, $ \Omega_\eps(K)$ is smooth in  $K \in [-\Delta,0]$ and in $\eps$, $ \Omega_\eps(K)=\Omega_0(\It(h)+K))+O(\eps)$ and $ \Omega_\eps(K)$ can be extended smoothly to $K\ge 0$ so that its leading order term in $\eps$ coincides with $\Omega_0(\It(h)+K))$.   Let 
 \begin{equation}\label{eq:omegaofadef}
     \omega^\eps_{a}(K) \coloneqq \left\{\begin{array}{ll}
        \Omega_\eps(K)  &  a=0 \\
      \Omega_\eps(K)+ |\mathcal{J}^\eps_{R}(K)|      &  a =1 \\
         \Omega_\eps(K) - |\mathcal{J}^\eps_{1}(K)|  &    a =R.
     \end{array}\right.
 \end{equation}

The time reversal symmetry implies that (see Section \ref{sec:timereversal})
\begin{equation}\label{eq:phi1rsymmetric}
    \phi_{1R}^\eps(K) = \pi-\frac{1}{2}\Omega_\eps(K)-\frac{1}{2} |\mathcal{J}^\eps_{R}(K)|+\frac{1}{2}|\mathcal{J}^\eps_{1}(K)|.
\end{equation}  
We are now ready to state the main Theorem:
\begin{theorem}\label{thm:returnmap}
For $h>h^w$, for sufficiently small $\eps$ and $(\phi,K) \in \mathcal{B^\eps}\setminus \sigma^\eps_{cor}$, the local return map $\mathcal{F}_\eps:\Sigma_h\to\Sigma_h$ of the step perturbed system near the first tangential curve  $\sigma^\eps_{tan}$ is of the following piecewise smooth symplectic form\begin{equation} \label{eq:returnmapfull}
    \mathcal{ F}_\eps:
\begin{cases}
    \bar{K} &= K + \eps f(\bar{\phi};\eps) + G_{a,K}(\phi,K,\eps),\\
    \bar{\phi} &= \phi +  \omega^\eps_a(K)+   G_{a,\phi}(\phi,K,\eps),\quad \ \phi\in \mathcal{J}^\eps_a(K),\ a\in\{0,1,R\},
\end{cases}
\end{equation} where $(\phi,K)$ are the normal coordinates \eqref{eq:kphidef},  $f(\phi;\eps)$ is the odd function \eqref{eq:defoff}, the intervals $\mathcal{J}^\eps_a(K), a\in\{0,1,R\}$ are defined by  \eqref{eq:mathcalJdef} and the translation vector is defined by \eqref{eq:omegaofadef}. The  remainder terms $G_{a,K}(\phi,K,\eps),G_{a,\phi}(\phi,K,\eps)$ are small in $\eps,\sqrt{-K}$ as detailed in Appendix \ref{sec:proofreturnmap} (see \eqref{eq:gcorner}, \eqref{eq:gcornerphi} there).
\end{theorem}
The proof, in Appendix  \ref{sec:proofreturnmap}, employs regular perturbation methods as in  \cite{PnueliRomKedarIntegrability}  for approximating the trajectories for $\phi\in \mathcal{J}^\eps_a(K), a\in\{0,1,0u\}$, and uses, additionally, the fact that $|\mathcal{J}^\eps_1(K)|$ is small near the onset. For  $\phi\in \mathcal{J}^\eps_R(K)$  the near-tangent analysis of \cite{PnueliRomKedarTangency} is utilized. 

\subsection{Truncated models of the return map}

We define the following two-parameters family of piecewise smooth symplectic maps:
\begin{equation}
    \label{eq:trunceps1eps2}
\mathbf{F}_{\eps_1,\eps_2}^{trun}:
\begin{cases}
    \bar{K} &= K + \eps_1 f(\bar{\phi})\\
    \bar{\phi} &= \phi + \omega^{\eps_2}_a(K),\ \ \ \ \phi \in \mathcal{J}_a^{\eps_2}(K),
\end{cases}
\end{equation}
where $f$ is an odd function, and the translation vector and the intervals are defined by   \eqref{eq:omegaofadef} and by \eqref{eq:phi1rsymmetric},  for any  smooth function  $ \Omega_{\eps_2}(K)$ and intervals $\mathcal{J}^{\eps_2}_{R}(K), \mathcal{J}^{\eps_2}_{1}(K)$ satisfying $\{|\mathcal{J}^{\eps_2}_{R}(K)|+|\mathcal{J}^{\eps_2}_{1}(K)|\}_{K \le 0}< 2\pi$, $\mathcal{J}^{\eps_2}_{R}(K)|_{K>0}=\emptyset$, $\mathcal{J}^{\eps_2}_{1}(K)|_{K\ge0}=\emptyset$.   
This map is reversible, with the time reversal  symmetry $\mathcal{R}_\eps$ of Theorem \ref{thm:reversalinkphi}
.

The map $\mathbf{F}_{\eps_1,\eps_2}^{trun}$   is a composition of  $\mathcal{T}_{\eps_2}$:  

\begin{equation}
    \label{eq:taueps2}
\mathcal{T}_{\eps_2}:
\begin{cases}
    \bar{K} &= K \\
    \bar{\phi} &= \phi +  \omega^{\eps_2}_a(K),\quad \ \phi\in \mathcal{J}^{\eps_2}_a(K),\ a\in\{0,1,R\},
\end{cases}
\end{equation}
a  family of IEM on the cylinder, near the transition between a rotation to a $3$-IEM,  and the smooth near identity symplectic transformation: \begin{equation}\label{eq:defP}
     P_{\eps_1}(\phi,K)=(\phi,K+ \eps_1 f(\phi)).
 \end{equation}
More generally, one can take any family of interval exchange maps $\mathcal{T}(\phi,K)$ and any time periodic function with zero mean, $f(\phi)$, and study the resulting perturbed dynamics of the area preserving, piecewise symplectic invertible map, $\mathcal{P}_{\eps} \circ \mathcal{T} $,  see~\cite{PaziRK}. 

Next, we show that the family of truncated maps,   $\mathbf{F}_{\eps_1,\eps_2}^{trun}$ of \eqref{eq:trunceps1eps2} satisfies the same time reversal symmetry as the HIS return map.  

\begin{definition}
    The family $\mathcal{T}_{\eps}$  is called $\phi$-symmetric if  $\mathcal{R}_2 \mathcal{T}_\eps= \mathcal{T}_\eps ^{-1}\mathcal{R}_2$.  
\end{definition}

\begin{lemma}
For any three functions $ \Omega_\eps(K),|\mathcal{J}^\eps_{R}(K)|, |\mathcal{J}^\eps_{1}(K)|$ satisfying $|\mathcal{J}^\eps_{R}(K)|+|\mathcal{J}^\eps_{1}(K)|< 2\pi$  the map $\mathcal{T}_{\eps}$  with  $\omega^{\eps}_{J_a}(K)$ of the form \eqref{eq:omegaofadef} and the intervals $\mathcal{J}^{\eps}_1(K)$, $\mathcal{J}^{\eps}_R(K)$, $\mathcal{J}^{\eps}_0(K)$ placed in this order on each circle with $\phi_{1R}^\eps(K)$ given by \eqref{eq:phi1rsymmetric} is  $\phi$-symmetric.
\end{lemma}
\begin{proof} By  \eqref{eq:phi1rsymmetric}
 the left boundaries of the three intervals, $(\mathcal{J}^{\eps}_1(K),\mathcal{J}^{\eps}_R(K),\mathcal{J}^{\eps}_0(K))$, are  (mod $2\pi$): $\phi_{01}^\eps(K)= \pi-\frac{1}{2} \Omega_\eps(K)-\frac{1}{2} |\mathcal{J}^\eps_{R}(K)|-\frac{1}{2}|\mathcal{J}^\eps_{1}(K)|$, $   \phi_{1R}^\eps(K)=\pi-\frac{1}{2} \Omega_\eps(K)-\frac{1}{2} |\mathcal{J}^\eps_{R}(K)|+\frac{1}{2} |\mathcal{J}^\eps_{1}(K)|$ , $\phi_{R0}^\eps(K)= \pi-\frac{1}{2} \Omega_\eps(K)+\frac{1}{2} |\mathcal{J}^\eps_{R}(K)|+\frac{1}{2}|\mathcal{J}^\eps_{1}(K)|$.

 By   \eqref{eq:omegaofadef}, direct computations shows that indeed
 $ \bar\phi_{10}^\eps(K)= \phi_{R0}^\eps(K)+\Omega_\eps(K)=\pi+\frac{1}{2}\Omega_\eps(K)+\frac{1}{2} |\mathcal{J}^\eps_{R}(K)| +\frac{1}{2}|\mathcal{J}^\eps_{1}(K)|= -\phi_{01}^\eps(K) + 2\pi$ and similarly 
 $ \bar\phi_{R1}^\eps(K)= \phi_{01}^\eps(K)+\Omega_\eps(K)+|\mathcal{J}^\eps_{R}(K)| =\pi+\frac{1}{2}\Omega_\eps(K)+\frac{1}{2} |\mathcal{J}^\eps_{R}(K)| -\frac{1}{2}|\mathcal{J}^\eps_{1}(K)|= -\phi_{1R}^\eps(K) + 2\pi$ and similarly
 $ \bar\phi_{0R}^\eps(K)= \phi_{1R}^\eps(K)+\Omega_\eps(K)-|\mathcal{J}^\eps_{1}(K)| =\pi-\frac{1}{2} |\mathcal{J}^\eps_{R}(K)|+\frac{1}{2}\Omega_\eps(K)-\frac{1}{2} |\mathcal{J}^\eps_{1}(K)|= -\phi_{R0}^\eps(K) + 2\pi$. 
 
 So, we verified that for such a family of symmetric IEM the intervals $\mathcal{\bar J}^\eps_R(K),\mathcal{\bar J}^\eps_0(K),\mathcal{\bar J}^\eps_1(K)$ are the reflections in $\phi$ of the intervals  $\mathcal{ J}^\eps_R(K),\mathcal{ J}^\eps_0(K),\mathcal{ J}^\eps_1(K)$, or, in formula   $\bar \phi_{ab}^\eps(K)= \mathcal{R}_2 \phi_{ba}^\eps(K)$. Notice that the reflection is interval wise and not point wise. Yet, this property implies that if $\theta \in \mathcal{J}^{\eps}_a(K)$ then $-\theta \in \mathcal{\bar J}^{\eps}_a(K)$, and hence   $   \mathcal{R}_2 \mathcal{T}_\eps= \mathcal{T}_\eps ^{-1}\mathcal{R}_2$.
\end{proof}

In general, any  IEM which reverses the intervals order on the circle is symmetric with respect to the interval mid-point, and thus, a family of such IEMs on a circle always has a symmetry line which is $K$-dependent. The choice  \eqref{eq:phi1rsymmetric} shifts this symmetry line to the origin. See \cite{PaziRK} for a more general settings.  

Recall that here $f(\phi)$ is periodic and odd, so $   \mathcal{P}_\eps$ also admits a time reversal symmetry with respect to a reflection in $\phi$:  $\mathcal {P}^{-1}_\eps(\phi,K)=\mathcal{R}_2 \circ \mathcal{P}_\eps(-\phi,K)=\mathcal{R}_2 \circ \mathcal{P}_\eps\mathcal\circ \mathcal{R}_2 (\phi,K)$ ).  Hence:
\begin{lemma}\label{lem:truncreturnmap}
 Provided  $\mathcal{T}_{\eps_2}$  is a $\phi$-symmetric family of IEM and $f$ is odd, for any $(\eps,\eps_2)$ the truncated map  $\mathbf{F}_{\eps,\eps_2}^{trun}$ is piecewise symplectic and obeys the same time reversal symmetry as the return map $\mathcal{F_\eps}$:  $(\mathbf{F}_{\eps,\eps_2})^{-1}_\eps(\phi,K)=\mathcal{R}_\eps \circ \mathbf{F}_{\eps,\eps_2}\circ \mathcal{R}_\eps $ where    $\mathcal{R}_\eps=\mathcal{R}(\phi,K)=(-\phi,K-\eps f(\phi)) $ of Theorem \ref{thm:reversalinkphi}. Thus, the singularity lines of the map and of its inverse are also related, for any $\eps_2$ by $\mathcal{R}_\eps \phi_{ab}^{\eps,\eps_2} =\bar \phi_{ba}^{\eps,\eps_2} $.
\end{lemma}

Setting in \eqref{eq:trunceps1eps2}  $\eps_1=\eps_2=\eps$, choosing $f$ to be defined by \eqref{eq:defoff},  and the intervals to be defined by the corner singularity curves of Theorem  \ref{thm:cornersing}, the return map  \eqref{eq:returnmapfull} is of the form: $\mathcal{F}_\eps=\mathbf{F}_{\eps,\eps}^{trun}+(G_{a,K},G_{a,\phi})$, namely, with this choice, $\mathbf{F}_{\eps,\eps}^{trun}$ is a truncation of the map  \eqref{eq:returnmapfull}. 
Since $ \mathbf{F}_{\eps_1,\eps_2}^{trun} =\mathcal{P}_{\eps_1} \circ \mathcal{T}_{\eps_2} $  is a composition of a smooth vertical shear ($\mathcal{P}_{\eps_1} $)  with a family of  horizontal circle exchange maps, and the latter is a discontinuous generalization of a horizontal shear,  
an analogous result to \cite{berger2024generators} for the return maps of HIS  is needed for justifying this truncation. The truncated map  $\mathbf{F}_{\eps,\eps}^{trun}$ may be computed to first order in $\eps, K$ by perturbation methods. A simpler model of the same symmetric form, which does not require further computations, is the map $\mathbf{F}_\eps\coloneqq\mathbf{F}_{\eps,0}^{trun}$. Since the singularity lines of  $\mathbf{F}_{\eps,0}^{trun}$ can be found explicitly this map is more convenient for numerical investigation and we propose it is a "good enough" model for studying numerically the dynamics.

\subsection{Hovering dynamics in the model map}
We study the map  $\mathbf{F}_\eps\coloneqq\mathbf{F}_{\eps,0}^{trun}$ with the forcing function  $f(\bar \phi;\eps)=\sin (\bar \phi)$ (this choice may be thought of as the first Fourier mode of $f(\bar \phi;\eps)$ of \eqref{eq:defoff}) and the intervals  $ \mathcal{J}_a(K)$ which are the leading order approximations in $(\eps,K)$ to $ \mathcal{J}_a^\eps(K)$:
\begin{equation}
     \label{eq:truncatedmap}
\mathbf{F}_\eps:
\begin{cases}
    \bar{K} &= K + \eps \sin(\bar{\phi})\\
    \bar{\phi} &= \phi + \omega_{a}(K) \ \ \phi \in \mathcal{J}_a(K).
\end{cases}
\end{equation}
Then  $\omega_{a}(K)$ is the resulting  leading order approximation in $K$ to $\omega^{\eps_2=0}_{a}(K)$ of \eqref{eq:omegaofadef}:
\begin{equation}\label{eq:omegaofalead}
     \omega_a(K)= \left\{\begin{array}{ll}
       \Omega_0+\tau_0 K &  a =0\\
         \Omega_0 + \tau_0 K+Heavi(-K)\cdot(\tau_1K+2(\pi-\tw))       &  a =1\\
          \Omega_0+\tau_0 K- 2\lambda\sqrt{\max(0,-K)}   &    a =R.
     \end{array}\right. 
 \end{equation}
 where $Heavi(x)$ is the Heaviside function.
All the parameters are defined in section \ref{sec:parametersdef}. We show in section \ref{sec:truncatedmapnum} that it is sufficient to study  the map for the parameter set:
\begin{equation}
    \label{eq:parameterset}\mathcal{P} \coloneqq \{(\eps,\Omega_0,\tau_0,\tau_1,\lambda,\tw)|\eps \in R,\Omega_0 \in [0, 2\pi), \tau_0\in \{\pm 1,0\},\tau_1 \in R ,\lambda>0 , \tw \in (0,\pi)\} 
\end{equation} and that for this range of parameters the map is well defined for $(\phi,K)\in S^1\times [-\Delta , \Delta], \Delta <\KMIN(\lambda,\tw,\tau_1)$ with   $\KMIN(\lambda,\tw,\tau_1)$  given by  \eqref{eq:kmindef}. 
Appendix \ref{sec:calculationofparam}  includes explicit expressions of these, including their dependence on $h$, for the quadratic and Tan potentials.

For positive $K$ the map  $\mathbf{F}_\eps$ is simply the scaled and shifted standard map,  $\mathbf{F}^{st}_\eps$:
$$ \label{eq:shiftstandardmap}
\mathbf{F}^{st}_\eps:
\begin{cases}
    \bar{K} &= K + \eps \sin(\bar{\phi})\\
    \bar{\phi} &= \phi +  \Omega_0+\tau_0 K. 
\end{cases}
$$
Orbits of $\mathbf{F}^{st}_\eps$ that reside only in the non-impacting regions ($J_0\cup J_{0u}$) are non-impacting orbits of $\mathbf{F}_\eps$.  In particular, for $|\eps|<0.97$, rotational invariant curves of $\mathbf{F}^{st}_\eps$ may fully reside in the upper half plane,  namely in $J_{0u}$, or, there may be curves that reside in $J_0 \cup J_{u0}$. 

Let  $k_\eps^{u0}(\phi;\Omega_0,\tau_0), \phi \in S^1$ denote the  infimum of the invariant curves residing fully in $J_{u0}$ (so   $\forall \phi, k_\eps^{u0}(\phi;\Omega_0,\tau_0)\ge 0$), with rotation number $\rho_{u0}(\eps;\Omega_0,\tau_0)$. Then, the region above it is invariant. It corresponds to orbits that never impact the step nor hover above it. In it, the map is identical to that of the standard map. Hence, switching the sign of $\eps$ does not alter the qualitative nature of the dynamics there, $k_{-\eps}^{u0}(\phi;\Omega_0,\tau_0)=k_{\eps}^{u0}(\phi+\pi;\Omega_0,\tau_0)$ and  $\rho^{u0}_\eps(\Omega_0,\tau_0)=\rho^{u0}_{-\eps}(\Omega_0,\tau_0)$.

The hovering case emerges when there are additional invariant curves residing in  $J_0 \cup J_{u0}$ that lie below $k_\eps^{u0}(\phi;\Omega_0,\tau_0)$. Denote the infimum of them by  $k_\eps^{hover}(\phi;\Omega_0,\tau_0,\tau_1,\lambda,\tw)$. The band  between these two curves is an invariant set in which the dynamics is smooth. The non-resonant orbits in this smooth band visit both $J_0$ and $J_{u0}$ and do not impact the step. The corresponding orbits of the oscillators-step flow both hover above the step and alongside the step without ever hitting it.  
In particular, in this case, there are initial conditions in $J_0$ which, under the unperturbed dynamics do impact the step after a finite number of iterations, whereas in the perturbed dynamics they never impact the wall:
\begin{definition}
The hovering set $\HOV(\eps)$  consists of all initial conditions $(\phi,K)$ that impact the step by the unperturbed dynamics ( $\exists n \ F_0 ^n(\phi,K) \notin J_0\cup J_{0u}$) and are non-impacting by the perturbed dynamics ($\forall n \ F_\eps^n(\phi,K) \in J_0\cup J_{0u}$) .  
\end{definition}
The condition  $\forall n \ F_\eps^n(\phi,K) \in J_0\cup J_{0u}$  implies that $(\phi,K)\in J_0\cup J_{0u}$, and the condition $\exists n \ F_0 ^n(\phi,K) \notin J_0\cup J_{0u}$ implies that $(\phi,K)\notin  J_{0u}$, namely the hovering set is a subset of $J_0$ for which the perturbed dynamics is simpler than the unperturbed one.

In Section \ref{sec:truncinvcurves} we establish:
\begin{theorem}
    \label{thm:hovering}
Given a  Diophantine rotation number $\frac{\Omega_0}{2\pi}\in(0,1)$, $\lambda>0,\tau_0 \ne 0 $ and  $\tw  \in (0,\pi)$, there exists  $\eps_c(\Omega_0,\tau_0,\tw)>0$ such that the hovering set of  $\mathbf{F}_\eps$ is of positive measure for all $\eps \in (0,\eps_c(\Omega_0,\tau_0,\tw)) $,  this measure is monotonically increasing in $\eps$  in this interval, and, for non-positive values of  $\eps$, $\eps \in (-\eps_c(\Omega_0,\tau_0,\tw),0]$, the hovering set is empty. 
\end{theorem}
We then find a one parameter family of maps with a critical curve of a given rotation number (up to order $\eps^2$). This allows to estimate the monotone dependence of their hovering set on the parameter:
\begin{theorem} \label{thm:hoveringc}  Given a  $0<\nu<\frac{1}{2}$ and a Diophantine rotation number $\frac{\Omega_G}{2\pi} \in (\nu,1-\nu)$, for any $|c|<|\tau_0|$, for sufficiently small $\eps>0$, the critical curve of the map $F_{\eps}$ with the parameters $\Omega_0=\Omega_G-\frac{c}{2}\eps,\tw=\cos^{-1}(\frac{c}{ \tau_0}\sin( \frac{\Omega_G }{2}-\frac{c\eps }{4})),\tau_0,\tau_1,\lambda$ has rotation number which is $O(\eps^2)$ close to $\Omega_G$ and its minimum is  $O(\eps^2)$ close to
$K(c)=- \frac{\eps}{2\sin \frac{\Omega_0 }{2}} (1-\frac{c}{\tau_0}\sin \frac{\Omega_0 }{2})$.  In particular, the hovering set has a positive measure along this family and  is monotone in $c/\tau_0$. 
\end{theorem}

Below the critical curve  $k_\eps^{hover}(\phi;\Omega_0,\tau_0)$, mixed dynamics, with stability islands and chaotic dynamics, of  impacting and non-impacting trajectory segments arise. Notice that for a Diophantine $\frac{\Omega_0}{2\pi}$, for sufficiently small $\eps$,  $\mathbf{F}^{st}_\eps$ has no resonant islands of small period near $K=0$, so the hovering set does not include resonances that lie below   $k_\eps^{hover}(\phi;\Omega_0,\tau_0)$ (in contrast, for larger $\eps$ or near-rational $\Omega_0$, the resonant islands of the standard map near $K=0$ may reside in $J_0 \cup  J_{u0}$ independent of the location or existence of $k_\eps^{hover}(\phi;\Omega_0,\tau_0,\tau_1,\lambda,\tw)$). 

The destruction of   $k_\eps^{hover}(\phi;\Omega_0,\tau_0)$ proves that the maps  $\mathbf{F}_\eps$  and  $\mathbf{F}_{-\eps}$ have different dynamics below $k_{\eps}^{u0}(\phi;\Omega_0,\tau_0)$ (and identical dynamics, up to a shift by $\pi$, above this curve, 
as $\mathbf{F}^{st}_\eps(\phi,K)=\mathbf{F}^{st}_{-\eps}(\tilde \phi=\phi+\pi,K)$ and thus $k_{-\eps}^{u0}(\phi;\Omega_0,\tau_0)=k_{\eps}^{u0}(\phi+\pi;\Omega_0,\tau_0)$).

Section \ref{sec:truncinvcurves} includes the proofs of the above theorems and Section \ref{sec:truncatedmapnumact} includes numerical simulations demonstrating the existence and destruction of the hovering set (with $\eps$ as large as 0.92). The numerical simulations also demonstrate that the chaotic and resonance zones below the critical curve are much larger and visible when compared to the dynamics above it, see also \cite{PaziRK}.

\section{The model of the truncated map} \label{sec:truncatedmapnum}

We establish first that the map $\mathbf{F}_\eps$ of  \eqref{eq:truncatedmap}  is well defined in a finite band around the tangency circle, and that it is sufficient to consider the parameter set $\mathcal{P}$ of  \eqref{eq:parameterset}. Then, in Section \ref{sec:standardmap}, we find the location of the extrema of invariant curves of the shifted standard map $\mathbf{F}^{st}_\eps(\phi,K)$, and in Section \ref{sec:truncinvcurves} we prove Theorems \ref{thm:hovering} and \ref{thm:hoveringc}. In Section \ref{sec:truncatedmapnumact} we present numerical simulations of the map.

The truncated map model of \eqref{eq:truncatedmap}, $\mathbf{F}_\eps:
    \bar{K} = K + \eps \sin(\bar{\phi}), 
    \bar{\phi} = \phi + \omega_{a}(K), \  \phi \in \mathcal{J}_a(K)$,  with $ \mathcal{J}_a(K)$ corresponding to the unperturbed intervals: 
\begin{eqnarray}\label{eq:Jdeftrunc}
K \ge 0 : \ & &  \nonumber\\
    \mathcal{J}_{u0}(K) &=& [-\pi,\pi] ,  \nonumber\\
    & &  \\
K < 0 : \ && \nonumber\\
    \mathcal{J}_0(K)&=& \left(-\tw-\frac{1}{2}(\Omega_0  +(\tau_0-\tau_1) K+\lambda\sqrt{-K}),\tw-\frac{1}{2}(\Omega_0+(\tau_0+\tau_1) K)-\lambda\sqrt{-K}\right),  \nonumber\\
     \mathcal{J}_1(K)&=&\left(\tw-\frac{1}{2}(\Omega_0+(\tau_0+\tau_1) K)-\lambda\sqrt{-K},\tw-\frac{1}{2}(\Omega_0+(\tau_0+\tau_1) K)+\lambda\sqrt{-K}\right),  \nonumber\\
    \mathcal{J}_R(K)&=&\left(\tw-\frac{1}{2}(\Omega_0+(\tau_0+\tau_1) K)+\lambda\sqrt{-K},2\pi-\tw-\frac{1}{2}(\Omega_0  +(\tau_0-\tau_1) K)+\lambda\sqrt{-K}\right),  \nonumber
\end{eqnarray}
and the corresponding truncated translation vector   \eqref{eq:omegaofalead}, is well defined for $K \in (-\KMIN(\lambda,\tw,\tau_1),\KMIN(\lambda,\tw,\tau_1))$ for any $\eps,\Omega_0,\tau_0,\tau_1$ and $\tw \in (0,\pi),\lambda>0$ where:
\begin{equation} \label{eq:kmindef}
 \KMIN(\lambda,\tw,\tau_1) \coloneqq \begin{cases}
         -\min((\pi/ \lambda)^2 ,  2(\pi-\tw)/\tau_1) & \tau_1>0 \\
         -\min((\pi/ \lambda)^2 ,  -2\tw/|\tau_1|) & \tau_1<0.
    \end{cases}
\end{equation}
Indeed, since $(|\mathcal{J}_0(K)|,|\mathcal{J}_1(K)|,|\mathcal{J}_R(K)|) = (2 \tw -2\lambda\sqrt{-K}-\tau_1 K,\ 2\lambda\sqrt{-K},\ 2(\pi-\tw)+\tau_1 K)$ the intervals lengths are positive and smaller than $2\pi$ for this range.

Notice that    
$\mathbf{F}_\eps(\phi,K;\Omega_0,\tau_0,\tau_1,\lambda)=\mathbf{F}_{-\eps}(\phi+\pi,K;\Omega_0,\tau_0,\tau_1,\lambda)$, that
$\mathbf{F}_\eps(\phi,K;\Omega_0,\tau_0,\tau_1,\lambda)=\mathbf{F}_{\eps}(\phi,K;\Omega_0+4\pi,\tau_0,\tau_1,\lambda)$
 and that for $\tau_0 \ne 0$, the change of variables $K \rightarrow |\tau_0| K$ produces the same map with properly rescaled parameters:  $\eps \rightarrow |\tau_0|\eps, \tau_0  \rightarrow \mathrm{sign}\tau_0,  \tau_1  \rightarrow \tau_1/|\tau_0|,  \lambda  \rightarrow \lambda/\sqrt{|\tau_0}|$.  Thus, by rescaling, it is sufficient to consider the parameters in $\mathcal{P}$ of \eqref{eq:parameterset}.

\begin{remark}
The map obeys the time reversal symmetry  $\mathcal{R}_\eps    \mathbf{F}_\eps  = {\mathbf{F}_\eps}^{-1} \mathcal{R}_\eps$ with $\mathcal{R}_\eps(\phi,K)=(-\phi,K-\eps \sin(\phi))$, see Lemma \ref{lem:truncreturnmap}. Replacing the $\sin$ function by any smooth periodic function produces a well defined invertible piecewise symplectic map. Yet, in general, if the function is not odd, such a map does not necessarily possess a time reversal symmetry. For example, the map in which the term  $\sin  \bar \phi$ is replaced by $\sin (\bar \phi +\phi^*)$  obeys the time reversal symmetry $\mathcal{R}_\eps$ if and only if $(\phi^* \in \{0,\pi\},  \mod{2 \pi})$, and taking $\phi^*=\pi$ is the same as reversing the sign of $\epsilon$. 
\end{remark}

\subsection{Invariant curves of the shifted standard map \label{sec:standardmap}}

 Recall that for any  $\Omega_0$ and finite  $\Delta >0$ (and $\tau_0 \in \{\pm  1\}$), there exists a finite $\eps_c(\Omega_0, \tau_0,\Delta)$ such that for all $\eps \in (-\eps_c(\Omega_0,  \tau_0,\Delta),\eps_c(\Omega_0,  \tau_0,\Delta))$  there exists a set, $\mathcal{\hat K}_\eps (\Omega_0, \tau_0)\subset [-\Delta,\Delta]$, of positive measure, such that for any $k\in \mathcal{\hat K}_\eps$ the map $\mathbf{F}^{st}_\eps$ has  a smooth invariant curve with rotation number $\rho(k) \coloneqq  \Omega_0+\tau_0 k $ (so, we define  $k(\rho;\Omega_0,\tau_0) \coloneqq  \frac{\rho - \Omega_0}{\tau_0} $). These KAM curves are graphs of the form $\{(\phi,K)| K=k_\eps(\phi;\rho), \phi \in [-\pi,\pi]\}$  and the smooth function $k_\eps(\phi;\rho)$ is $\eps$ close to the constant function  $k_0(\phi;\rho,\Omega_0,\tau_0)=k(\rho;\Omega_0,\tau_0)=\langle k_\eps(\phi;\rho) \rangle_\phi$.  Let
$$
\phi_{m1}(k;\Omega_0,\tau_0) \coloneqq -\frac{\Omega_0}2 -\frac{\tau_0}2 k,\quad 
\phi_{m2}(k;\Omega_0,\tau_0) \coloneqq \pi -\frac{\Omega_0}2 -\frac{\tau_0}2 k.
$$
For $|\eps|<\tilde \eps_c(\Omega_0, \tau_0,\Delta)$,  the KAM curves intersect transversely the lines 
\begin{equation}
\label{def:L1L2} L1(\Omega_0,\tau_0)\coloneqq(\phi_{m1}(k),k), \ L2(\Omega_0,\tau_0)\coloneqq(\phi_{m2}(k;\Omega_0,\tau_0),k); \  k \in [-\Delta,\Delta]
\end{equation}  at the unique points,   $(\phi_{m1}(k_{m1}(\rho)),k_{m1}(\rho)),(\phi_{m2}(k_{m2}(\rho)),k_{m2}(\rho))$ with $k_{m1,m2}(\rho)=k(\rho;\Omega_0,\tau_0)+O(\eps)$.

Denote the positive measure sets $\mathcal{ K}_{i,\eps}\coloneqq\{k|k=k_{mi}(\rho(k)), k\in \mathcal{\hat K}_\eps\}, i=1,2$. Since $\tau_0 \in \{\pm  1\} \ne 0$, these relations are invertible: for sufficiently small $\eps$, for any $k_i \in \mathcal{\hat K}_{i,\eps}, i=1,2$ there exist  unique $\rho_i (k_i), i=1,2$ such that for each $i$ the initial condition $(\phi_{mi}(k_i),k_i)$ belongs to the KAM curve with rotation $\rho_i(k_i) $ (in general, for $k\in \mathcal{ K}_{1,\eps}\cap \mathcal{ K}_{2,\eps} $, the initial conditions $(\phi_{m1}(k),k)$ and  $(\phi_{m2}(k),k)$ belong to different KAM curves: $\rho_1(k)-\rho_2(k) = O(\eps)\ne 0 $, see below). We denote the corresponding graphs by  $k_{\eps,1}(\phi;\rho_1(k))$ and  $k_{\eps,2}(\phi;\rho_2(k))$, namely,    $k_{\eps,1}(\phi_{m1}(k);\rho_1(k))=k$ and  $k_{\eps,2}(\phi_{m2}(k);\rho_2(k))=k$.
 
We establish next that the intersection points of these KAM curves with $L1,L2$ are $\eps$-close in $\phi$ (and thus $\eps^2$ close in $K$) to the global minimum /
maximum  
of the graphs $k_{\eps,1}(\phi;\rho_1)$ and  $k_{\eps,2}(\phi;\rho_2)$: 
\begin{theorem}\label{thm:minmaxinvriantcurves}
For sufficiently small  $|\eps|$, for  $k \in \mathcal{ K}_{1,\eps}$ satisfying $\eps \sin\frac{\Omega_0 +\tau_0 k}{2}>0$ (respectively, $\eps \sin\frac{\Omega_0 +\tau_0 k}{2}<0$), the global minimum (respectively,  global maximum) of the graph  $k_{\eps,1}(\phi;\rho_1(k))$ is attained at an angle which is $O(\eps)$ close to $\phi_{m1}(k)$ and action which is  $O(\eps^2)$ close to $k$. For  $k \in \mathcal{ K}_{2,\eps}$, for the graph  $k_{\eps,2}(\phi;\rho_2(k))$ the opposite results hold. 
\end{theorem}
\begin{proof}
Notice that the map  $\mathbf{F}^{st}_\eps(\Omega_0=0,\tau_0=1)$  is the standard map:
\begin{equation}
    \mathbf{F}^{st}_\eps(0,1)=
\begin{cases}
    \bar{K} &= K + \eps \sin\bar{\phi},\\
    \bar{\phi} &= \phi + K.
\end{cases}
\end{equation}
So we show this property first for the standard map, and then rescale and shift $K$ to establish the same property for general parameter values $\Omega_0,\tau_0$. 

For $|\eps|<0.9716$ the standard map has rotational invariant curves \cite{greene1979method}, hence, there exists $\eps^*$ such that for   $|\eps|<\eps^*<0.9716$, there exists a $C>0$ such that for $|K|> C \sqrt{\eps}$ the standard map has a positive measure set of invariant smooth KAM curves, $\mathcal{ K}_{1,\eps} (\Omega_0=0, \tau_0=1)$ that intersect the line $L1(\Omega_0=0,\tau_0=1)=( -K/2, K)$ transversely.  
Starting at a point $( \phi_0= -K_0/2,K_0) \in L1\cap \mathcal{ K}_{1,\eps} (0, 1) $, belonging to the invariant curve  $k_{\eps,1}(\phi;\rho_1(K_0))$ (so  $k_{\eps,1}(-K_0/2;\rho_1(K_0))=K_0$), we calculate its iterations:
\begin{equation}
\begin{array}{ll}
    K_0 &= K_0,\\ 
    \phi_0 &= -K_0/2 \\
    K_1 &= K_0 + \eps \sin \frac{K_0}{2},\\
    \phi_1 &= \frac{K_0}{2}\\
    K_2 &= K_0 + \eps \sin \frac{K_0}{2} + \eps \sin \left(\frac{3K_0}{2} + \eps \sin\frac{K_0}{2}\right) \\ \phi_2 &= \frac{3K_0}{2} + \eps \sin(\frac{K_0}{2})\\
    K_3 &= K_0 + \eps \sin \frac{K_0}{2} + \eps \sin \left( \frac{3K_0}{2} + \eps \sin\frac{K_0}{2}\right) + \\
    & \quad \eps \sin \left(\frac{5K_0}{2} + 2\eps \sin(\frac{K_0}{2})+ \eps \sin \left( \frac{3K_0}{2} + \eps \sin\frac{K_0}{2}\right)\right), \\
    \phi_3 &= \frac{5K_0}{2} + 2\eps \sin(\frac{K_0}{2})+ \eps \sin \left( \frac{3K_0}{2} + \eps \sin\frac{K_0}{2}\right)
\end{array}
\end{equation}and, more generally 
\begin{equation}\label{eq:sumknphin}
\begin{array}{ll}
    K_n &=K_0 + \sum\limits_{i=1}^{n}\eps \left( \sin \frac{(2i-1)K_0}{2}\right) + \mathcal{O}(n \eps^2)\\
    &=K_0+ \eps \sin^2\frac{nK_0}{2}\mathrm{cosec}\frac{K_0}{2}+ \mathcal{O}(n\eps^2) \\
    \phi_n &=\frac{(2n-1)K_0}{2} +  \mathcal{O}(n\eps)=nK_0-\frac{K_0}{2}+  \mathcal{O}(n\eps),
\end{array}
\end{equation}
where we used formula A361.7 of \cite{gradshteyn2014table}:
$$
\sum\limits_{i=1}^{n}\sin \frac{(2i-1)K_0}{2} = \sin^2\frac{nK_0}{2}\mathrm{cosec}\frac{K_0}{2}.
$$
A similar summation formula for a  general $\phi_0$ can be easily found, yet, starting at $-K_0/2$ provides a convenient way to establish that the line L1 is close to the extrema of the invariant curves. In fact, this should also follow from the symmetries of the standard map.
Since the initial conditions belong to the invariant curve $k_{\eps,1}(\phi;\rho_1(K_0))$, we know that $K_n =k_{\eps,1}(\phi_n;\rho_1(K_0))$ and that $k_{\eps,1}(\phi;\rho)$ is a smooth graph associated with the rotation number $\rho_1(K_0)$. We conclude that for $K_0 $ which is bounded away from $0$ (mod $2\pi$):\begin{equation}\label{eq:k1phin}
\begin{array}{ll}
   k_{\eps,1}(\phi_n,\rho_1(K_0))  &=   K_0+ \eps \sin^2\frac{\phi_n+\frac{K_0}{2}  }{2}\mathrm{cosec}\frac{K_0}{2}+ \mathcal{O}(n\eps^2)\\
    &= K_0+\frac {\eps}{2\sin\frac{K_0}{2}}  (1-\cos(\phi_n+\frac{K_0}{2}  ))+\mathcal{O}(n\eps^2).
\end{array}
\end{equation}
Here the assumption that the trajectory lies on a smooth invariant curve (or, alternatively, belongs to a periodic island of finite period $N>1$, see remark below) is essential, allowing to interpolate the curve from a finite number of iterations (otherwise, for order  $\mathcal{O}(1/\eps)$ iterations,  the correction terms accumulate, and the finite $n$ approximation fails).

The first statement of the theorem for the case $\Omega_0=0,\tau_0=1$ follows: since $ k_{\eps,1}(\phi,\rho_1)$ is smooth and $ k_{\eps,1}(\phi_n,\rho_1)$ is given by  \eqref{eq:k1phin}  up to $\mathcal{O}(\eps)$ correction in the angle variable and  $\mathcal{O}(\eps^2)$ in $K$,  for $\eps \sin\frac{K_0}{2} >0$ the minimum is achieved near $\phi=-K_0/2$ and the maximum near  $\phi=\pi-K_0/2$ and the opposite statement follows when $\eps \sin\frac{K_0}{2}<0$.

Notice that shifting $\phi$ by $\pi$ is equivalent to reversing the sign of $\eps$ in the map, so, initial conditions starting on the line $L2$ and belonging to a KAM curve lie on the graph of
\begin{equation}
      k_{\eps,2}(\phi_n,\rho_2) =K_0- \frac {\eps}{2\sin\frac{K_0}{2}}(1+\cos(\phi_n+\frac{K_0}{2}  ))+ \mathcal{O}(n\eps^2)
\end{equation}
(so $k_{\eps,2}(\pi- K_0/2,\rho_2) = K_0 $ whereas  $k_{\eps,2}(- K_0/2,\rho_2) = K_0 -\dfrac {\eps}{\sin\frac{K_0}{2}} $ ) completing the proof for the standard map case.

Now we establish the results for general parameter values. Setting $(\phi',K') =(\phi, \Omega_0 +\tau_0 K) $ and $\eps'=\tau_0 \eps$, (so $K=(K'-\Omega_0)/\tau_0)), \eps=\eps'/\tau_0$) brings the map   $\mathbf{F}^{st}_\eps$  to the standard form without changing the rotation number. Hence the existence of the positive measure sets $\mathcal{ K}_{1,\eps} (\Omega_0, \tau_0),\mathcal{ K}_{2,\eps} (\Omega_0, \tau_0)$  in a strip of size $2 \Delta$ immediately follows. Since
\begin{equation}
      k'_{\eps',1}(\phi_n,\rho) =K_0'+ \eps'\frac {1-\cos(\phi_n+\frac{K_0'}{2}  )}{2\sin\frac{K'_0}{2}}  + \mathcal{O}(n\eps'^2),
\end{equation}
we get that $k_{\eps,1}(\phi_n,\rho)= ((k'-\Omega_0)/\tau_0) $ becomes:
\begin{equation}\label{eq:keps1}
      k_{\eps,1}(\phi_n,\rho_1(K_0))=  K_0+ \eps  \frac {1-\cos(\phi_n+\frac{\Omega_0 +\tau_0 K_0}{2}  )}{2\sin\frac{\Omega_0 +\tau_0 K_0}{2}} + \mathcal{O}(n \tau_0\eps^2)
\end{equation}
and 
\begin{equation}\label{eq:keps2}
      k_{\eps,2}(\phi_n,\rho_2(K_0))=  K_0-\eps  \frac {1+\cos(\phi_n+\frac{\Omega_0 +\tau_0 K_0}{2}  )}{2\sin\frac{\Omega_0 +\tau_0 K_0}{2}} + \mathcal{O}(n \tau_0\eps^2).
\end{equation}
Hence, the minima/maxima of the curves
$(\phi,k_{\eps,1}(\phi;\rho_1))$  and, respectively, $(\phi,k_{\eps,2}(\phi;\rho_2))$ that cross the lines  $L1(\Omega_0,\tau_0)$ and, respectively, $L2(\Omega_0,\tau_0)$ at height $K_0$ occur close to the lines $ (-(\Omega_0 +\tau_0 K)/2, K)$  and  $ (\pi -(\Omega_0 +\tau_0 K)/2, K)$ which are exactly the lines $L1(\Omega_0,\tau_0)$ and $L2(\Omega_0,\tau_0)$, as claimed. 
\end{proof}
Note that while the maximum of $k_{\eps,1}(\phi;\rho_1(K_0))$  does not belong to $L2$, it is $\eps$-close to it in $\phi$ (it is realized at $(\phi=\pi-\frac{\Omega_0 +\tau_0 K_0}{2}, K=K_0+\eps  \frac {1}{\sin\frac{\Omega_0 +\tau_0 K_0}{2}}  )$ ), with no contradiction to the above Theorem.

According to KAM theory the error terms in the above expressions also depend on how badly the rotation rate is approximated by rationals.  The rotation rates along the invariant curves can be approximated by averaging:

\begin{lemma}
Provided that $(\Omega_0 +\tau_0 K_0)/\pi$ is badly approximated by rationals, for sufficiently small $\eps$  and for   $K_0 \in \mathcal{ K}_{1,\eps}\cap \mathcal{ K}_{2,\eps} \ne \emptyset$, the rotation rates along the invariant curves  $k_{\eps,i}(\phi;\rho_i(K_0))$ are 
\begin{equation}\label{eq:rotationscaled}
\rho_1(K_0) \coloneqq \Omega_0 + \tau_0 K_0+  \frac{\tau_0 \eps}{2} \mathrm{cosec}\frac{\Omega_0 +\tau_0 K_0}{2}+ \mathcal{O}(\tau_0^2 \eps^2).  
\end{equation}
and  
\begin{equation}\label{eq:rotationscaled2}
   \rho_2(K_0) = \Omega_0 + \tau_0 K_0-  \frac{\tau_0 \eps}{2} \mathrm{cosec}\frac{\Omega_0 +\tau_0 K_0}{2} + \mathcal{O}(\tau_0^2 \eps^2).  
\end{equation}
In particular,  the difference in the rotation numbers of the KAM curves that cross the lines $L1$ and $L2$ at the same height, $K_0 \in \mathcal{ K}_{1,\eps}\cap \mathcal{ K}_{2,\eps} $, is  \begin{equation}\label{eq:rotationsdiff}
  \rho_1(K_0)- \rho_2(K_0) =  \eps \ \tau_0  \  \mathrm{cosec}\frac{\Omega_0 +\tau_0 K_0}{2} + \mathcal{O}(\tau_0^2 \eps^2).  
\end{equation}
\end{lemma}
\begin{proof}
Averaging formula \eqref{eq:keps1} and  \eqref{eq:keps2}  in $\phi$, which provides the leading order approximation to the rotation rate when  $(\Omega_0 +\tau_0 K_0)/\pi$ is badly approximated by rationals, leads to formula \eqref{eq:rotationscaled},\eqref{eq:rotationscaled2}. In particular, notice that $\rho_1(K_0)\ne \rho_2(K_0)$ and their difference is just \eqref{eq:rotationsdiff} (indeed, for a positive $\sin\frac{\Omega_0 +\tau_0 K_0}{2}$ the curve $  k_{\eps,1}(\phi,\rho_1(K_0)) $ is above the curve $  k_{\eps,2}(\phi,\rho_2(K_0)) $:  $k_{\eps,1}(\phi_{m2}(K_0);\rho_1(K_0))=K_0+\eps\mathrm{cosec}\frac{\Omega_0 +\tau_0 K_0}{2} >  k_{\eps,2}(\phi_{m2}(K);\rho_2(K))= K_0$).    
\end{proof}


Formula \eqref{eq:sumknphin} is valid, for a finite $n$, for any initial condition on the lines $L1$ or $L2$. If the orbit that starts on these lines is $N$-periodic or if it belongs to an island of stability with a finite period $N$ and width $d(N;\eps)\ll \eps$ (so $N$ is not too small), by applying  \eqref{eq:sumknphin} to $n=1,\dots,N-1$,  we obtain that the initial condition must also belong, up to order $\eps^2$ terms, to the global minimal/maximal island of this island chain. 

\subsection{Hovering orbits  for Diophantine rotation numbers \label{sec:truncinvcurves}}

Using the above results, the curve  $k_\eps^{u0}(\phi;\Omega_0,\tau_0)$, the lowest invariant curve which resides in the upper half plane, is easily found, up to order $\eps^2$:
\begin{lemma}
For a  Diophantine $\frac{\Omega_0}{2\pi} \in (0,1) $, $\tau_0\ne 0$,   and sufficiently small $\eps>0$, the curve    $k_\eps^{u0}(\phi;\Omega_0,\tau_0)$ is given, up to order $\eps^2$, by $ k_{\eps,1}(\phi,\rho_1(0))$ whereas for small negative $\eps$ it is given, up to order $\eps^2$, by  $ k_{\eps,2}(\phi,\rho_2(0))=k_{-\eps,1}(\phi+\pi,\rho_1(0))$. 
\end{lemma}
\begin{proof}
Since $\rho_1(0)=\Omega_0+O(\eps)$, for sufficiently small $|\eps|$, there is a positive set of preserved curves of  $\mathbf{F}^{st}_\eps$  that are close to $ k_{\eps,1}(\phi,\rho_1(0))$ and, similarly, to  $ k_{\eps,2}(\phi,\rho_2(0))$. Since, for  $\Omega_0 \in (0,2 \pi) $ and $\eps>0$  the minimum of  $ k_{\eps,1}(\phi,\rho_1(0))$ occurs at $\phi=-\Omega_0/2+O(\eps),K=0+O(\eps^2)$, we conclude that the lowest preserved KAM curve with a non-negative minimum, is, for sufficiently small $\eps$,  $\eps^2 $ close to the curve  $ k_{\eps,1}(\phi,\rho_1(0))$. 

The same argument follows for the case of negative $\eps$ with the minimum of   $ k_{\eps,2}(\phi,\rho_2(0))$ occurring near $\phi=\pi-\Omega_0/2+O(\eps),K=0+O(\eps^2)$. Moreover, by the symmetry   $\mathbf{F}^{st}_\eps(\phi,K)=\mathbf{F}^{st}_{-\eps}(\phi+\pi,K)$ it follows that 
 $ k_{\eps,2}(\phi,\rho_2(K_0))=k_{-\eps,1}(\phi+\pi,\rho_1(K_0))$. 
\end{proof}
The above lemma is demonstrated in Figure  \ref{fig:smalleps}, see Section \ref{sec:truncatedmapnumact} for details.

Comparing the observations regarding the locations of the minima of the invariant curves of the smooth map  $\mathbf{F}^{st}_\eps$  with the position of the singularity lines of the truncated near-tangency family of perturbed interval exchange maps, $\mathbf{F}_\eps$, leads to the identification of two distinct scenarios:
\begin{lemma} \label{lemma:mininj0jr} Let $\Omega_0 \in (0,2\pi)$ and let $\Delta< \min ((\frac{\pi-\tw}{\lambda})^2,\KMIN(\lambda,\tw,\tau_1)) $. Then, for all  $K_0 \in \mathcal{ K}_{1,\eps}\cap [-\Delta,\Delta]$ such that the minimum of the invariant curves  of  $\mathbf{F}^{st}_\eps$  is negative  (i.e. $\min_\phi k_{\eps,1}(\phi,\rho_1(K_0))<0$), there exists $\eps_c$ such that for all $\eps \in (0,\eps_c)$, the minimum resides in $J_0$ when  $\eps \sin\frac{\Omega_0 +\tau_0 K_0}{2}>0$ (the \textbf{Hovering case}) and in $J_R$ when $\eps \sin\frac{\Omega_0 +\tau_0 K_0}{2}<0$ (the \textbf{impacting case}).
\end{lemma}
    \begin{proof}
Recall that  the midpoints of the intervals $(\mathcal{J}_0(K),\mathcal{J}_1(K),\mathcal{J}_R(K))$  are, correspondingly,   $-\frac{1}{2}(\Omega_0  +\tau_0 K),\tw-\frac{1}{2}(\Omega_0+(\tau_0+\tau_1) K),\pi-\frac{1}{2}(\Omega_0  +\tau_0 K)+\lambda\sqrt{-K}) )$ and that their lengths are $(2 \tw -2\lambda\sqrt{-K}-\tau_1 K,2\lambda\sqrt{-K},2(\pi-\tw)+\tau_1 K)$, namely are positive for the range of parameters we consider and $K\in (-\KMIN(\lambda,\tw,\tau_1),0)$, see \eqref{eq:kmindef}.

For this range of $K$ values, the line $L1$ is exactly at the midpoint of  $\mathcal{J}_0(K)$ and the line $L2$ is $\lambda\sqrt{-K}$ close to the midpoint of $\mathcal{J}_R(K)$. So, for $\Delta < (\frac{\pi-\tw}{\lambda})^2$ the $L2$ line resides in the interior of $J_R$. 
Hence, provided $\eps_c \ll \min((\frac{\pi-\tw}{\lambda})^2,\tw)$, Theorem \ref{thm:minmaxinvriantcurves} implies the lemma.
\end{proof}
For larger values of $K$ the line $L2$ can cross to other regions and the implications of this are left for future studies. 

We are now ready to prove Theorem \ref{thm:hovering}:
\begin{proof}
Recall that for all $K_0\in \mathcal{K}_{\eps,1}$:
\begin{equation}
      k_{\eps,1}(\phi,\rho_1(K_0))=  K_0+   \frac {\eps}{2\sin\frac{\Omega_0 +\tau_0 K_0}{2}} ( 1-\cos(\phi+\frac{\Omega_0 +\tau_0 K_0}{2}  ))+ \mathcal{O}(\eps^2).
\end{equation}
Hence, for $\eps>0$, the curve  $k_{\eps,1}(\phi,\rho_1(K_0))$ crosses the tangency line $K=0$ if and only if $K_0+O(\eps^2) \in (-\frac {\eps}{2\sin\frac{\Omega_0 }{2}},0)$.  Let $C\eqqcolon -\frac{2 K_0 \sin \frac{\Omega_0 }{2}}{\eps} \in (0,1)$,  so  $K_0(C)=-C\frac{\eps}{2\sin \frac{\Omega_0 }{2}}$. Then, by \eqref{eq:keps1},  for any $C \in (0,1)$ the horizontal line $K=0$ is crossed at 
\begin{equation}\begin{array}{ll}
      \phi^1_{1,2}(K_0(C))&=   -\frac{\Omega_0 +\tau_0 K_0(C)}{2}  +\arccos( 1 -\frac{-2K_0(C)}{\eps}\sin \frac{\Omega_0 +\tau_0 K_0(C)}{2}+ \mathcal{O}(\eps))\\
      &=   -\frac{\Omega_0 +\tau_0 K_0(C)}{2}  +\arccos( 1 -\frac{-2K_0(C)}{\eps}\sin \frac{\Omega_0 }{2}+ \mathcal{O}(\eps))\\
     &= -\frac{\Omega_0 }{2}  +\arccos( 1 -C) + \mathcal{O}(\eps).
\end{array}
\end{equation}
Thus,  $\phi^1_{1,2}(K_0(C))|_{C=1-\cos( \tw) }= -\frac{\Omega_0 }{2} \pm \tw$ (a more precise formulation of the intersection of $   k_{\eps,1}(\phi,\rho_1(K_0))$ with the corner singularity line $(\phi_{01}(K),K)$ leads to order $\eps^2$ corrections). We thus define \begin{equation}\label{eq:kcrit}
\tilde K_c(\eps, \tw,\Omega_0):= K_0(C)|_{C=1-\cos( \tw) }=-\eps \frac{1-\cos( \tw)}{2\sin \frac{\Omega_0 }{2}}. 
\end{equation}
The above calculations show that for sufficiently small $\eps$ there exists an $A>0$ such that provided the set $B_\eps\coloneqq(K_c(\eps, \tw,\Omega_0)-B\eps^2,K_c(\eps, \tw,\Omega_0)+ B\eps^2) \cap \mathcal{K}_{\eps,1}$ is non-empty, it includes a value $K_c=K_c(\eps, \tw,\Omega_0)$ with a corresponding invariant curve $ k_{\eps,1}(\phi,\rho_1(K_c))$ and rotation number 
\begin{equation}\label{eq:rhocrit}\rho_c\coloneqq\rho_1(K_c) = \Omega_0   +\frac{\eps \tau_0\cos( \tw)}{2\sin \frac{\Omega_0 }{2}} +  \mathcal{O}(\tau_0^2 \eps^2),  
\end{equation}
which is non-impacting and passes $\eps^2$ close and above the boundaries of $J_0$ at $K=0$. For a Diophantine $\Omega_0/2\pi$ and sufficiently small $\eps<\eps_c(\Omega_0,\tau_0,\tw)$, $|B_\eps|\approx 2B\eps^2>0$, so such a curve exists, and thus, for all $\eps\in (0,\eps_c(\Omega_0,\tau_0,\tw))$ the size of the hovering set increases, to leading order, linearly in $\eps$;
For this range of $\eps$ values the $\HOV(\eps)$ consists of all initial conditions that are below the tangency line $K=0$ and above $ k_{\eps,1}(\phi,\rho_c) $: \begin{equation}
       \{(\phi,K)| K \in [k_{\eps,1}(\phi,\rho_c),0), \phi \in [ \phi_1,  \phi_2]\}  \end{equation} where $k_{\eps,1}(\phi_{1,2},\rho_c)=0$ with $\phi_{1,2} =-\frac{\Omega_0 }{2} \pm \tw+O(\eps)$, so, by  \eqref{eq:kcrit}:
\begin{equation}\label{eq:hovsetsize}
  |\HOV(\eps;\Omega_0,\tw)|=-\int_{\phi_1}^{\phi_2}    k_{\eps,1}(\phi,\rho_1(K_c(\eps)))d \phi=  \frac {\eps }{\sin\frac{\Omega_0 }{2}} ( \sin(\tw)-\tw\cos( \tw))+ \mathcal{O}(\eps^2).
\end{equation}
Since  $\frac{d |\HOV(\eps;\Omega_0,\tw)|}{d\tw}=\tw \sin \tw >0$ for all $\tw \in (0,\pi)$ and $|\HOV(\eps;\Omega_0,\tw=0)|=0$, we established the first claim, that for $s>0$, $ |\HOV(\eps;\Omega_0,\tw)|$  is monotonically increasing with $\eps$ for all  $\eps \in(0, \eps_c(\Omega_0,\tau_0,\tw))$. 

The second claim follows as here, for $\eps < 0$, by Lemma \ref{lemma:mininj0jr},  when the minimum of $k_{\eps,1}$ or of $k_{\eps,2}$  is negative it is close to their crossings with $L2$, namely, it is in the region $J_R$.  Thus the curves do not belong to $J_0\cup J_{0u}$. In fact, the lowest invariant curve that remains in  $ J_{0u}$ is  $k_{\eps,2}(\phi,\rho_2(0))=-\eps  \frac {1+\cos(\phi+\frac{\Omega_0 }{2}  )}{2\sin\frac{\Omega_0 }{2}} + \mathcal{O}(n \tau_0\eps^2)$ so the smooth band is empty for negative $\eps$.  Taking a Diophantine $\Omega_0$ implies that $\eps_c(\Omega_0,\tau_0,\tw)$ can be chosen so that for $|\eps|<\eps_c(\Omega_0,\tau_0,\tw)$ the standard map has no low order resonances close to $K=0$, thus, in particular, there can be no resonances that reside only in $J_0$. Hence the hovering set is empty. For $\eps=0$ the hovering set is empty by definition.
\end{proof}

The persistence of the invariant curves of  $\mathbf{F}^{st}_\eps(\Omega_0,\tau_0)$ that cross the $L1$ line at $K_0$,   $k_{1,\eps}(\phi,\rho_1(K_0))$, depend sensitively on the numerical properties of their rotation number,   $\rho_1(K_0)$, as does the accuracy of the approximation of $\rho_1(K_0)$ which is achieved by averaging  $k_{1,\eps}(\phi,\rho_1(K_0))$ over $\phi$. 
Hence, as in KAM theory, we may expect that for larger $\eps$ values, setting $ \rho_c(\eps,\Omega_0,\tw)$ to a badly approximated irrational,  $\Omega_G$ (e.g.,  the Golden mean or its equivalents), will lead to persistence of the curve $k_c$ for larger values of $\eps$ when compared to varying $\eps$ for a fixed set of parameters, as formulated in Theorem \ref{thm:hoveringc}, which is proved next:

\begin{proof}
Let  $\Omega_0(\eps,c)=\Omega_G-\frac{c}{2}\eps, |c|\le |\tau_0|$ and  let $   \cos( \tw (\eps,c)) = \frac{2(\Omega_G-\Omega_0)\sin \frac{\Omega_0 }{2}}{\eps \tau_0}=\frac{c}{ \tau_0}\sin \frac{\Omega_0 }{2}=\frac{c}{ \tau_0}\sin( \frac{\Omega_G }{2}-\frac{c\eps }{4})$, so, for $|c/\tau_0|<1$ the angle $\tw$ is well defined. Then, by \eqref{eq:rhocrit},  we obtain that 
$\rho_c(\eps,\Omega_0(\eps,c),\tw(\eps,c))=\Omega_G+O(\tau_0^2\eps^2)$. Thus, by Theorem \ref{thm:hovering}, for sufficiently small $\eps$, there is an invariant curve of the truncated map with a minimum which is close to its crossing with $L1$ with a $K$ value which is $\eps^2$ close to $\tilde K_c(\eps, \tw (\eps,c),\Omega_0 (\eps,c))=-\eps \frac{1-\frac{c}{ \tau_0}\sin\frac{\Omega_0 }{2}}{2\sin \frac{\Omega_0 }{2}}$. 
Since   $\frac{\Omega_G}{2\pi} \in (\nu,1-\nu)$, we get that   $\frac{\Omega_0(\eps,c)}{2\pi} \in (\nu-\frac{c}{2}\eps,1-\nu-\frac{c}{2}\eps)$, so for sufficiently small $\eps$ in $(0,\frac{\nu}{|c|})$, we obtain that  $\frac{\Omega_0(\eps,c)}{2\pi} \in (\frac{\nu}{2},1-\frac{\nu}{2})$, and hence that $\sin \frac{\Omega_0 }{2} $ is positive and is bounded from below:  $\sin \frac{\Omega_0 }{2} > \sin \frac{\pi \nu}{2} >0 $. It follows that for $|c/\tau_0|<1$ the minimum of the critical curve, which is $\eps^2$ close to $\tilde K_c(\eps, \tw (\eps,c),\Omega_0 (\eps,c))$, is negative and depends monotonically on $c/\tau_0$.  
\end{proof}
Notice that the limits $c \rightarrow \pm \tau_0$, at which $\tw \rightarrow \{0,\pi\}$ are singular, since the lengths of some of the intervals approach zero. 
This assertion works beautifully for $\eps \le 0.3$, see details below, and in particular Figure \ref{fig:hoverinnc}; The lowest invariant critical curves (orange curve  Figure \ref{fig:hoverinnc}a and blue curve in Figure \ref{fig:hoverinnc}b) have rotations which are only $0.02$ different from the predicted golden mean value. 

\subsection{Numerical simulations\label{sec:truncatedmapnumact}}
\begin{figure}
    \centering
    \includegraphics[width=0.9\textwidth]{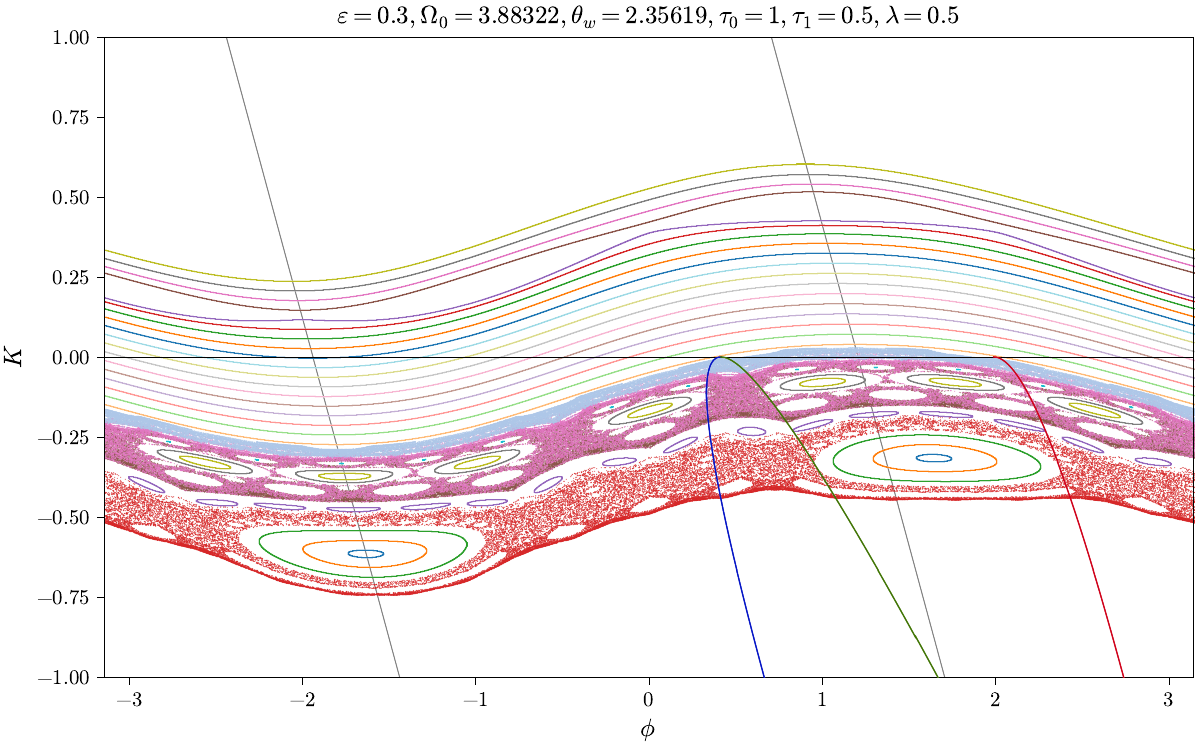}
    \includegraphics[width=0.9\textwidth]{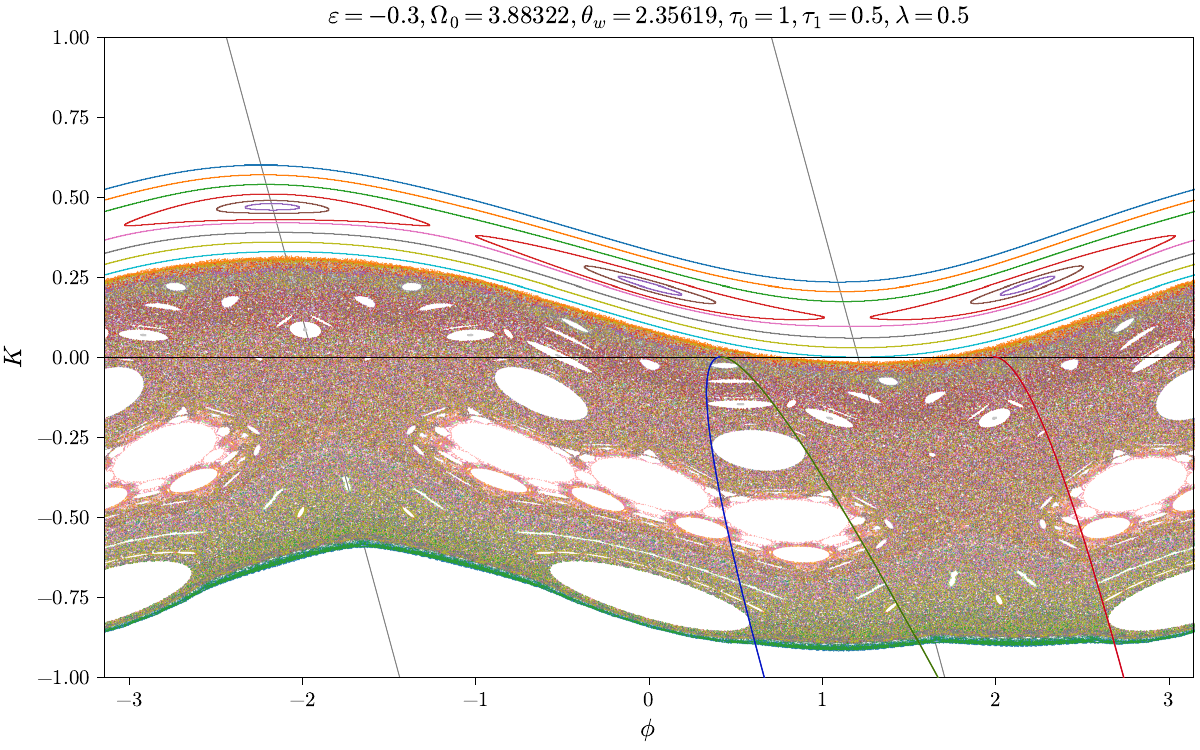}
    \caption{Dynamics for small $\pm \eps$ values. Thirty evenly spaced initial conditions along the $L1$ line \eqref{def:L1L2}  are iterated 50,000 times for $\eps=0.3$ (upper figure) and  $\eps=-0.3$ (lower figure).}
    \label{fig:smallepsmany}
\end{figure}

Figure \ref{fig:smallepsmany} shows trajectories (hereafter, 50,000 iterates\footnote{Simulations are limited to avoid escape beyond the lower limit,  $K=-\KMIN$.}) of the maps $F_{\pm \eps}(\Omega_0,\tw, \allowbreak \tau_0,\tau_1,\lambda)$ of \eqref{eq:truncatedmap} with 30 initial conditions, with evenly spaced $K$ values along the $L1$ line. The figure reveals the dramatic affect of the discontinuity lines on the dynamics; it is observed that for $|\eps|=0.3$ the non-impacting region exhibits almost integrable dynamics: only the 3-resonant island is observed and neither chaotic regions nor small islands are seen. On the other hand, in the impacting region large stochastic regions are observed, as well as many high-period elliptic islands. The dramatic effect of changing the sign of $\eps$ is also apparent, demonstrating the existence of hovering set for positive $\eps$ and its absence for negative $\eps$.

\begin{figure}
    \centering
    \includegraphics[width=0.9\textwidth]{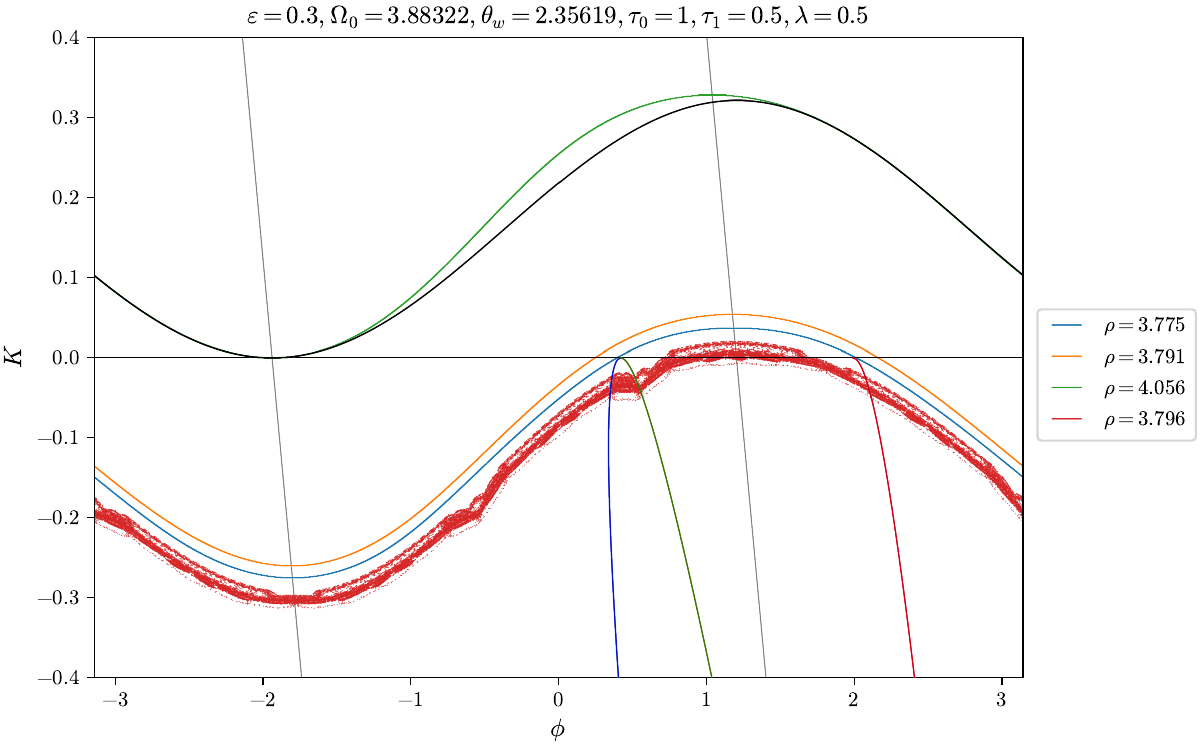}
    \includegraphics[width=0.9\textwidth]{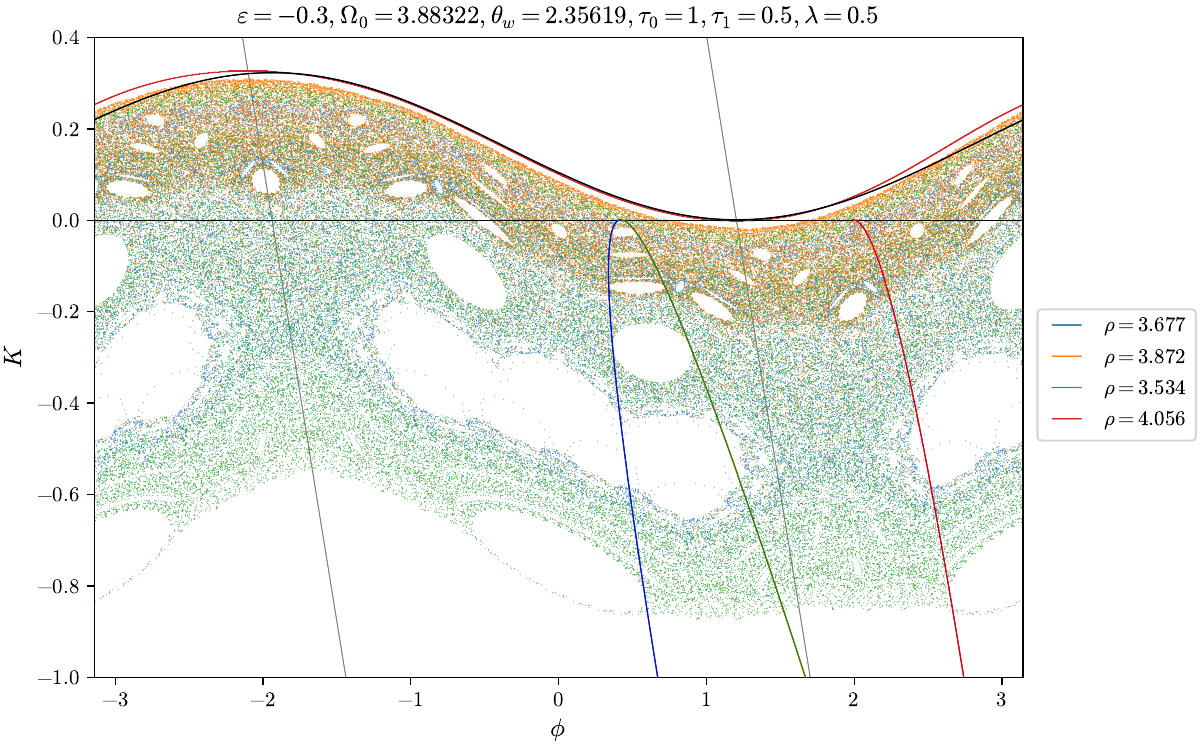}
    \caption{Critical curves for  $|\eps|=0.3$.  The first 3 initial conditions are on $L1$: the blue trajectory starts at   $K_0=\tilde K_c$, the orange at  $K_0=\tilde  K_c+0.05\cdot|\eps|$, and the green at $K_0=0$. The red trajectory is on $L2$  starting at $(K_0=0,\phi_0=\pi-\Omega_0/2)$. The label shows the approximated rotation number for each trajectory, $\rho =  \langle \omega_a \rangle  $. In light gray we draw the tangency curve $K=0$ and the lines $L1,L2$. The black curves are the leading order approximations to the upper critical curves, $k_{1,\eps=0.3}(\phi,\rho_1(0))$ and $k_{2,\eps=-0.3}(\phi,\rho_2(0))$. }
    \label{fig:smalleps}
\end{figure}

\begin{figure}
    \centering
    \includegraphics[width=0.7\textwidth]{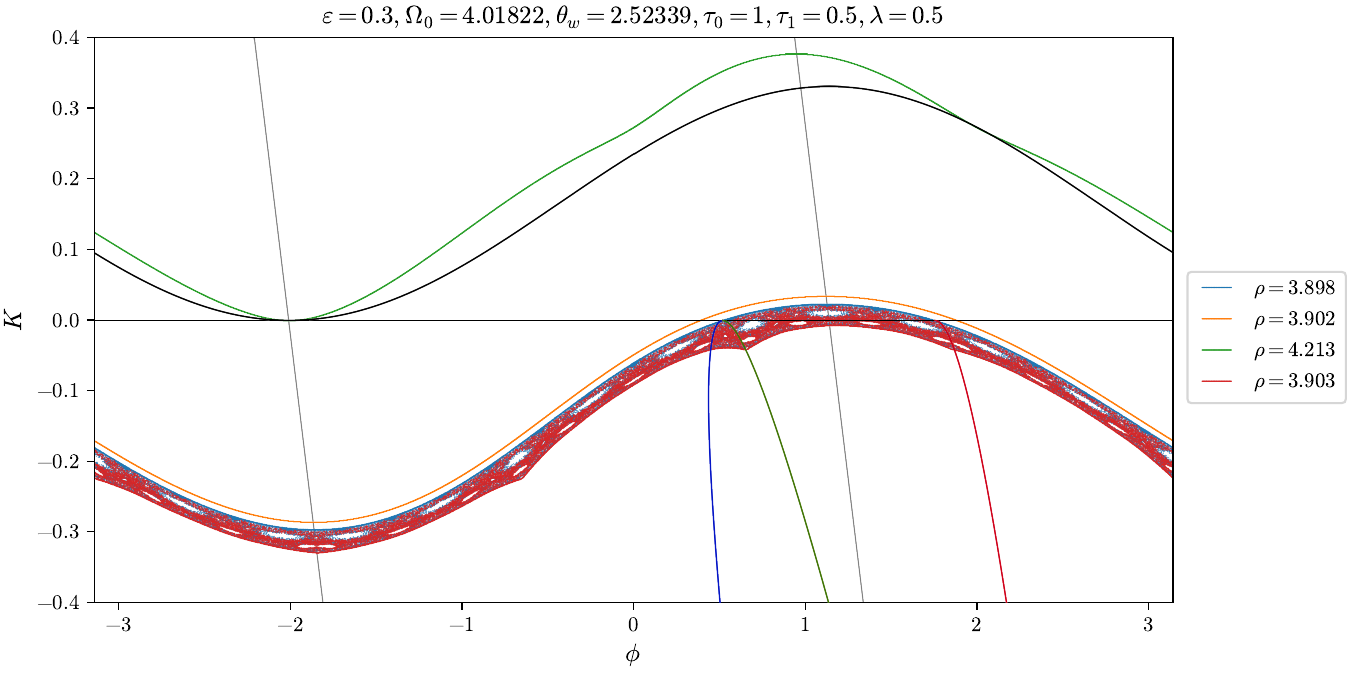}
    \includegraphics[width=0.7\textwidth]{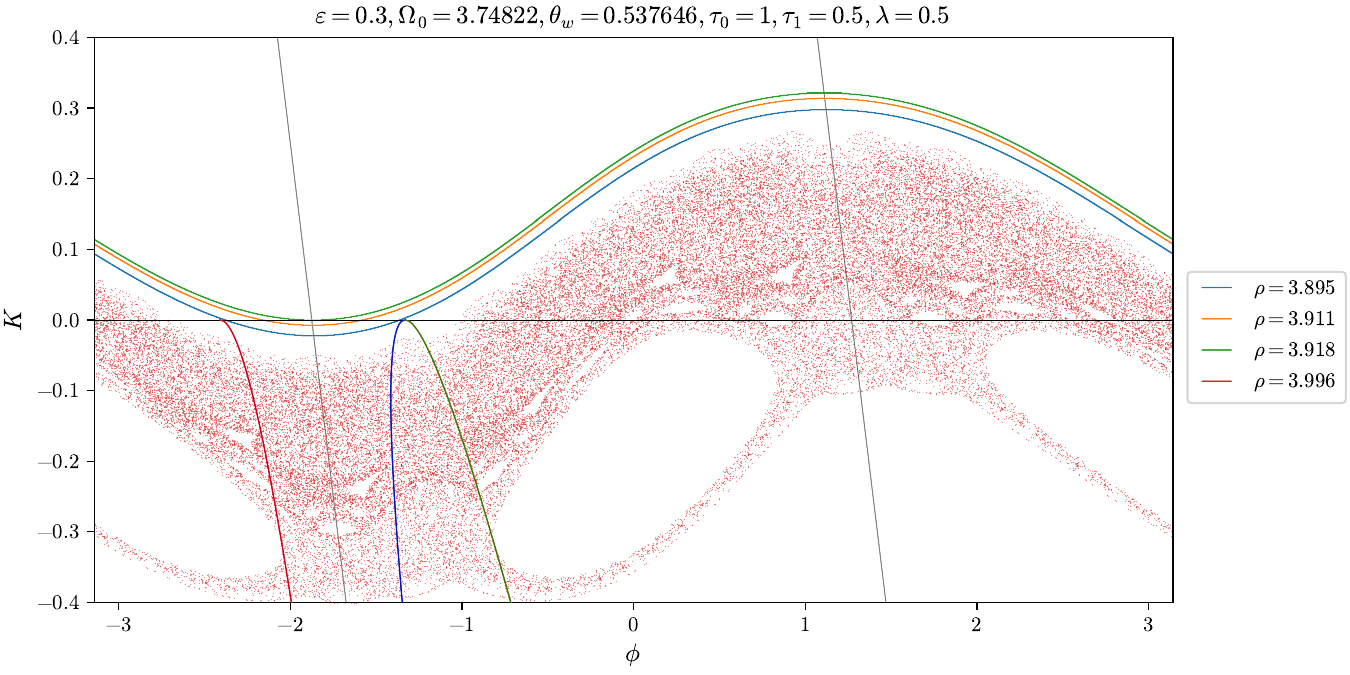}
    \includegraphics[width=0.7\textwidth]{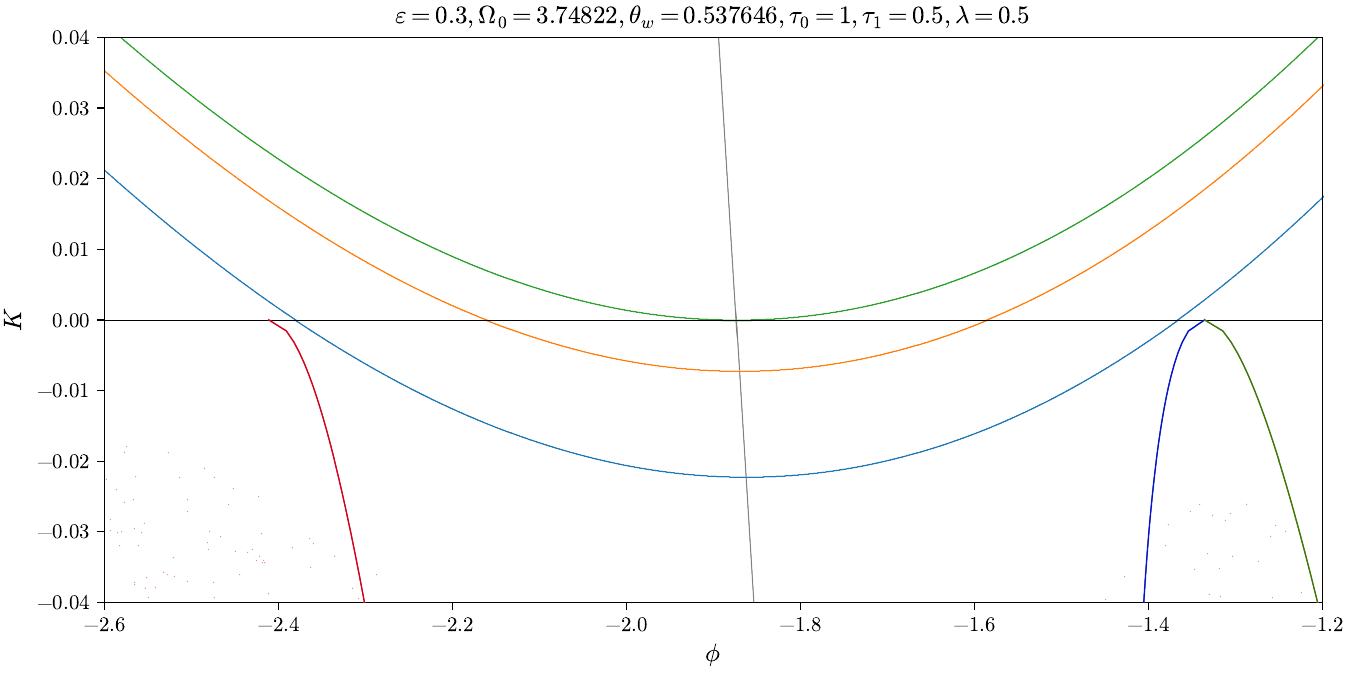}
    \caption{ The critical curve for $\eps=0.3$ and $\Omega_0(\eps,c;\Omega_G),\tw(\eps,c,\Omega_G)$ a) Near maximal hovering set ($c =-0.9$) b) Near minimal hovering set ($c = 0.9$) c) details of b), showing the existence of a small hovering set. }
    \label{fig:hoverinnc}
\end{figure}

\begin{figure}
    \centering
    \includegraphics[width=0.3\textwidth]{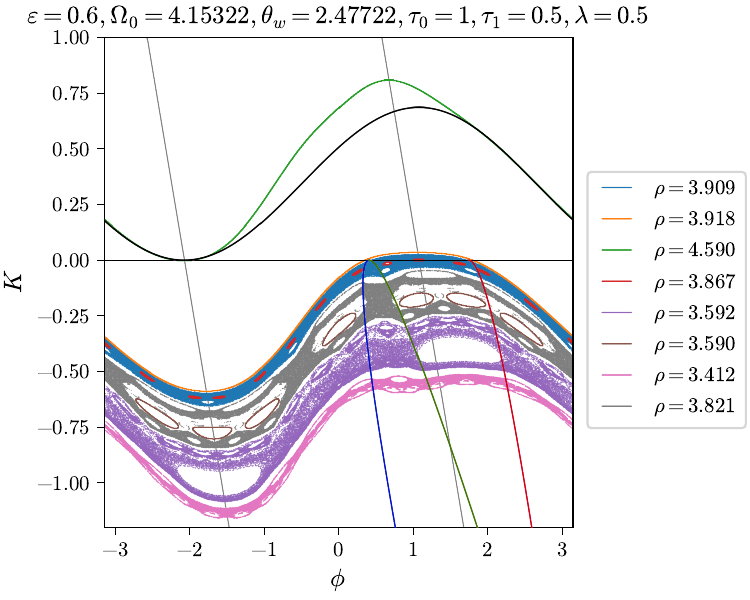}
    \includegraphics[width=0.3\textwidth]{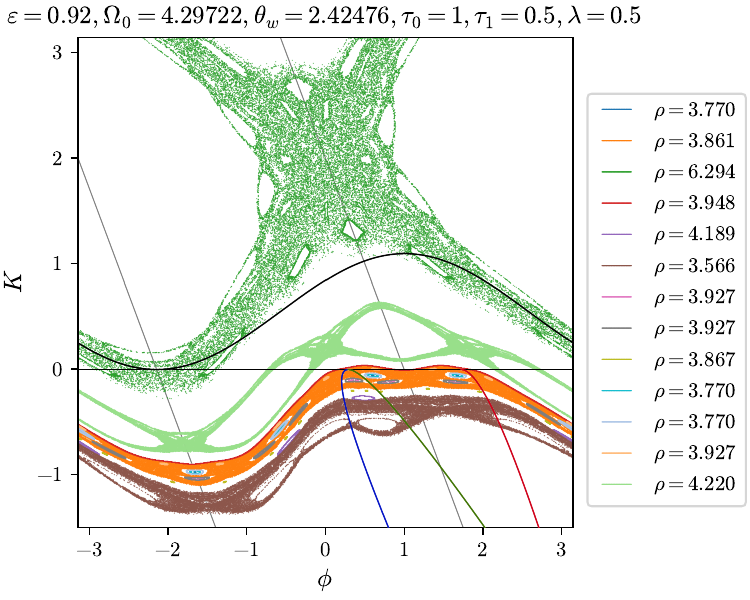} \includegraphics[width=0.3\textwidth]{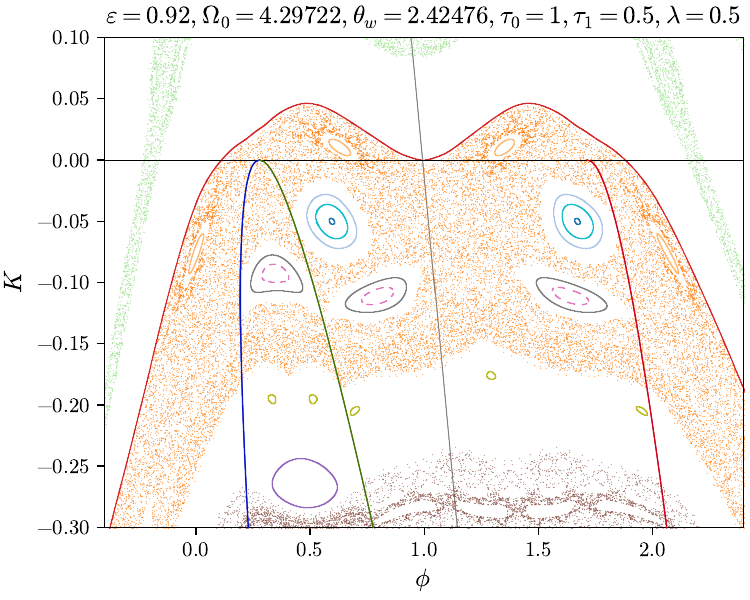} \\
    \includegraphics[width=0.3\textwidth]{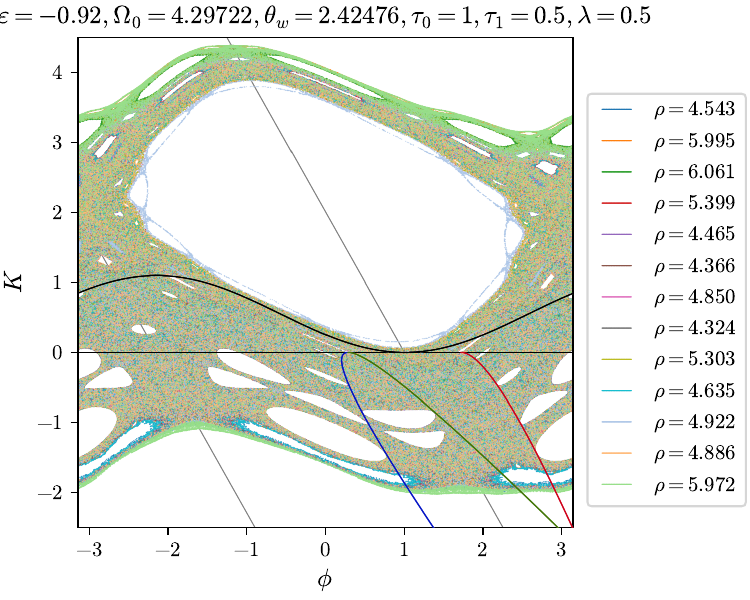}
    \includegraphics[width=0.3\textwidth]{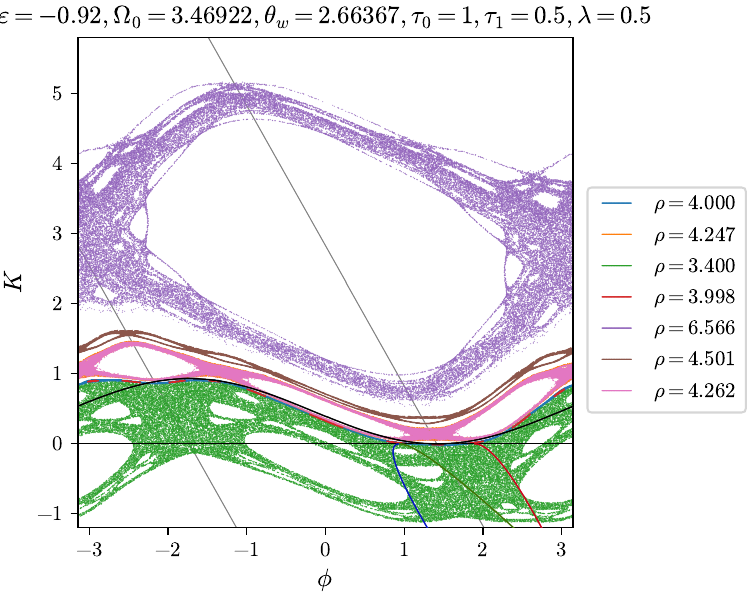}
    \caption{ The critical curves for $\Omega_0(\eps,c;\Omega_G),\tw(\eps,c,\Omega_G)$ with $c=-0.9$, $\Omega_G=2\pi\frac{\sqrt{5}-1}{2}$,   (a) $\eps=0.6$ (b) $\eps=0.92$  (c) zoom in on islands for  $\eps=0.92$ of (b), (d) $\eps=-0.92$, with $\Omega,\theta$ as in (b)  (e) $\eps=-0.92$ with the corresponding $\Omega_0(\eps,c;\Omega_G),\tw(\eps,c,\Omega_G)$. The upper main island chain boundary is due to the tangency curve, whereas the lower larger islands are cut by the corner singularity lines. the black curve is the leading order approximation to $k_{1,\eps}(\phi,\rho_c(0))$. }
    \label{fig:epsbig}
\end{figure}

In Figures \ref{fig:smalleps}-\ref{fig:epsbig}, we demonstrate more accurately the role of the critical curves  $ k_\eps^{u0}(\phi;\Omega_0,\tau_0)$ and of $ k_{\eps,1}(\phi,\rho_1(K_c))$ by plotting first the trajectories of the  following four selected initial conditions (in some of the plots we also add additional trajectories to demonstrate resonant structures). The first two (blue and orange) are on $L1$ and are close, for positive $\eps$, to the predicted position of the critical curve: the  blue trajectory starts at $K_0=\tilde K_c$ of \eqref{eq:kcrit} and the orange starts at  $K_0=\tilde  K_c+0.05\cdot|\eps|$. The other two initial conditions are at the intersection of the $L1$ and $L2$ curves with the tangency line $K=0$. The green trajectory starts at $(\phi_0=-\Omega_0/2, K_0=0$),  and the red one at $(K_0=0,\phi_0=\pi-\Omega_0/2)$. We also plot, in black,  the tangency line $K=0$ and the extrema lines $L1,L2$. In some of the graphs we plot in black the leading order approximation to $ k_\eps^{u0}(\phi;\Omega_0,\tau_0)$. For small $\eps>0$ (respectively, $\eps<0$)  this curve approximates the green (respectively, red) trajectory.  For each trajectory on these plots,  $\{(\phi_j,K_j)\}_{j=0}^{n-1}$  with $n=50,000$, the label shows their approximate rotation number, namely  $\rho =  \langle \omega_a \rangle =\frac{1}{n}\sum_{j=0}^{n-1} \omega_{a}|_{\{a:\phi_j \in \mathcal{J_a(K_j)}\}}  $. For trajectories that remain in $J_0\cup J_1$, and thus coincide with the standard map trajectories, this is simply $\rho =\Omega_0+\tau_0 \langle K \rangle$, so the rotation numbers of such rotational invariant curves are monotone increasing with their heights. The properties of the rotation numbers for the non-smooth dynamics are left for future studies.

Figure \ref{fig:smalleps}a,b shows the simulations for  
\begin{equation*}
    (\eps,\Omega_0,\tw,\tau_0,\tau_1,\lambda)=(\pm0.3,\Omega_G,3\pi/4,1,0.5,0.5)
\end{equation*}
where $\Omega_G=2 \pi \frac{\sqrt{5}-1}{2}=3.88322\dots$ is the Golden mean.  Figure \ref{fig:smalleps}a demonstrates that for positive $\eps\le 0.3$ the predictions regarding the critical curve are quite accurate; the first two trajectories produce invariant curves that are just above the corner singularity lines and reside in $J_0\cup J_1$. The predicted rotation at the critical curve, from \eqref{eq:rhocrit}, is $\rho_c=3.7694\dots$, and the numerical rotation of  the blue curve is $\rho=3.791$, showing a very good agreement with the prediction that the rotations should agree to order $O(\eps^2)$. Similarly, we see that the green curve, the initial condition that starts at the intersection of $L1$ with the tangency curve traces quite closely the black curve, which is the leading order approximation, \eqref{eq:k1phin}, for $k_{1,\eps}(\phi,\rho_1(0))$ (at $\eps=0.1$ they are indistinguishable).  
Figure \ref{fig:smalleps}b shows, using the same scheme for the initial conditions, the trajectories of the map $F_{-\eps}(\Omega_0,\tw,\tau_0,\tau_1,\lambda)$, namely the map with the opposite sign of $\eps$. As expected, there is no hovering set and the critical curve is destroyed. The lowest invariant curve which we can detect is the red trajectory,     $k_{\eps=-0.3}^{u0}(\phi;\Omega_0,\tau_0)=k_{2,\eps=-0.3}(\phi,\rho_2(0))$, which is just the $\pi$-shifted green trajectory of Figure \ref{fig:smalleps}a (up to the different vertical scales of the two figures),  $k_{\eps=0.3}^{u0}(\phi;\Omega_0,\tau_0)=k_{1,\eps=0.3}(\phi,\rho_1(0))$.

Similarly, setting $c =-0.9$ and increasing $\eps$ while tuning $(\Omega_0,\tw)$ so that the critical curve has rotation which is $O(\eps^2)$ close to $\Omega_G$, we obtain that the hovering set increases in size for $\eps$ as large as 0.92, see Figure \ref{fig:epsbig} a,b (a few additional trajectories are added to demonstrate the dynamics). Here the critical curves do not start at the predicted value, and their shape is strongly deformed, yet, their rotation is close to the golden mean.  Switching the sign of $\eps$ in \ref{fig:epsbig} c demonstrates a dramatic effect: while the hovering set is of order one for positive $\eps$ it disappears for negative $\eps$.

Figures \ref{fig:smallepsmany}-\ref{fig:epsbig} contrast the discontinuous dynamics below the critical curve with the smooth dynamics above it: 
\begin{itemize}
\item The discontinuities increase the size of the chaotic layers and of the resonances phenomena: Resonant impacting islands of high period, surrounded by chaotic layers, are abundant and visible  much more than the non-impacting resonances, see Figure \ref{fig:smallepsmany}. We have checked that impacting resonances and impacting chaotic layers are visible for $\eps$ as small as $\eps=0.01$, at which the standard map appears indistinguishable from being integrable.
\item There are islands of stability of all dynamical types, i.e. the periodic orbits and their surrounding quasi-periodic orbits have periodic  itinerary which may include all symbols or only part of the symbols.
\item  The boundaries of the resonant islands may be associated with either the smooth dynamics (via homoclinic or heterocilinc tangles of hyperbolic periodic orbits), or due to the singularities of the map. In the latter case the island is abruptly cut by either a tangency to a corner singularity curve, or by a tangency to the singular-tangency segment $\sigma^\eps_{tan-R}$, see Figure  \ref{fig:epsbig}d.   Such singular boundaries appear to enlarge the chaotic layer surrounding the islands.
\end{itemize}

\section{Discussion}\label{sec:discussion}

The return map for a class of near-pseudo integrable Hamiltonian impact systems of the oscillator-step type near the onset of impacts has been derived.  The resulting map is area preserving, piecewise symplectic and inherits a time-reversal symmetry from the mechanical form of the Hamiltonian impact system. The form of the map is of a perturbed family of interval exchange transformation on the circle, with higher order corrections. A truncated model for the map that respects the same symmetry as the return map is studied analytically and numerically.

A central finding of this study is the existence of hovering, non-resonant orbits for an open set of parameter values. Specifically, we showed that for small but non-zero perturbation ($\eps \neq 0$), there exists a set of orbits with phase space measure of order $O(\eps)$ that consistently avoid impacts with the step, despite being aperiodic and passing infinitely often both above and laterally to the step. This non-resonant hovering behavior cannot occur in the uncoupled case ($\eps = 0$). Furthermore, this orbit set is destroyed by reversing the sign of $\eps$.
This striking behavior holds across a range of perturbation amplitudes.

The key mechanism behind this phenomenon is the presence of a critical invariant curve in the return map, the last KAM torus of the smooth perturbed flow that does not intersect the singularity set of the impact system. In the non-resonant case, we derived an asymptotic expansion for this curve in the small $\eps$ limit, showing that it remains $\eps$-close to the unperturbed torus tangent to the right side of the step (the blue torus in Figure \ref{fig:two-tang}). Since the full return map and the truncated map are $C^r$-close  in the regular region, $J_0 \cup J_{0u}$, and since their singularity sets are also close, similar conclusions extend to the full return map \eqref{eq:returnmapfull}, identifying the upper boundary of the impact zone. Similarly, the lower boundary of the impact zone is an invariant torus which is $\eps$-close to the unperturbed tangent torus  to the upper side of the step (the red tangent torus of Figure \ref{fig:two-tang}).
The phase space region between these two tori contains all the non-smooth dynamics. Our analysis focused on the onset of this region, deriving the return map in neighborhoods of the tangent tori. Numerical simulations of the truncated map suggest that no additional rotational invariant curves exist right below the critical curve, indicating the absence of further phase space division of the impact zone near the critical tori.  

Establishing the potential existence of global connecting orbits that traverse the entire impacting region, or, conversely, of a dividing invariant curve in the impact zone,
remains an open problem, both for the truncated map and the full return map. This may be approached via global numerical simulations of the full return map. Alternatively,  it can be approached by constructing local return maps inside the impact region,  assembling transition chains for their truncated version, and establishing that the correction terms cannot destroy them. These questions are both technically challenging and conceptually intriguing, and we leave them for future investigation.

Finally, we emphasize that the study of hovering and related phenomena at the onset of non-smooth dynamics in perturbed pseudo-integrable systems opens the door to a larger program: understanding the intricate phase space structure of Hamiltonian impact systems beyond the smooth KAM regime. These studies may offer new perspectives on the dynamics in billiards that correspond to small perturbations of pseudo-integrable tables \cite{richens1981pseudointegrable,Vladimir2014JournalofModernDynamics}, and more generally, to the dynamics in invertible piecewise smooth maps, such as isometries \cite{ashwin2005invariant}. The richness of the dynamics uncovered here suggests that many more surprising behaviors await discovery, and we hope this work stimulates further explorations in this direction.

\section*{Acknowledgements} The work was funded by Israel Science Foundation (grant 787/22).

\section*{Data declaration} Data sharing not applicable to this article as no datasets were generated or analysed during the current study.

\printbibliography

\appendix

\section{Details of the derivation of the return map \label{sec:appendixA}}

 In Section \ref{sec:constructret}, the iso-energy return map $   \mathcal{ F}_\eps$ of \eqref{eq:returnmapfull} is constructed as a composition of three maps, introducing the proper coordinates on intermediate sections. In Section \ref{sec:proofofcornersing}, we find the asymptotic form of the singularity curves and identify the dynamics in the different regions, thus proving Theorems \ref{thm:cornersing} and \ref{thm:regionsJdyn}. In Section\ \ref{sec:timereversal} the time-reversal symmetries of the return map (associated with the time-reversal symmetry of mechanical Hamiltonian systems) and of the singularity lines in various coordinates are explained. In Section \ref{sec:proofreturnmap} we use these constructions to build the return map and prove Theorem \ref{thm:returnmap}, in which the return map is derived up to correction terms that are piecewise smooth and whose order is established.
 
\subsection{Auxiliary sections and the Poincaré first return map}\label{sec:constructret}

Following \cite{PnueliRomKedarTangency}, to construct the return map to $\Sigma_h$ for near-tangent orbits we define auxiliary local sections near the step, $ \Sigma_h^{*,\pm}$, which are crossed by all near-tangent initial conditions of  $\Sigma_h$, importantly, by both impacting and non-impacting segments:
\begin{definition}\label{def:star_section} The iso-energy star sections $\Sigma_h^{*,\pm}$ are  2D-sections of the energy level set that are unions of two sections $\Sigma^{w,\pm}_h$ and $\Sigma_h^>$:
\begin{eqnarray}
    \Sigma_h^{*,\pm} &\coloneqq& \Sigma^{w,\pm}_h\cup\Sigma_h^>,\label{eq:star-sec}
    \\
    \Sigma^{w,\pm}_h &\coloneqq&\{(q,p), H(q,p;\eps) = h,\ q_1= q_1^w,\ \pm p_1\geq 0\}\label{eq:Sigmahw}
    \\
    \Sigma^>_h &\coloneqq& \{(q,p), H(q,p;\eps) = h,\ p_1= 0, \qw_1\leq q_1<0\}
\end{eqnarray}
\end{definition}
Let $\Phi_t^{\eps,im}$  denote the flow associated with the HIS defined by \eqref{eq:perturbedhamil} and \eqref{eq:stepdef}  and  $\Phi_t^{\eps,sm}$ the flow associated with the corresponding smooth system (i.e. when there is no step) of the perturbed Hamiltonian \eqref{eq:perturbedhamil}. Then, the return map to  $\Sigma_h$ is of the form:
\begin{equation}
\mathcal{F}_\eps  =\Phi_{t^+}^{\eps,sm} \circ \Phi_{\Delta t}^{\eps,im} \circ \Phi_{t^-}^{\eps,sm}\coloneqq \mathcal{F}^{sm,+} \circ \mathcal{F}^{step}_\eps  \circ  \mathcal{F}^{sm,-}_\eps  
\end{equation}
where  $t^-=t^-(z_0)$  is the travel time between  $ \Sigma_h$ and , $ \Sigma_h^{*,-}$ , $\Delta t= \Delta t (z_-= \mathcal{F}^{sm,-}_\eps z_{0} )$ is the travel time between $ \Sigma_h^{*,-}$ and  $ \Sigma_h^{*,+}$ (in particular $\Delta t=0$ for initial conditions that reflect from the right boundary of the step and for initial conditions that cross $ \Sigma_h^>$, namely avoid $ \Sigma_h^{w,-}$ ), and  $t^+= t^+ (z_+= \mathcal{F}^{step}_\eps z_- )$ is the travel time between $ \Sigma_h^{*,+}$ and  $ \Sigma_h$:
\[ \Sigma_h \longrightarrow_{\Phi_{t^-}^{\eps,sm}} \Sigma_h^{*,-}\longrightarrow_{\Phi_{\Delta t}^{\eps,im}} \Sigma_h^{*,+}\longrightarrow_{\Phi_{t^+}^{\eps,sm}}\Sigma_h. \]
As the smooth flow is near integrable and $q_1^w<0$, it is clear that  $t^\pm$ are finite and are $C^0$-close, with a square-root singularity near tangency,  to the unperturbed travel time. So, the segments $\Phi_{t^\pm}^{\eps,sm}$ of the return map are well approximated by perturbation theory which takes into account the singular travel-time-dependence on initial conditions near tangent trajectories; Also, notice that for initial conditions in the interior of $J_0^\eps $, the return time to $\Sigma_h$, $T_1^\eps = t^-+t^+ + \Delta t$, depends smoothly on initial conditions. 

In  \cite{PnueliRomKedarTangency}, the HIS corresponding to impacts from the infinite wall at $q_1=q_1^w$ was analyzed. There, crossing  of   $\Sigma^{w,-}_h$ (or, trivially, of   $\Sigma^{>}_h$), leads to a reversal of $p_1$, and the passage time from $\Sigma^{*,-}_h$ to $\Sigma^{*,+}_h$ is instantaneous. Denoting by $\mathcal{F}_\eps^{tan}$ the corresponding return map of the wall system (Eq. (4.2) in  \cite{PnueliRomKedarTangency}), we conclude that: 
\begin{equation}
\mathcal{F}_\eps^{tan} z_0 =\Phi_{t^+}^{\eps,sm} \circ \mathcal{R}_1 \circ \Phi_{t^-}^{\eps,sm}z_0\coloneqq \mathcal{F}^{sm,+}_\eps \circ \mathcal{R}_1 \circ  \mathcal{F}^{sm,-}_\eps z_{0}. 
\end{equation}
 In particular,  for segments that pass to the right of the step (initial conditions in $J_{0u}$ ) and for segments that reflect from the right side of the step (initial conditions in $J_R$), $\mathcal{F}_\eps$ is identical to  $\mathcal{F}_\eps^{tan}$, as in Theorem 2 in \cite{PnueliRomKedarTangency}.  By smoothness,  segments that pass above the step without impacting it  (initial conditions in $J_{0}$) have the same leading order behavior as those in  $J_{0u}$. 
 
 Hence, to construct $   \mathcal{ F}_\eps$, we need to find the corner-singularity curves and to compute  $\mathcal{F}^{step}_\eps z_{-}\coloneqq\Phi_{\Delta t}^{\eps,im}z_{-}=z_{+}$  for initial conditions in  $J_1$, where, hereafter   $ z_{\pm} \in \Sigma_h^{*,\pm}$.

\begin{figure}[H]
    \centering
    \resizebox{0.7\textwidth}{!}{\input{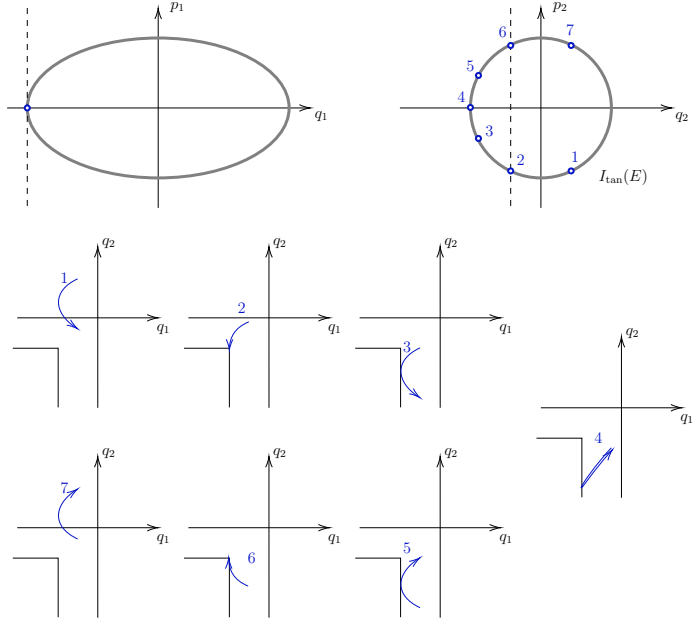}}
    \caption{The unperturbed tangency line $\Xi^0_{wall}(h)$ on $\Sigma^{*,-}_h$ and the corresponding trajectories in the $(q_1,\ q_2)$ plane. The points $1$ and $7$ correspond to the passage above the corner; the points $2$ and $6$ correspond to tangencies at the corner (from which the corner singularity curves emerge) and the points $3,\,4,\,5$ belong to $\sigma^0_{tan-R}$ and correspond to tangencies to the right side of the step at  $q_2<\qw_2$.  }
    \label{fig:Sigma_0_wall}
\end{figure}

The section  $\Sigma_h^{*,-}$ includes both impacting segments to the wall $q=q_1^w$ (those segments that cross  $\Sigma_h^{w,-}$) and non-impacting segments (those that cross  $\Sigma_h^>$), with the tangent curve
$\{(q_1 = \qw_1, p_1 = 0, q_2, p_2): p_2^2/2 + V_2(q_2)+\eps V_c(\qw_1, q_2) = h - \hw_1\}
    \}$
dividing this 2D section to its two parts\footnote{Indeed, due to the nesting property of the potential $V_2(q_2)$, for $\eps$ small enough, $\Xi_{wall}^\eps$ is a closed curve (for $\eps = 0$ this curve coincides with the angle trajectory of the Hamiltonian $H_2$ on $h-\hw_1$ level and is defined by the action $\It(h)$ (see \eqref{eq:Itan})) }. 
Each point on this curve corresponds to a 
trajectory that passes tangentially to the line $q_1 = \qw_1$ on the $(q_1, p_1)$-plane, and has different angles on $(q_2,p_2)$ plane (on Figure \ref{fig:Sigma_0_wall}, the unperturbed case is shown).

Lemmas 3.2 and 3.3 in \cite{PnueliRomKedarTangency} prove that for sufficiently small $\eps\geq 0$, 
 $\Sigma_h^{*,-}$ is well-defined, and can be parameterized by $(q_2, p_2)$, thus by $(\theta, I)=S_2(q_2,p_2)$, and that in these coordinates, the projection of the dividing tangent curve to the $(q_2,p_2)$-plane, denoted by $\sigma^\epsilon_{wall}(h)$, is well-defined and can be represented by a graph in the action-angle variables:
\begin{equation}\label{eq:Itanstar}
I = I_{tan}^{\eps,-}(\theta).
\end{equation}
This function is smooth and is $C^r$ close to $\It(h)$, satisfying the equality:
\begin{equation}\label{eq:XiwallepsI}
   H_2( I_{tan}^{\eps,-}(\theta))+\eps V_c(\qw_1, q_2( \theta,I_{tan}^{\eps,-}(\theta))) = h - \hw_1.
\end{equation}
The circle $\sigma^\epsilon_{wall}(h)$ divides the parameterized plane  $\Sigma^{*,-}_h$ to its two parts: its exterior correspond to crossings of  $\Sigma^>_h$ , namely to segments that turn back smoothly to $\Sigma_h$ at $q_1>q_1^w$ (so   $\mathcal{F}^{step}_\eps|_{\Sigma^{>}_h}=Id $) , whereas its interior correspond to segments that cross   $\Sigma^{w,-}_h$.  The tangency curve, $\sigma^\eps_{tan}(h)\in \Sigma_h$, is exactly the pre-image of this curve:   $\sigma^\eps_{tan}(h) = (\mathcal{F}^{sm,-})^{-1}\sigma^\epsilon_{wall}(h) $.

The wall tangency curve,   $\sigma^\epsilon_{wall}(h)$, is symmetric (independent of the symmetries of the potentials);
Rewriting \eqref{eq:XiwallepsI} in terms of $(q_2,p_2)$ at the tangency curve, leads to the equation 
$
   \frac{p_2^2}{2}+V_2( q_2)+\eps V_c(\qw_1, q_2) = h - \hw_1$
which is satisfied, for $h>h_1^w$ and for sufficiently small $\eps$, for the proper range of $q_2$, by $\pm p_2(q_2)$. Since  $S_2(q_2,\pm p_2)=(\pm \theta,I)$, this implies  that \begin{equation}\label{eq:Itanwallsymmetry}
    I_{tan}^{\eps,-}(\theta)=I_{tan}^{\eps,-}(-\theta).
\end{equation}
This symmetry, together with the time reversal symmetries of the smooth flow, implies that while  the tangency curve $\sigma^\eps_{tan}(h)\in \Sigma_h$ is not necessarily symmetric, its image is a symmetric reflection of this curve: $\overline I_{tan}^{\eps}(-\theta) = I_{tan}^{\eps}(\theta)$.

On $\Sigma_h^{*,-}$, we use the wall coordinates $(\theta,\rw)$, where $\rw$ measures the action distance to the tangency curve at   $\Sigma_h^{*,-}$;  For any $z_-^w=(q_1=q_1^w,p_1 ^\eps(q_2,p_2;h)<0,q_2,p_2)\in \Sigma_h^{w,-} $ , defining  $(\theta, I)=S_2(q_2,p_2)$ (with our usual convention that $\theta=0$ at $(q_2=q_{2,max}(I),p_2=0)$, so that  $S_2(q_2,-p_2)=(-\theta,I)$), the action distance $\rw$ is defined by
\begin{equation} \label{eq:rhow}
\rw    = I^{\eps,-}_{tan}(\theta) -  I, \ 
\end{equation} so $\rw>0$ for $z_-^w=z_-^w (\theta,\rw)\in \Sigma_h^{w,-} $. Similarly, for any  $z_-^>=(q_1=q_{1,\min}^\eps(q_2,p_2;h),p_1=0,q_2,p_2)\in \Sigma_h^{>} $ , with  $(\theta, I)=S_2(q_2,p_2)$, $\rw$ is again defined by Eq. \eqref{eq:rhow}, where here  $\rw<0$ and  $z_-^>=z_-^> (\theta,\rw)\in \Sigma_h^{>} $. The tangency curve divides between these two regions and corresponds to $\rw=0$. 

Let\footnote{here and hereafter the numbers in inequalities, e.g. the "$\frac{1}{2}$",  are arbitrary and are introduced to reduce notation, they may be replaced by any constant $c<1$ or, respectively, $c>1$.} $\rw_{max}(h)= \frac{1}{2} \min(\It(h)-H_2^{-1}(h_2^w),H_2^{-1}(h)-\It(h))$. Then,
\begin{equation}
    \label{eq:rhomaxh} h_2^w<H_2(\It(h)-\rw_{max}(h))<H_2(\It(h)+\rw_{max}(h))<h.
\end{equation} It follows that for $|\rw|<\rw_{max}(h)$, the unperturbed circle in the $(q_2,p_2)$ plane that corresponds to the action $I=(\It(h)-\rw)$ is well defined and intersects transversely the line $q_2=q_2^w$ at exactly two points. Thus, there exists $\eps_0$ such that for all $|\eps|<\eps_0$  and $\theta \in [-\pi,\pi)$ the values \begin{equation}\label{eq:defqpstar}
   (q_2^{\eps,-}(\theta, \rw),p_2^{\eps,-}(\theta, \rw))\coloneqq S_2^{-1}(\theta,  I^{\eps,-}_{tan}(\theta) - \rw)
 \end{equation}are uniquely defined, depend smoothly on both $(\rw,\epsilon)$ and correspond to a closed symmetric curve in the $(q_2,p_2)$-plane,  $\{(q_2^{\eps,-}(\theta, \rw),p_2^{\eps,-}(\theta, \rw))=S_2^{-1}(\theta,  I^{\eps,-}_{tan}(\theta) - \rw), \theta \in [-\pi,\pi]\}$,  that intersects the line $q_2=q_2^w$ transversely, see Figure \ref{fig:wallsection}; By \eqref{eq:Itanwallsymmetry},  $S_2^{-1}(-\theta,  I^{\eps,-}_{tan}(-\theta) - \rw)=S_2^{-1}(-\theta,  I^{\eps,-}_{tan}(\theta) - \rw)$, and thus
 \begin{equation}\label{eq:q2starsym}
   (q_2^{\eps,-}(-\theta, \rw),p_2^{\eps,-}(-\theta, \rw))= (q_2^{\eps,-}(\theta, \rw),-p_2^{\eps,-}(\theta, \rw)).
\end{equation}
 
For  $\rw > 0$ this curve resides inside the wall tangency curve and corresponds to phase space values at the wall section   $z_-^w (\theta,\rw)\in \Sigma_h^{w,-} $. It follows that for all $|\eps|<\eps_0$ there
are unique symmetric angles at which this curve crosses transversely the line $q_2=q_2^w$, namely, for these angles   $z_-^w (\pm\theta,\rw) $ corresponds to the corner point with negative horizontal momentum and opposite vertical momenta. Next, we calculate these values, finding from them the corner singularity curves.

\subsection{The corner singularity curves at the wall section.}

The division of $\Sigma_h^{w,-} $ to the three different dynamical regimes $(J^{\epsilon,-}_R,J^{\epsilon,-}_0, J^{\epsilon,-}_1)$, the wall images of  $(J^\epsilon_R,J^\epsilon_0, J^\epsilon_1)$  under $\mathcal{F}^{sm,-}$:  $\mathcal{F}^{sm,-}J_a^\eps=J_a^{\eps,-} \in \Sigma_h^{w,-}$, is found next. 
Denote the corner-singularity curves at the wall section by  $\sigma_{ab}^{\eps,-} = (\mathcal{F}^{sm,-})^{-1}\sigma_{ab}^{\eps,-} ,\ ab \in \{R0,01,1R\}$.


\begin{lemma}\label{lemma:border}
For all $h>h_\eps^w$ there exists $\Delta>0$, such that for sufficiently small $\eps $, the corner-
singularity curves at the wall,  $\sigma_{ab}^{\eps,-},\ ab \in \{R0,01,1R\}$  can be represented, in the wall coordinates  $(\theta, \rw)$,  by  non-intersecting graphs of the form $\sigma_{ab}^{\eps,-}= \{(\theta_{ab}^{\eps,-} (\rw), \rw)|     \rw \in [0,\Delta]  \}$. Moreover, $\theta_{R0}^{\eps,-} $ and , $\theta_{1R}^{\eps,-} $   depend smoothly on $\eps,\rw$ for all $\rw\ge 0$, whereas, near $\rw=0$, $\theta_{01}^{\eps,-} $ depends smoothly on $\eps,\sqrt{\rw}$  and smoothly on $\eps,\rw$ for positive $\rw$ which is bounded away from zero. Finally,  $(\theta, \rw)\in J^{\eps,-}_b$ for all $ \theta \in (\theta_{ab}^{\eps,-} (\rw),\theta_{bc}^{\eps,-} (\rw))$. In particular, the order on the cylinder of the dynamical regions $J_b^{\eps,-}$ at the wall is $(J^{\epsilon,-}_R,J^{\epsilon,-}_0, J^{\epsilon,-}_1)$. 
\end{lemma}

\begin{proof}
Recall that  $J^{\epsilon,-}_R$ corresponds to orbits segments that impact the right side of the step, so, on  $\Sigma_h^{w,-} $,   $J^{\epsilon,-}_R$   corresponds to $z_-^w$ with $q_2<q_2^w$.   
Segments that cross the line $q_1=q_1^w$ above the step, with  $q_2>q_2^w$, and turn around and return to $\Sigma_h^{w,+}$ without impacting the upper part of the step belong to $J^{\epsilon,-}_0 $ and those that return  to $\Sigma_h^{w,+}$ after  impacting the upper part of the step once belong to $J^{\epsilon,-}_1 $; we show below that there exists a finite $\rw_b (h) $ such that for  $\rw \in (0,\rw_b(h))$ and sufficiently small $\eps$ no additional impacts can occur, so we hereafter set $\Delta<\rw_b (h)$. 

We first find the borders of $J^{\epsilon,-}_R$ and show they correspond to the borders with  $J^{\epsilon,-}_{1}$ on the left and   $J^{\epsilon,-}_{0}$ on the right,  namely they correspond to the wall-corner singularity curves   $\sigma_{1R}^{\eps,-}$ and $\sigma_{R0}^{\eps,-}$ which occur at $q_2=q_2^w$ with opposite signs of $p_2$. Notably, at $\Sigma_h^{w,-}$ the dependence of these curves on $\rw$ is smooth (the non-smooth dependence of  the singularity curves of $\mathcal{F}_\eps$ on $K$ in \eqref{eq:cornersingbordrho} is associated with the non smooth dependence of the travel times from/to to $\Sigma_h$ near $\rw=0$, namely of  $t^{\pm}(\theta,\rw)$ ).
We then find the border between $J^{\epsilon,-}_{0}$ and $J^{\epsilon,-}_{1}$,   namely $\sigma_{01}^{\eps,-}$, which corresponds to the $(q_2,p_2)$ value that crosses   $\Sigma_h^{w,-} $ at $q_2>q_2^w$ and return to  $\Sigma_h^{w,-} $ at  $q_2^w$, see Figure \ref{fig:Sigma_0_wall}.  In both cases we first find the angles $\theta_{ab}^{\eps,-}(\rw)$ that hit the corner for $I= I^{\eps,-}_{tan}(\theta) - \rw$ and then show that the nearby angles correspond to the correct dynamics (belong to $J_a^{\eps,-}$ to the left and to  $J_b^{\eps,-}$ to the right).

\paragraph{The corner-singularity angles separating $J_R$ and $J_0\cup J_1$.}

Rearranging \eqref{eq:perturbedhamil} and utilizing the parametrization of $\Sigma_h^{w,-}$ by  $z_-^w (\theta,\rw)\in \Sigma_h^{w,-}$, we obtain:
\begin{equation}\label{eq:pstarex}\begin{array}{ll}
   \frac{(p_1^\eps)^2}{2}  & =h-[ V_1(\qw_1) + H_2( I^{\eps,-}_{tan}(\theta) - \rw)  \\
     & \ \ \qquad +  \eps V_c(\qw_1,q_2^{\eps,-}(\theta,  \rw))] \\
     &\eqqcolon \Upsilon(\theta,\rw;h,\eps).    
\end{array}  
\end{equation}
For $h>h^w$, $I^{0,-}_{tan}(\theta)=\It(h)>0$  (see \eqref{eq:Itan}) , hence, for sufficiently small $\eps$ the right hand side $\Upsilon(\theta,\rw;h,\eps)$ is smooth in all its variables including $\rw,\eps$. Let $ \hat \Upsilon(\rw;h) \coloneqq \Upsilon(\theta,\rw;h,\eps=0)=h-h_1^w-H_2(\It(h)-\rw)$, so, for sufficiently small $\eps$, $\Upsilon(\theta,\rw;h,\eps)=\hat \Upsilon(\rw;h)+O(\eps)$ where the $O(\eps)$ are smooth functions of $(\theta,\rw;h,\eps)$. 

By the definition of $\It(h)$, $\hat \Upsilon(\rw=0;h)=0$, so some care is needed to keep the right hand side of \eqref{eq:pstarex}  positive. For any fixed  $c>0 $, for $\rw \in (c,\rw_{max}(h))$,  by \eqref{eq:rhomaxh} and the monotonicity of $H_2(I)$, we indeed conclude that  $ \hat \Upsilon(\rw;h)  >\hat \Upsilon(c;h)>0$. Hence, for sufficiently small $\eps$, $\Upsilon(\theta,\rw;h,\eps)=\hat \Upsilon(\rw;h)+O(\eps)>0$. Since $H_2'(\It(h))>0$,  the same conclusion applies for $c$ which approaches zero slower than $\eps$.   
Now consider the small $\rw$ behavior. By the definition of $I^{\eps,-}_{tan}(\theta) $, $\Upsilon(\theta,0;h,\eps)=0 $ for all  small $\eps$ for which $I^{\eps,-}_{tan}(\theta) $ is defined. Since  $\frac{\partial \hat \Upsilon}{\partial \theta }=0$, we conclude that $\frac{\partial \Upsilon}{\partial \theta }=O(\eps \rw)$ and that $\frac{\partial \Upsilon}{\partial \rw }|_{\rw=0}=H_2'(I^{\eps,-}_{tan}(\theta))+O(\eps)=\omega_2(\It(h))+O(\eps)> 0$. Thus, for small  $\rw\ge 0$  and  all $\theta$ \begin{equation}
      \Upsilon(\theta,\rw;h,\eps)=\rw \cdot (\omega_2(\It(h))+O(\eps,\rw)) >0.
  \end{equation} 
 
\begin{figure}
    \centering
    \resizebox{0.8\columnwidth}{!}{\input{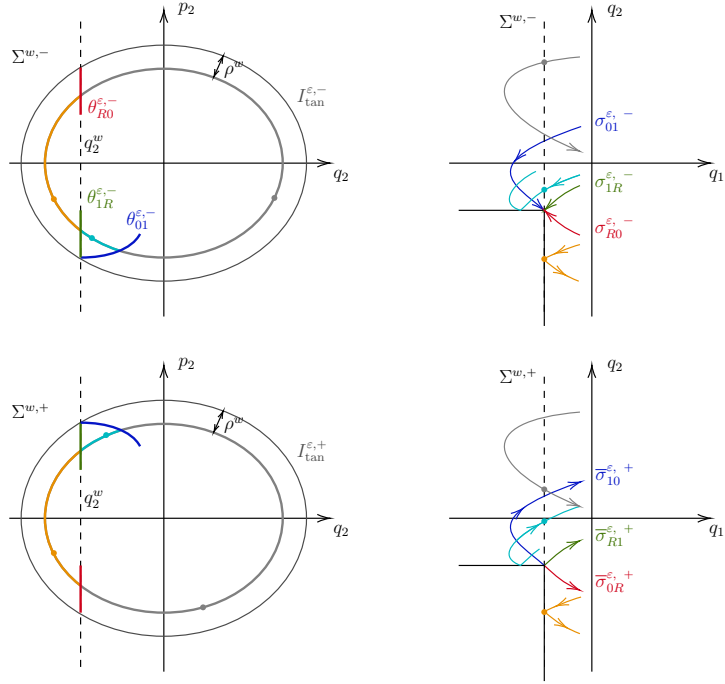}}
  \caption{The structure of the wall sections.  The left column shows the structure of the regions at the wall sections phase plane (for simplicity, for the unperturbed system). In particular the corner singularity curves (red, blue and green curves) and  a fixed $\rw$ circle are shown. The  yellow,  gray and cyan arcs on the $\rw$-circle belong to $J_R, J_0$ and $J_1$ respectively. On the right column, a few trajectory segments corresponding to the $\rw$-circle are shown in the configuration space; the red, blue and black segments correspond to the intersection of the circle with the corner singularity curves. The yellow ($J_R$), gray ($J_0$) and cyan ($J_1$) trajectory segments intersection with the wall sections are shown as solid dots on both columns.  }    \label{fig:wallsection}
\end{figure}
Hence,  for all sufficiently small  $\eps$, for a fixed  $\rw \in [0,\rw_{max}(h)]$, the incoming horizontal velocity at  $\Sigma_h^{w,-} $  is monotone increasing in $\rw$ and either bounded away from zero, or, for small $\rw$, proportional to $\sqrt{\rw}$:
$$\begin{array}{ll}\label{eq:p1epsUpsilon}
   p_1^{\eps,-}   (\Upsilon(\theta,\rw;h,\eps)) & = -\sqrt{2\Upsilon(\theta,\rw;h,\eps)} \\
     &=\begin{cases}
         -\sqrt{2\rw\omega_2(\It(h))
+\mathcal{O}((\rw)^2,\eps\rw))} &  \rw < c  \\
          -\sqrt{2 \hat \Upsilon(\rw;h)} +\mathcal{O}(\eps) &  \rw>c.
     \end{cases}
\end{array}
$$

Recall that for any $\rw \in [0,\rw_{max}(h)]$ the curve with a fixed $\rw$ intersects the line  $q_2=q_2^w$
transversely, at two symmetric points with opposite vertical momenta. For any $\rw$ in this interval the equation $q_2^{\eps,-}(\theta,\rw)=q_2^w<0$ has two isolated solutions,  $\theta_{R0}^{\eps,-}(\rw)<0$, and  $\theta_{1R}^{\eps,-}(\rw)=-\theta_{R0}^{\eps,-}(\rw)>0$ where  $p_2^{\eps,-}(\theta_{R0}^{\eps,-}(\rw),\rw)=-p_2^{\eps,-}(\theta_{1R}^{\eps,-}(\rw),\rw)>0$, see Figure \ref{fig:wallsection}. Moreover, since $\frac{\partial q_2^{\eps,-}(\theta, \rw) }{\partial \theta}|_{q_2^w} =\frac{\partial q_2^{0,-}(\theta,\rw) }{\partial \theta}|_{q_2^w} +O(\eps) $,  $p_2^{0,-}(\theta_{R0}^{0,-}(\rw),\rw)= \frac{\partial q_2^{0,-}(\theta,\rw) }{\partial \theta} \omega_2(\It(h)-\rw)>0$, and $\frac{d }{d \rw} q_2^{\eps,-}(\theta_{R0}^{\eps,-}(\rw)), \rw)\equiv 0$, we conclude that
\begin{equation}\label{eq:dthetadrhor0}
    \frac{d}{d\rw}\theta_{R0}^{\eps,-}(\rw)=-\frac{d}{d\rw}\theta_{1R}^{\eps,-}(\rw) =-\frac{\frac{\partial q_2^{\eps,-}(\theta, \rw) }{\partial \rw}|_{(\theta_{R0}^{\eps,-}(\rw),\rw)}}{\frac{\partial q_2^{\eps,-}(\theta, \rw) }{\partial \theta}|_{(\theta_{R0}^{\eps,-}(\rw),\rw)}}>0
\end{equation} and that \begin{equation}\label{eq:dq2dthet}\frac{\partial q_2^{\eps,-}(\theta, \rw) }{\partial \theta}|_{(\theta_{R0}^{\eps,-}(\rw),\rw)}>0, \frac{\partial q_2^{\eps,-}(\theta, \rw) }{\partial \theta}|_{(\theta_{1R}^{\eps,-}(\rw),\rw)}<0\     
\end{equation}
Since the equation   $q_2^{\eps,-}(\theta,\rw)=q_2^w $ has a solution also for negative $\rw$ (corresponding to circles parameterizing the section $\Sigma_h^>$), the leading order behavior for small $\rw$ is regular and of the form:
\begin{align}\label{eq:thetar0expan}
    \theta_{R0}^{\eps,-}(\rw) &=   -\theta_{1R}^{\eps,-}(\rw)= \theta_{R0}^{\eps,-}(0)+a \rw  + O((\rw)^2), 
 \\ &\\
 a  &= -\frac{\frac{\partial q_2^{\eps,-}(\theta, \rw) }{\partial \rw}|_{(\theta_{R0}^{\eps,-}(0),0)}}{\frac{\partial q_2^{\eps,-}(\theta, \rw) }{\partial \theta}|_{(\theta_{R0}^{\eps,-}(0),0)}}=-\frac{\frac{\partial q_2^{0,-}(\theta, \rw) }{\partial \rw}|_{(\theta_{R0}^{0,-}(0),0)}}{\frac{\partial q_2^{0,-}(\theta, \rw) }{\partial \theta}|_{(\theta_{R0}^{0,-}(0),0)}} + O(\eps)\\ &=
 - \frac{\omega_2(\It(h)-\rw)\frac{\partial q_2^{0,-}(\theta, \rw) }{\partial \rw}|_{(\theta_{R0}^{0,-}(0),0)}}{p_2^{0,-}(\theta_{R0}^{0,-}(\rw),\rw)} + O(\eps)\\&=-\frac{d\tw}{dI}+O(\eps)=\frac{1}{2} \tau_1 +O(\eps)>0
 \end{align} 
 where for the last line we used \eqref{eq:deftau1} and the fact that at $\eps=0$ the wall phase $\theta_{1R}^{\eps=0,-}(\rw)$ is simply $\tw(\It(h)-\rw)$.

\paragraph{Behavior near the separating angles - the corner-singularity curves }
Next, we assert that there exists a positive $\rw_{R}(h)\le \rw_{max}(h)$ such that, for sufficiently small $\eps$,    $\sigma_{R0}^{\eps,-}=\{(\theta^*,\rw)| \theta^*=\theta_{R0}^{\eps,-}(\rw),\rw \in [0,\rw_R(h)]\}$ whereas    $\sigma_{1R}^{\eps,-}=\{(\theta^*,\rw)| \theta^*=\theta_{1R}^{\eps,-}(\rw),\rw \in [0,\rw_R(h)]\}$.

For all $\rw \in [0,\rw_{max}(h)]$,  since $q_2^w<0$ and  $h>h^w_\eps$,  for sufficiently small $\eps$, both $|\Theta_u(\rw)|$ and $|\Theta_R(\rw)|$ are bounded away from zero. For $\theta \in \Theta_u(\rw)=(\theta_{R0}^{\eps,-}(\rw),\theta_{1R}^{\eps,-}(\rw))$ the initial condition   $z_-^w(\theta,\rw) \in \Sigma_h^{w,-}$ crosses the line $q_1=q_1^w$ above the step, whereas for the complimentary angles,   $\theta \in \Theta_R(\rw)\coloneqq [-\pi,\pi] \setminus \Theta_u(\rw)$ the  initial condition   $z_-^w(\theta,\rw)$ impacts the right side of the step; 
Indeed, by \eqref{eq:dq2dthet},  at  $\theta_{R0}^{\eps,-}(\rw)$ (respectively, at $ \theta_{1R}^{\eps,-}(\rw)$)  the function $q_2^{\eps,-}(\theta^*,\rw)$ is monotone increasing (respectively, decreasing) in $\theta$ and, by its smooth dependence on $\eps$, for sufficiently small $\eps$ it intersects the line  $q_2=q_2^w$ only at these two end points. Since the function is smooth, it follows that  $q_2^{\eps,-}(\theta^*,\rw)>q_2^w$ for all $\theta^* \in \Theta_u(\rw)$ and   $q_2^{\eps,-}(\theta^*,\rw)\le q_2^w$ for all  $\theta^* \in \Theta_R(\rw)$.  Since  for all $\rw>0$ we established that $p_1^{\eps,-}   (\Upsilon) <0$  (see \eqref{eq:p1epsUpsilon}), it follows that $  J_R^{\eps,-} = \{\theta \in \Theta_R(\rw),\rw \in (0, \rw_{max}(h)]\}$.

We complete this part of the proof  by establishing that there exists $\rw_1\le \rw_{max}(h)$ such that for   $\rw \in [0,\rw_1]$, initial conditions that  cross  $\Sigma_h^{w,-}$ with angles in  the interior of  $\Theta_u(\rw)$ , bounded away of   $\Theta_u(\rw)$'s  boundary (see below), cross subsequently  $\Sigma_h^{w,+}$ without hitting the step. We then establish the same result for  initial conditions that  cross  $\Sigma_h^{w,-}$ just above $q_2^w$ , namely with angles close to the left boundary of   $\Theta_u(\rw)$,  $\theta_{R0}^{\eps,-}(\rw)$. Finally,  we establish that those that cross close to the right boundary, namely near  $\theta_{1R}^{\eps,-}(\rw)=-\theta_{R0}^{\eps,-}(\rw)$,  hit the upper part of the step exactly once before crossing  $\Sigma_h^{w,+}$.  This proves that  $\theta_{1R}^{\eps,-}(\rw),\theta_{R0}^{\eps,-}(\rw)$ provide, respectively, parameterization of $\sigma^{\eps,-}_{1R},\sigma^{\eps,-}_{R0}$ as claimed.

First, note that there exists a finite $\rw_1(h)$, such that for all $\rw \in [0,\rw_1]$,  for sufficiently small $\eps$, both $|\Theta_u(\rw)|$ and $|\Theta_R(\rw)|$ are bounded away from zero, and that $\Delta t^\eps(\theta^*,\rw)$, the passage time of the smooth flow between   $\Sigma_h^{w,-}$ and   $\Sigma_h^{w,+}$, is bounded, monotone, and small in $\rw$ (i.e., for any $\delta>0$ there exist $\rw_1,\eps_1$ such that $\Delta t^\eps(\theta^*,\rw)<\delta$ for all  $\rw \in [0,\rw_1]$,  $|\eps|\le \eps_1 $, to avoid cumbersome notation we do not insist on this formal setting).

Indeed, since  $\dot p_1|_{\Sigma_h^{w,-}}=-V_1'(q_1^w)+O(\eps)>0$ the unperturbed and perturbed passage times are close. Thus, and since $ p_1^{\eps,-}=p_1(\theta^*,\rw)  $ is monotone in $\rw$,  one can choose $\rw_1(h)$ such that the dependence of the travel time on $\rw$ is regular and monotone in $\rw$, so, say, for all $\rw \in [0,\rw_1]$  and sufficiently small $\eps$,   $\Delta t^\eps(\theta^*,\rw)<2 \Delta t^0(\rw_1)$, where $\Delta t^0(\rw)$ denotes the passage time of the unperturbed smooth flow between these sections. The near tangent orbits that cross  $\Sigma_h^{w,-}$ satisfy  $p_1(\theta^*,\rw) \propto -\sqrt{\rw}$, hence, for sufficiently small $\eps$, the passage time $\Delta t^\eps(\theta^*,\rw)$, is of order  $\sqrt{\rw}$   (see Eq. 4.24 in  \cite{PnueliRomKedarTangency}, with $\Delta t^\eps = t_\star+\hat t_\star$):
\begin{equation}\label{eq:Deltadepend}\begin{array}{ll}
  \Delta t ^\eps(\theta^*;\rw) 
    & = \Delta t^0 (\rw)(1+\mathcal{O}(\sqrt{\rw},\eps)) \\
     & = (2\frac{\sqrt{2\rw\omega_2(\It(h))}}{-V_1'(q_1^w)}+O(\rw))(1+\mathcal{O}(\sqrt{\rw},\eps)),
\end{array}  
\end{equation}where $\mathcal{O}(\sqrt{\rw},\eps)$ denotes smooth functions of all the arguments (i.e. of ($\sqrt{\rw},\eps,\theta$)) which vanish at  $\sqrt{\rw}=\eps=0$).

For small $\rw_1$, the flow above the step is simple to calculate; Since $\dot p_2$ is finite, along the segment  $ t\in[0,\Delta t^\eps]$ the  $p_2$  component of  the segment $\Phi_{t}^{\eps,sm} z_-^w$ changes, at most, by $O(\Delta t^\eps)$. Namely,  there exists a $C>0$ such that $|p_2^{\eps,sm}(t)-p_2^{\eps,-}|< C \Delta t^\eps$ for all   $ t\in[0,\Delta t^\eps]$. Since $p_2^{\eps,-}$ is also bounded,  there exist $\tilde C(\rw_1), \hat C(\rw_1)$ such that $|q_2^{\eps,sm}( t)-q_2^{\eps,-}(\theta^*,\rw)|< \tilde C \Delta t^\eps <  \hat C \Delta t^0(\rw) $ for all    $ t\in[0,\Delta t^\eps]$. In particular, notice that near $\rw=0$, Eq. \eqref{eq:Deltadepend} implies that for all sufficiently small  $\eps$,  $\Delta t^\eps=O(\sqrt{\rw})$ and thus that for all   $ t\in[0,\Delta t^\eps]$, $|p_2^{\eps,sm}(t)-p_2^*|< C \sqrt{\rw}$  and  $|q_2^{\eps,sm}( t)-q_2^{\eps,-}|< \tilde C \sqrt{\rw}$.

In particular, if  $q_2^{\eps,-}(\theta^*,\rw)>q_2^w+2\hat C \Delta t^0(\rw)  $, for sufficiently small $\eps$ the trajectory segment between    $\Sigma_h^{w,-}$ and   $\Sigma_h^{w,+}$ remains bounded away from the step. Since  $h>h^w_\eps$, and thus $|\Theta_u(\rw)|$ is bounded away from zero, one can choose $\rw_1(h)>0$ for which, say,   $|\Theta_u(\rw_1)|>10\hat C \Delta t^0(\rw_1) $. Then, for all  $\rw \in [0,\rw_1]$, there is an open interval in $\Theta_u(\rw)$ for which this condition is satisfied.  Namely, there exists $C_1=C_1(\eps, h), \rw_1(h)$ such that for all sufficiently small  $\eps$ and $\rw\in [0, \rw_1]$ for all 
\begin{equation}\label{eq:J0interior}
    \theta^* \in (\theta_{R0}^{\eps,-}(\rw)+C_1\Delta t^0(\rw),\theta_{1R}^{\eps,-}(\rw)-C_1 \Delta t^0(\rw)) \subset \Theta_u(\rw),
\end{equation}
the trajectory passes above the step, implying that $  q_2^{\eps,-}(\theta^*,\rw)\in J_0^{\eps,-}$.


Now we examine the right and left boundaries of $\Theta_u(\rw)$.  

Left boundary of  $\Theta_u(\rw)$: Take $z_-^w(\theta^*,\rw)$ near $\theta_{R0}^{\eps,-}(\rw)$ with  $\theta^*=\theta_{R0}^{\eps,-}(\rw)+\Delta \theta $  slightly greater than $\theta_{R0}^{\eps,-}(\rw) $ (as $q_2^*$ is monotone increasing in $\theta^*$ there): fix a positive $\Delta_M \theta $ so that both $q_2$ and $p_2$ are monotone increasing on the segment  $\Delta\theta \in(0,\Delta_M \theta)$  for all   $\rw\in [0,\rw_1(h)]$.  This is possible since there exists $c_1=c_1(\rw_1,h)$ such that $p_2^{\eps,-}(\theta_{R0}^{\eps,-}(\rw),\rw)>c_1>0$ for all  $\rw\in [0,\rw_1]$. It follows that we can find a finite  $\Delta_M \theta $ so that  $q_2^{\eps,-}(\theta^*,\rw)>q_2^w $ and  $p_2^*(\theta^*,\rw)= p_2^{\eps,-}(\theta_{R0}^{\eps,-}(\rw),\rw)+O(\Delta_M \theta)  > \frac{1}{2} c_1$ on   $\Delta\theta \in(0,\Delta_M \theta)$ for all sufficiently small $\eps$. Thus, there exists $\rw_{2}=\rw(\Delta_M \theta,h,C_1)\le \rw_{1}$ such that for all  $\rw\in [0,\rw_2]$ ,  $p_2^{\eps,sm}(t)>\frac{1}{4} c_1>0$ for all $t\in [0,\Delta t^\eps(\theta^*,\rw)]$, so $q_2$ strictly increases on this time interval, crossing   $\Sigma_h^{w,+}$ at some $q_2(\Delta t)>q_2^*(\theta^*,\rw)>q_2^w$, and thus it does not impact the upper part of the step. So, for all $\Delta\theta \in(0,\Delta_M \theta)$ the initial condition $z_-^w(\theta_{R0}^{\eps,-}(\rw)+\Delta \theta,\rw)$ belongs to $J^{\epsilon,-}_0$ for all  $\rw\in [0,\rw_{2}]$. Since  $\Delta_M \theta $  is finite, by  \eqref{eq:J0interior} , we conclude that the curve $(\theta_{R0}^{\eps,-}(\rw),\rw)$ indeed corresponds to  $\sigma_{R0}^{\eps,-}$:   for all $\theta^* \in (\theta_{R0}^{\eps,-}(\rw),\theta_{1R}^{\eps,-}(\rw)- C_1 \Delta t^0(\rw) ) $  the initial point $z_-^w(\theta^*,\rw) \in \Sigma_h^{w,-}$ belongs to $J_0^{\eps,-}$.

Right boundary of  $\Theta_u(\rw)$: Take $z_-^w(\theta^*,\rw)$ near $\theta_{1R}^{\eps,-}(\rw)$ with $\theta^*$  slightly smaller than $\theta_{1R}^{\eps,-}(\rw) $: let  $\theta^*=\theta_{1R}^{\eps,-}(\rw)-\Delta \theta $ with  $\Delta \theta \in (0,\Delta_L)$ so that $q_2^{\eps,-}(\theta^*,\rw)>q_2^w $ and  $p_2^{\eps,-}(\theta^*,\rw)= p_2^*(\theta_{1R}^{\eps,-}(\rw),\rw)+O(\Delta \theta)  <- \frac{1}{2} c_1<0$ (recall that  $p_2^{\eps,-}(\theta_{1R}^{\eps,-}(\rw),\rw)<-c_1<0$). 
Then there exists  $\rw_{3}\le \rw_{2}$ such that for all  $\rw\in [0,\rw_{3}]$ and   $\Delta \theta \in (0,\Delta_L)$, the smooth flow has a strictly negative vertical velocity $p_2^{\eps,sm}(t)=p_2^{\eps,-}(\theta^*,\rw)+O(t)<- \frac{1}{4} c_1<0$ on the time interval  $\Delta t^\eps(\theta^*,\rw)$. Hence,  $q_2^{\eps,sm}(t)$ is strictly decreasing on this time interval, and thus there exist $0<\Delta_L'<\Delta_L$ for which, for all  $\Delta \theta \in (0,\Delta_L')$, $q_2^{\eps,-}(\theta^*,\rw)-q_2^w$ is smaller than, say, $\Delta t^\eps(\theta^*,\rw) c_1/8$, so there exists $t^{hit} <\Delta t^\eps$ such that  $\Phi_{t^{hit}}^{\eps,sm}(q_2^*,p_2^*)=(q_1^w +p_1^*t^{hit}+O((t^{hit})^2 ),p_1^*+O(t^{hit}),q_2^w,p_2^*+O(t^{hit}))$, namely, the trajectory hits the upper part of the step with strictly negative vertical velocity, $p_2^{hit}<- \frac{1}{2} c_1+O(t^{hit})$.  After this impact the trajectory must cross   $\Sigma_h^{w,+}$ at some $q_2(\Delta t)>q_2^w$ with no additional impact of the step; Indeed, by the impact rule, just after $t^{hit}$, at $t^{hit+}$, we have 
\begin{equation}\label{eq:thit}
\begin{array} {ll}
 \Phi_{t^{hit+}}^{\eps,im}(q_2^*,p_2^*)&=\mathcal{R}_2 \Phi_{t^{hit}}^{\eps,sm}(q_2^*,p_2^*) \\
    &=(q_1^w +p_1^*t^{hit}+O((t^{hit})^2 ),p_1^*+O(t^{hit}),q_2^w,-p_2^*+O(t^{hit}))
\end{array}
\end{equation}
where $-p_2^*>\frac{1}{2} c_1+O(t^{hit})>0$, and, since $t^{hit},\Delta t^\eps $ are sufficiently small compared with $c_1/2$ for all  $\rw\in [0,\rw_{3}]$, $p_2$ remains positive on the time interval $\Delta t^\eps -t^{hit}$ and thus $q_2$ increases on this segment and thus it cannot impact the upper part of the step again. So indeed,  $z_-^w(\theta^*,\rw)\in J^{\epsilon,-}_1$ for all $\rw\in [0,\rw_{3}]$ and  $\Delta \theta \in (0,\Delta_L')$. 

Setting $\rw_R=\rw_3\le\rw_2$,  we proved that for all   $\rw\in [0,\rw_{R}]$, for sufficiently small $\eps$ the curves $(\theta_{R0}^{\eps,-}(\rw),\rw)$ and $(\theta_{1R}^{\eps,-}(\rw),\rw)$ corresponds to $\sigma_{R0}^{\eps,-}$  and   $\sigma_{1R}^{\eps,-}$, the right-incoming corner singularity lines at the wall. 

\paragraph{The corner-singularity angle that separates $J_0$ and $J_1$.}

Finally, we find $(\theta_{01}^{\eps,-} (\rw), \rw)$, the boundary between  $J^{\epsilon,-}_0$ and  $J^{\epsilon,-}_1$. By the above, for all $\rw \in  [0,\rw_{3}] $, the angle $\theta_{01}^{\eps,-} (\rw)$ must belong to the interval $(\theta_{1R}^{\eps,-}(\rw)-C_1 \Delta t^0(\rw),\theta_{1R}^{\eps,-}(\rw)) $, where $C_1$ of \eqref{eq:J0interior} is chosen so that $z_-^w(\theta_{1R}^{\eps,-}(\rw)-\Delta \theta,\rw) \in J^{\epsilon,-}_0$  at $\Delta \theta = C_1  \Delta t^0(\rw)$ for all $\rw \in [0,\rw_3]\subset [0,\rw_1]$.

For a fixed positive $\rw$, examine the one-parameter family of initial conditions $z_-^w(\theta_{1R}^{\eps,-}(\rw)-\Delta \theta,\rw)$ with small $\Delta \theta$; Take, for any given $C_1>0$,  a $\rw_{4}\le\rw_3$ such that for all $\rw\le\rw_{4}$   for all  $\Delta \theta \in [0, C_1 \Delta t^0(\rw)]$,   it is still true that  $q_2^{\eps,-}(\theta_{1R}^{\eps,-}(\rw)-\Delta \theta,\rw) $  is monotone increasing in $\Delta \theta$ with order one derivative (it follows from  \eqref{eq:dq2dthet} that such a $\rw_4$ exists), and  that  $p_2^{\eps,sm}(t)<-\frac{1}{4} c_1<0$ for all $t \in [0,\Delta t^\eps]$ (recall that $\Delta t^\eps$  is small for  $\rw \in  [0,\rw_{4}] $).

 Let  $q_2^{\eps,art}(\Delta \theta,\rw)$  denote the $q_2$ components of $\Phi_{\Delta t^\eps}^{\eps,sm}(z_-^w(\theta_{1R}^{\eps,-}(\rw)-\Delta \theta,\rw))$. 
  By the smooth dependence of  $\Phi_{t}^{\eps,sm}$  on i.c. and time, and the dependence of $\Delta t^\eps$ on its arguments, for any fixed $\rw \in  [0,\rw_{4}] $, the curve $\{q_2^{\eps,art}(\Delta \theta,\rw), \Delta \theta \in [0, C_1 \Delta t^0(\rw)]\}$, is smooth in $\Delta \theta$.  By the choice of $C_1$,   $q_2^{\eps,art}( C_1 \Delta t^0(\rw),\rw)>q_2^w$ whereas  
 $q_2^{\eps,art}(0,\rw)< q_2^w$, indeed, $ q_2^{\eps,art}(0,\rw)-q_2^w =\int_0^{\Delta t^\eps}   p_2^{\eps,sm}(t;\theta_{1R}^{\eps,-}(\rw),\rw) dt$ and the integrand, the vertical momentum, is strictly negative on this time interval. Next, we show next that the curve $q_2^{\eps,art}(\Delta \theta,\rw)$ is monotone in $\Delta \theta$. To this aim we
 integrate the equations of motion on the small time interval $\Delta t^\eps$: \begin{equation}\label{eq:q2artDeltat}\begin{array}{ll}
     q_2^{\eps,art}(\Delta \theta,\rw) &- q_2^{\eps,-}(\theta_{1R}^{\eps,-}(\rw)-\Delta \theta,\rw)\\
      &=    \int_0^{\Delta t^\eps (\theta_{1R}^{\eps,-}(\rw)-\Delta \theta;\rw)}   p_2^{\eps,sm}(t;\theta_{1R}^{\eps,-}(\rw)-\Delta \theta,\rw) dt 
      \\
      & =   \Delta t^\eps (\theta_{1R}^{\eps,-}(\rw)-\Delta \theta;\rw)   p_2^{\eps,sm}(\Delta t';\theta_{1R}^{\eps,-}(\rw)-\Delta \theta,\rw),\\
       & =   \Delta t^\eps (\theta_{1R}^{\eps,-}(\rw)-\Delta \theta;\rw)   p_2^{\eps,-}(\theta_{1R}^{\eps,-}(\rw)-\Delta \theta,\rw)(1+O(\Delta t^\eps(\cdot))),
 \end{array}
 \end{equation}
where, for the third line we use the mean value theorem, with some $\Delta t'\in (0,\Delta t (\theta_{1R}^{\eps,-}(\rw)-\Delta \theta;\rw))$ and for the fourth one the fact that $p_2^{\eps,sm}(t)$ is bounded away from zero and depends smoothly on its arguments. Notice that  by \eqref{eq:Deltadepend} the right hand side dependence on  $\Delta \theta$ is of higher order in $(\sqrt{\rw},\eps)$, so, the choice of $\rw_4$ insures the monotone dependence of  $q_2^{\eps,-}(\theta_{1R}^{\eps,-}(\rw)-\Delta \theta,\rw)$ on  $\Delta \theta$, as claimed.

Hence, the curve  intersects transversely the line $q_2=q_2^w$ at a unique positive $\Delta\theta^*(\rw)$.  For   $\Delta\theta \in [0,\Delta\theta^*(\rw))$ we showed that $q_2^{\eps,art}(0,\rw)<q_2^w$ so the upper part of the step is hit before returning to   $\Sigma_h^{w,+}$ whereas for   $\Delta\theta \in (\Delta\theta^*(\rw), C_1 \Delta t^0(\rw)]$ the return to   $\Sigma_h^{w,+}$ occurs with $q_2^{\eps,art}(\Delta\theta,\rw)>q_2^w$. 

Defining $  \theta_{01}^{\eps,-}(\rw)= \theta_{1R}^{\eps,-}(\rw)-\Delta\theta^*(\rw)$, let  $\vartheta =\Delta \theta^* (\rw)-\Delta \theta =   \theta_{1R}^{\eps,-}(\rw)-\theta_{01}^{\eps,-}(\rw)-\Delta \theta$, and $z_-^w(\theta_{01}^{\eps,-}(\rw)+ \vartheta,\rw)=z_-^w(\theta_{1R}^{\eps,-}(\rw)-\Delta \theta,\rw)$.  
We showed above that  $z_-^w(\theta_{01}^{\eps,-}(\rw) ,\rw) $,  hits the step exactly at the corner point,   that for   $\vartheta \in [\theta_{1R}^{\eps,-}(\rw)-\theta_{01}^{\eps,-}(\rw)- C_1 \Delta t^0(\rw),0)$ the segment emanating from $z_-^w(\theta_{01}^{\eps,-}(\rw)+\vartheta ,\rw) $  does not hit the step, namely it is in $ J^{\epsilon,-}_0$ (so, by \eqref{eq:J0interior}, we conclude that  for all  $\theta \in (\theta_{R0}^{\eps,-}(\rw),\theta_{01}^{\eps,-}(\rw))  $ we have that $z_-^w( \theta,\rw) \in J^{\epsilon,-}_0$), and that  for   $\vartheta \in (0,\theta_{1R}^{\eps,-}(\rw)-\theta_{01}^{\eps,-}(\rw)]$, the segment emanating from $z_-^w(\theta_{01}^{\eps,-}(\rw)+\vartheta ,\rw) $ hits the upper part of the step. We now show that on this segment  there is a single impact. Indeed, since we established that $p_2$ is strictly negative on this segment, after the reflection it becomes strictly positive, so it remains strictly positive at the remaining (small) flight time. Hence, there can be no additional impacts with the upper part of the step. Namely, for all  $\theta \in (\theta_{01}^{\eps,-}(\rw),\theta_{1R}^{\eps,-}(\rw))  $  we have that $z_-^w(\theta,\rw) \in J^{\epsilon,-}_1$.  We thus established that for all  $\rw \in  [0,\rw_{4}] $ the corner singularity curve $\sigma_{01}^{\eps,-}$ is given by the curve $(\theta_{01}^{\eps,-}(\rw),\rw)$. 
 
\paragraph{Asymptotic form for the corner-singularity angles.}

 Since  $q_2^{\eps,art}(\Delta\theta^*(\rw),\rw)=q_2^w$  and $q_2^{\eps,-}(\theta_{1R}^{\eps,-}(\rw),\rw)=q_2^w$, and since, by the mean-value theorem, there exists a $\Delta \theta'\in(0 ,\Delta \theta)$ such that $q_2^{\eps,-}(\theta_{1R}^{\eps,-}(\rw)-\Delta \theta,\rw)=q_2^{\eps,-}(\theta_{1R}^{\eps,-}(\rw),\rw)-\Delta \theta \frac{\partial q_2^{\eps,-}(\theta_{1R}^{\eps,-}(\rw)-\Delta \theta',\rw)}{\partial \theta}$   and since $q_2^{\eps,-}(\theta,\rw)$ depends smoothly on $\theta$ we obtain, using \eqref{eq:q2artDeltat}: 
\begin{equation}\begin{array}{ll}
     \Delta \theta^*(\rw) &=   \Delta t (\theta_{1R}^{\eps,-}(\rw)-\Delta \theta^*;\rw)   \frac{p_2^{\eps,sm}(0;\theta_{1R}^{\eps,-}(\rw)-\Delta \theta,\rw)(1+O(\Delta t)) }{ \frac{\partial q_2^{\eps,-}(\theta_{1R}^{\eps,-}(\rw)-\Delta \theta',\rw)}{\partial \theta}}  \\
      &=   \Delta t^0 (\rw)\omega_2(\It(h))(1+\mathcal{O}(\sqrt{\rw},\eps) ).
 \end{array}
 \end{equation}
This computation provides the natural approximation: the shift in the angle corresponds, to leading order, to the unperturbed  flight time from    $\Sigma_h^{w,-}$ and   $\Sigma_h^{w,+}$ times the unperturbed frequency. 
Combining the above analysis with \eqref{eq:thetar0expan},  and \eqref{eq:Deltadepend} shows that
 for sufficiently  small $\rw$  this curve can be approximated by 
\begin{equation}\label{eq:theta01expan}
    \theta_{01}^{\eps,-}(\rw)= \theta_{1R}^{\eps,-}(\rw)-
  (2\frac{\sqrt{2\rw\omega_2^3(\It(h))}}{-V_1'(q_1^w)}+O(\rw))(1+\mathcal{O}(\sqrt{\rw},\eps))
\end{equation} 
showing the square root singularity of  $\theta_{01}^{\eps,-}(\rw)$ .

\paragraph{Non-intersecting division to regions.}

 In summary, we have shown that for $h>h^w_\eps$, there exists  $\Delta = \rw_4(h)>0$, such that for all   $\rw \in  [0,\Delta] $, for sufficiently small $\eps$, the curves  $(\theta_{1R}^{\eps,-}(\rw),\rw)$,  $(\theta_{R0}^{\eps,-}(\rw),\rw)$  and  $(\theta_{01}^{\eps,-}(\rw),\rw)$ are monotone in $\rw$, non-intersecting, and separating the  regions $(J^{\epsilon,-}_R,J^{\epsilon,-}_0, J^{\epsilon,-}_1)$ on the cylinder, in this order, completing the proof of Lemma \ref{lemma:border}. 
 \end{proof}

 \subsection{Proofs of Theorems \ref{thm:cornersing} and \ref{thm:regionsJdyn}}\label{sec:proofofcornersing}

We establish next that the pre-images of the wall corner singularity curves to the section $\Sigma_h$ lead to formula \eqref{eq:cornersingbordrho}.   Since, in between the corner singularity lines, the map $\mathcal{F}^{sm,-}_\eps$ is smooth and symplectic, mapping  the regions  $J_b^{\eps}$ to  $J_b^{\eps,-}$, this completes the proofs of Theorems \ref{thm:cornersing} and \ref{thm:regionsJdyn}.

 Lemma 4.4 of  \cite{PnueliRomKedarTangency} implies that near the tangency curve, for $K\le 0$, the mapping from $(\phi, K)$  to $(\theta^w,\rw)$ is a homeomorphism which is smooth in $\sqrt{\rw},\sqrt{-K}$. 
 Since our map $\mathcal{F}^{sm,-}_\eps$ coincides with that used in  \cite{PnueliRomKedarTangency}, the same result applies here. 

Thus, the pre-images of the curves $(\theta^{\eps,-}_{ab}(\rw),\rw)$ are parametric curves of the form $(\phi(\theta^{\eps,-}_{ab}(\rw),\rw),K(\theta^{\eps,-}_{ab}(\rw),\rw))$. We first show that these parametric curves can be expressed as graphs of $\sqrt{-K}$,    and then that their asymptotic form are exactly  $(\phi^{\eps}_{ab}(\sqrt{-K}),K)$ of  \eqref{eq:cornersingbordrho}.  

To express the parametric curves as graphs of  $\sqrt{-K}$,  we need to show that   $\frac{d \sqrt{-K}(\theta^{\eps,-}_{ab}(\rw),\sqrt{\rw},\eps;h) }{d \sqrt{\rw}}\ne 0$. 
Since\footnote{see above Eq. (4.32)}
\begin{equation}\label{eq:thetaofthetawall}\begin{array}{ll}
   \sqrt{-K}(\theta^w,\sqrt{\rw},\eps;h)&= \sqrt{\rw}(1+\eps G_-(\sqrt{\rw},\theta^w,\eps;h)),
\end{array}     
\end{equation} 
we obtain
\begin{equation}\begin{array}{cc}
    \frac{d \sqrt{-K}(\theta^{\eps,-}_{ab}(\rw),\sqrt{\rw},\eps;h) }{d \sqrt{\rw}} =&  1+\eps G_-(\sqrt{\rw},\theta^{\eps,-}_{ab}(\rw),\eps;h)\\
     & +  \eps \sqrt{\rw} \frac{\partial G_-() }{\partial \sqrt{\rw}}+\eps \sqrt{\rw} \frac{\partial G_-() }{\partial \theta^w} \frac{d \theta^{\eps,-}_{ab}(\rw) }{d \sqrt{\rw}},
\end{array}
\end{equation}
 since $G_-()$ is bounded and smooth in its arguments, we only need to verify that $\eps \sqrt{\rw}\frac{d \theta^{\eps,-}_{ab}(\rw) }{d \sqrt{\rw}}$ remains small.   For $\theta^{\eps,-}_{R0}$ and $\theta^{\eps,-}_{1R}$ this follows directly from \eqref{eq:thetar0expan} as $\frac{\partial \theta^{\eps,-}_{R0,1R}(\rw) }{\partial \sqrt{\rw}}\propto \sqrt{\rw}$.  For   $\theta^{\eps}_{01}$, Eq. \eqref{eq:theta01expan} shows that there is a leading $\sqrt{\rw}$ term, so:
\begin{equation}
   \frac{\partial \theta^{\eps,-}_{01}(\rw) }{\partial \sqrt{\rw}} = -2\frac{\sqrt{2\omega_2(\It(h))^3}}{-V_1'(q_1^w)}(1+\mathcal{O}(\sqrt{\rw},\eps)), 
\end{equation}hence $\eps \sqrt{\rw}\frac{\partial \theta^{\eps,-}_{01}(\rw) }{\partial \sqrt{\rw}}=O(\eps \sqrt{\rw} )$. We conclude that indeed the corner singularity curves are graphs of the form  $(\phi^{\eps}_{ab}(\sqrt{-K}),K)$.  

Next we establish  \eqref{eq:cornersingbordrho}.  
Applying Eq. (4.29) of   \cite{PnueliRomKedarTangency} to the parametric curves we obtain:   
\begin{equation}
    \phi(\theta^{\eps,-}_{ab}(\rw),\rw)= \phi_{tan}(\theta^{\eps,-}_{ab}(\rw))-\frac{1}{2}\tau_0 K(\theta^{\eps,-}_{ab}(\rw),\rw) +\lambda \sqrt{\rw} + \mathcal{O}(\eps,\eps \sqrt{-K},K^2)
\end{equation} where $\phi_{tan}(\theta)$ denotes the pre-image of $\theta$ on the circle $\rw=0$, which is $\eps$ close to a rotation by $-\frac{1}{2}\Omega_0$, so we may write $\phi_{tan}(\theta^{\eps,-}_{ab}(\rw))=\phi_{tan}(\theta^{\eps,-}_{ab}(0))+(\theta^{\eps,-}_{ab}(\rw)-\theta^{\eps,-}_{ab}(0))+O(\eps)$.

By definition,  $ \phi^{\eps}_{R0}(0)=\phi_{tan}(\theta^{\eps,-}_{R0}(0))$ and  $ \phi^{\eps}_{1R}(0)=\phi_{tan}(\theta^{\eps,-}_{1R}(0))$, so, replacing $\rw $ by $-K$, we get  from \eqref{eq:thetar0expan} that
  $ \phi^{\eps}_{R0}(K)= \phi^{\eps}_{R0}(0)+\lambda \sqrt{-K} -\frac{1}{2}\tau_1 K -\frac{1}{2}\tau_0 K + \mathcal{O}(\eps,\eps \sqrt{-K},K^2)$
and
  $ \phi^{\eps}_{1R}(K)= \phi^{\eps}_{1R}(0)+\lambda \sqrt{-K} +\frac{1}{2}\tau_1  K -\frac{1}{2}\tau_0 K + \mathcal{O}(\eps,\eps \sqrt{-K},K^2)$
and from \eqref{eq:theta01expan} that
  $ \phi^{\eps}_{01}(K)= \phi^{\eps}_{1R}(0)+\lambda \sqrt{-K} +\frac{1}{2}\tau_1 K -\frac{1}{2}\tau_0 K -2\lambda \sqrt{-K} + \mathcal{O}(\eps,\eps \sqrt{-K},K^2)$, proving   \eqref{eq:cornersingbordrho}.  The dependence of $\phi^{\eps}_{ab}(0)$ on $\eps$ is smooth (since it corresponds to the angle along the tangent circle), and its asymptotic form follows from the construction of $\theta^{\eps,-}_{1R,R0}(0)$ (which are, asymptotically, just $\pm \tw$) and the fact that to leading order in $\eps$ the rotation on the tangent circle is $\Omega_0$.


\subsection{Time reversal symmetry and the corner singularity curves\label{sec:timereversal}}  

 Applying Lemma \ref{lem:timerevgen} to the region  $\mathcal{B}^\eps\subset \Sigma_h$, where the return map is well defined for all initial conditions that do not hit the corner, we establish:
\begin{remark} \label{thm:timereversal}The return map of the oscillators-step system obeys, for all initial conditions in $\mathcal{B}^\eps\setminus \{(\sigma_{01}^\eps\cup \sigma_{1R}^\eps\cup \sigma_{R0}^\eps)\}$, the time-reversal symmetry of Eq.   \eqref{eq:timereversal}.
\end{remark}

Next,  we prove Theorem \ref{thm:reversalinkphi} regarding the form of the reversal symmetry in the normal coordinates:
\begin{proof}
 Recall that $(\phi,K)=S(\theta,I)=(\theta,I-\It^\eps(\theta)), {S}^{-1}(\phi,K)=(\theta,I)=(\phi,K+\It^\eps(\phi))$
 so, the map in the $(\phi,K) $ coordinates is $(\bar \phi,\bar K)= S \mathcal{F}_\eps {S}^{-1} (\phi,K)={\mathcal{ F}_\eps}(\phi,K)$ and its inverse is ${\mathcal{ F}_\eps}^{-1} =  S \mathcal{F}_\eps^{-1} {S}^{-1} (\phi,K)$.

The time reversal symmetry in the $(\theta,I)$ coordinates is
 $\mathcal{R}_2 \mathcal{F}_\eps(\theta,I) =\mathcal{F}^{-1}_\eps \mathcal{R}_2(\theta,I)$, hence  $S \mathcal{R}_2 \mathcal{F}_\eps {S}^{-1}(\phi,K) =S \mathcal{F}^{-1}_\eps \mathcal{R}_2 {S}^{-1}(\phi,K)$.
 
  Notice that 
 $S \mathcal{R}_2= S (-\theta,I)=(-\theta,I-\It^\eps(-\theta))$ 
 whereas  $\mathcal{R}_2 S = \mathcal{R}_2 (\theta,I-\It^\eps(\theta))=(-\theta,I-\It^\eps(\theta))$ so 
 $S \mathcal{R}_2= P \mathcal{R}_2 S $ where $P$ is defined by \eqref{eq:defP} with  $\eps f(\phi)=\It^{\eps}(-\phi)-\It^{\eps}(\phi)$ (so indeed  $P(-\theta,I-\It^\eps(\theta))=(-\theta,I-\It^\eps(-\theta))$), and thus $ \mathcal{R}_2 S^{-1} (\phi,K) = S^{-1} P \mathcal{R}_2  (\phi,K) $.

 Notice that $ P \mathcal{R}_2 (\phi,K)=P(-\phi,K)=(-\phi,K-\eps f(\phi))=\mathcal{R}(\phi,K)$. 

 It follows that
$S \mathcal{R}_2 \mathcal{F}_\eps {S}^{-1} (\phi,K)= P \mathcal{R}_2 S \mathcal{F}_\eps {S}^{-1} (\phi,K)=\mathcal{R}\mathcal{F}_\eps (\phi,K)$ and  that 
$S \mathcal{F}^{-1}_\eps \mathcal{R}_2 {S}^{-1}(\phi,K)= S \mathcal{F}^{-1}_\eps S^{-1} P \mathcal{R}_2 (\phi,K)=   \mathcal{ F}_\eps^{-1} \mathcal{R}(\phi,K)$, and thus, by the time-reversal symmetry in the $(\theta,I)$ coordinates the time reversal symmetry in the new coordinates with  $\mathcal{R}(\phi,K) =P \mathcal{R}_2  (\phi,K)=(-\phi,K-\eps f(\phi)) $ is established.
\end{proof}
Consistency checks: Note that $\mathcal{R}$   is indeed an involution: $\mathcal{R} \mathcal{R}(\phi,K)=\mathcal{R}(-\phi,\,K-\eps f(\phi))=(\phi,\,K-\eps f(\phi)-\eps f(-\phi)) =(\phi,\,K)$, and that  since $f$ is odd and periodic $f(0)=f(\pi)=0$, so $\mathrm{Fix}(\mathcal{R})=\{(\phi,K)|\phi=0, \phi=\pi)$.  Also note that this relation is consistent with the symmetry found for the tangency curve. In the normal coordinates, the tangency curve is the circle $K=0$, and so its image under the time-reversal symmetry is as expected,  $\bar{\sigma}_{tan}^\eps$:  $\mathcal{R}\sigma^\eps_{tan} =\{(\phi,K)=\mathcal{R}(\phi,0) =(-\phi,-\eps f(\phi)), \phi \in[-\pi,\pi]\}= \{(\phi,\eps f(\phi)), \phi \in[-\pi,\pi]\}=   \bar{\sigma}_{tan}^\eps$.

While we cannot apply the symmetry rules to the corner-singularity curves at which the return map is not defined, we prove next Theorem \ref{thm:reverseJs}, namely, that the symmetry implies the corresponding reflection symmetry for the impacting regions $J_a^\eps$ and their images:

\begin{proof}
The order of the regions   $(J^\epsilon_R,J^\epsilon_0, J^\epsilon_1)$ follows from Theorem \ref{thm:regionsJdyn}. 

We show first that the images of the three regions under first return map,  $\bar{J}_a^\eps = \mathcal{F}_\eps(J_a^\eps)\ (a\in\{0,1,R\})$ ,  are the reflections in $\theta$ of the original regions; If
$(\theta, I)\in J_a^\eps$, $a\in\{0,1,R\}$, then $\mathcal{R}_2 (\theta, I)=(-\theta, I) \in \bar{J}_a^\eps$ , namely, there exists  a point $(\theta^*, I^*)\in  J_a^\eps$  such that
\begin{equation}
\label{eq:refl_sym}
(-\theta, I) = \mathcal{F}_\eps(\theta^*,I^*)\in \bar{J}_a^\eps .
\end{equation}
Indeed, given a $(\theta, I)\in J_a^\eps$, $a\in\{0,1,R\}$, its backward trajectory is also well defined and belongs to the same dynamical region. The backward trajectory of  $(\theta, I)$  is the forward trajectory of $(\theta^*, I^*)=\mathcal{R}_2 \mathcal{F}_\eps (\theta, I)=(-\bar \theta,\bar I)$, so  $(\theta^*, I^*)\in J_a^\eps$ and hence its image is in the image of $J_a^\eps$, namely   $\mathcal{F}_\eps (\theta^*, I^*) \in \bar{J}_a^\eps$. By the time reversal symmetry \eqref{eq:timereversal},  $\mathcal{F}_\eps (\theta^*, I^*)=(-\theta, I)$, proving the claim, and identifying $(\theta^*, I^*)$ as the reflection of the image of $(\theta,I)$.

Next we establish that these symmetries imply that the boundaries of the images of the regions, $\bar{J}_b^\eps$,  that correspond to the corner singularity curves of the inverse map,   $\mathcal{F}^{-1}_\eps$, are just the reflections of other corner singularity curves. 

We established that the images of each of the regions in the $(\theta,I)$ coordinates is simply its reflection in $\theta$. Thus,  if $\bar \sigma_{ab}^\eps$ is the left boundary of $\bar J_b^\eps$, so  $\bar J_a^\eps$ is its left neighbor, their pre-images are at the reverse order, so  $ \sigma_{ba}^\eps$ is indeed well defined and corresponds to the right boundary of  $J_b^\eps$.  In particular, we saw that  if $(\theta,I)\in J_b^\eps $ then  $(-\theta,I)\in \bar J_b^\eps $, hence, taking  $(\theta,I)\in J_b^\eps $ with $\theta$ approaching the right boundary of $J_b^\eps$ ( $\theta \rightarrow^+ \theta_{ba}^\eps(I)$)  implies that $-\theta \rightarrow^-  -\theta_{ba}^\eps(I)$ so 
 $(-\theta_{ba}^\eps(I),I)$ must be the left boundary of  $\bar J_b^\eps$, namely f $\bar \sigma_{ab}^\eps$.    

Since the transformation $S$ is smooth and symplectic this reversal of ordering also applies to the normal coordinates (in fact, notice that $S$ does not change the angle part, namely $\phi=\theta$ and  also  $\bar \phi =\bar \theta$  since
 ${\mathcal{ F}_\eps}(\phi,K)=(\bar \phi,\bar K)= S \mathcal{F}_\eps {S}^{-1} (\phi,K)=S (\bar \theta, \bar I)=(\bar \theta,\bar I-\It^\eps(\bar \theta))$)
 \end{proof}

 Notice that since $(abc) \in p_{cyc}(10R)$, we obtain, consistently, that  $(cba) \in p_{cyc}(R01)=p_{cyc}\circ p_{rev}(10R)$.

\subsection{The return map: proof of Theorem \ref{thm:returnmap}}\label{sec:proofreturnmap}

To compute the return map $\mathcal{F}_\eps$  in regions $J_0,J_R$ we utilize the results regarding the tangency return map of  \cite{PnueliRomKedarTangency}.  Then we calculate the return map for the region $J_1$, thus obtaining \eqref{eq:returnmapfull}, with the remainder terms:
\begin{equation}\label{eq:gcorner}
 G_{a,K}(\phi,K,\eps) \coloneqq \begin{cases}
    \mathcal{O}_{\mathcal{C}^r}(\eps K), & \phi\in \mathcal{J}^\eps_0(K),
\\ \mathcal{O}_{\mathcal{C}^{r-1}}(\eps \sqrt{-K}), & \phi\in \mathcal{J}^\eps_1(K),
\\ \mathcal{O}_{\mathcal{C}^{r-2}}(\eps \sqrt{-K}), & \phi\in \mathcal{J}^\eps_R(K),
\end{cases}
\end{equation}
and
\begin{equation}\label{eq:gcornerphi}
 G_{a,\phi}(\phi,K,\eps) \coloneqq \begin{cases}
    \mathcal{O}_{\mathcal{C}^r}(\eps ), &  \phi\in \mathcal{J}^\eps_0(K),
\\ \mathcal{O}_{\mathcal{C}^{r-1}}(\eps, \eps\sqrt{-K}), & \phi\in \mathcal{J}^\eps_1(K),
\\ \mathcal{O}_{\mathcal{C}^{r-2}}(\eps , \eps\sqrt{-K}), & \phi\in \mathcal{J}^\eps_R(K).
\end{cases}
\end{equation}
Hereafter,  $\mathcal{O}_{\mathcal{C}^{r-1}}(\epsilon \sqrt{-K}) $ denotes a $C^{r-1}$ function of $(\phi,\epsilon,\sqrt{-K})$ which vanishes  at $\epsilon \sqrt{-K}=0$ whereas $\mathcal{O}_{\mathcal{C}^{r-1}}(\epsilon, \epsilon \sqrt{-K}) $ denotes a $C^{r-1}$ function of $(\phi,\epsilon,\sqrt{-K})$  which is $O(\epsilon)$ at $K=0$.  

Theorem 2 of  \cite{PnueliRomKedarTangency} states that for the wall system, namely, the HIS with the Hamiltonian \eqref{eq:perturbedhamil} and a wall at $q_1=\qw_1$,  there exists a symplectic smooth change of coordinates such that near the tangency curve, this singular dependence can be expressed as a  square-root dependence in $K$:
\begin{equation}\label{eq:fwall}
\mathcal{F}_\eps^{tan}:
\begin{cases}
    \bar{K} &= K + \eps f(\bar{\phi}) + \mathcal{O}_{\mathcal{C}^{r-2}}(\eps G_{wall} (K)),\\
    \bar{\phi} &= \phi + \Omega_0 + G_{wall}(K)+\mathcal{O}_{\mathcal{C}^{r-2}}(\eps,\eps  G_{wall} (K),K^2),
\end{cases}    
\end{equation} 

where:  
\begin{equation}\label{eq:gwall}
 G_{wall}(K)=\begin{cases}
    \tau_0 K + \mathcal{O}_{\mathcal{C}^r}(K^2), & K \ge 0
\\ \tau_0 K -2 \lambda\sqrt{-K} + 
\mathcal{O}_{\mathcal{C}^{r-2}}((-K)^{3/2},K^2), & K<0
\end{cases}
\end{equation}
$\Omega_0$ is defined by \eqref{eq:omega0ofI}, $f$ is defined by \eqref{eq:defoff} and $\lambda $ is defined by \eqref{eq:deflambdah}. To establish this result,  it is established that for negative $K$ the mappings from the wall coordinates $(\theta,\rw)$ (see Eq. \eqref{eq:rhow}) to  $(\bar \phi,\bar K)$ and to  $(\phi,K)$ are regular in $\sqrt{\rw}$ and in $\sqrt{-K}$.


Since initial conditions with  $K<0$  belonging to $J_0^\eps$  do not impact, they have a smooth dependence on $K$, which is of the same form as those with $K>0$, so, by the same construction as in   \cite{PnueliRomKedarTangency}, which here amounts to regular perturbation theory, we obtained the map $\mathcal{F}_\eps$, \eqref{eq:returnmapfull}  for such initial conditions:
\begin{equation}\label{eq:mapj0}
    \mathcal{F}_\eps z(\phi,K)|_{z(\phi,K) \in J_0^\eps}=(K + \eps f(\bar{\phi}) + \mathcal{O}(\eps K,\eps^2), \phi + \Omega_0 + \tau_0 K+\mathcal{O}(\eps,\eps K,K^2))
\end{equation}  Initial conditions with  $K<0$  belonging to $J_R^\eps$  impact from the right side of the step, exactly as for the wall system, so $\mathcal{F}_\eps z(\phi,K)|_{z(\phi,K) \in J_R^\eps}=\mathcal{F}_\eps^{tan}z(\phi,K)|_{K<0}$ and thus, the return map is given by \eqref{eq:fwall}  :  \begin{equation}\label{eq:mapjr}
    \mathcal{ F}_\eps z(\phi,K)|_{z(\phi,K) \in J_R^\eps}= \begin{cases}
    \bar{K} &= K + \eps f(\bar{\phi}) + \mathcal{O}_{\mathcal{C}^{r-2}}(\eps \sqrt{-K}),\\
    \bar{\phi} &= \phi + \Omega_0 +  \tau_0 K - 2\lambda \sqrt{-K} + 
\mathcal{O}_{\mathcal{C}^{r-2}}(\eps,\eps \sqrt{-K},K^2),
\end{cases}
\end{equation}So the only computation needed for constructing the return map is for initial conditions belonging to $J_1^\eps$.

\paragraph{Return map for $J_1^\eps$}

\begin{lemma}
In $J_1^\eps$,  in the $(\phi, K) $ coordinates, the return map is of the form:
\begin{equation}    
\label{eq:mapj1}
\mathcal{ F}_\eps|_{z \in J_1^\eps }:
\begin{cases}
    \bar{K} &= K+ \eps f(\bar \phi) + \mathcal{O}_{\mathcal{C}^{r-1}}(\eps \sqrt{-K})\\
    \bar{\phi} &= \phi +  \Omega_{J_1}(\It(h)+K) +  \mathcal{O}_{\mathcal{C}^{r}}(\eps).
\end{cases}
\end{equation}
where $\Omega_{J_1}(\cdot)$ denotes the unperturbed change in the angle for orbits that reflect once from the upper part of the step, namely: 
\begin{equation}\label{eq:Thetaj1}
    \Omega_{J_1}(I)= \frac{2 \pi T_1(h-H_2(I))}{ T_2 (I)}+2 (\pi-\tw(I))=\Omega_0(I)+2 (\pi-\tw(I)).
 \end{equation} 
\end{lemma}
 \begin{proof}
Proof outline: we first construct the return map in the $(\theta,I)$ coordinates near a point of a single transverse impact with the upper part of the step, showing, as expected, that it is a smooth near-integrable twist map of the form \eqref{eq:mapj1itheta} . We then show that for small negative $K$,  close to the tangency with $\qw_1$, the change in the action of the map is close to zero, hence, when we return to the $(\phi,K)$ coordinates, the map becomes \eqref{eq:mapj1}.
     
\textbf{Return map in $(\theta,I)$ coordinates:} Initial conditions in $J_1^\eps$ undergo a single, transverse  impact (recall that $p_2$ is bounded away from zero near the step corner since $h>h^w_\eps$), hence, the return map dependence on initial conditions is smooth within this region (see \cite{PnueliRomKedarIntegrability}). Thus, similar to  \cite{PnueliRomKedarIntegrability}, we show next that the return map in the $(\theta,I)$ coordinates for initial conditions in  $J_1^\eps$  is of the form: 
\begin{equation}
\label{eq:mapj1itheta}
\mathcal{F}_\eps|_{z \in J_1^\eps }:
\begin{cases}
    \bar{I} &= I+ \eps \tilde f_1(\theta,I;\eps) \\
    \bar{\theta} &= \theta + \Omega_{J_1}(I) + \eps \tilde g_1(\theta,I;\eps)
\end{cases}
\end{equation}
where $\Omega_{J_1}(I)$ is given by \eqref{eq:Thetaj1} and $\tilde f_1,\tilde g_1$ denote the $C^r$ smooth corrections to the unperturbed impact dynamics due to the smooth perturbations, integrated backwards and forward from the impact point. 

In more details: we introduce the notation  $z^{im}=\Phi^{\eps,sm}_{t_-(z^{im})}z(\theta,I)=(q_1^{im},p_1^{im},\qw_2,p_2^{im})= (q_1^{im},p_1^{im},\tw_2(I^{im}),I^{im})$, where $t^\eps_-(z^{im})$ is the travel time of the backward smooth flow from $z_{im}$ to $\Sigma_h$ and  $t^\eps_+(z^{im})$ is the forward travel time to $\Sigma_h$. Integrating backwards from $z^{im}$ (with small negative $p_1^{im}, (q_1^{im}-\qw_1)$, see Section \ref{sec:proofofcornersing}) guarantees that  $z(\theta,I) \in J_1^\eps$. Notice that  $t^\eps_\pm(z^{im})$ are always bounded, and, since the impact with the upper wall is transverse for all $(\theta,I)\in J_1^\eps$, they
 depend smoothly on both $\eps$  and, within this region, on $(\theta,I)$.

Define:  $\eps \tilde f_1^-(\theta,I;\eps)  =   \int^0_{-t^\eps_-(z^{im})} \{I,H\}_{\Phi_t^{\eps,sm}z^{im}}dt $  and  $\eps \tilde f_1^+(\bar \theta, \bar I;\eps)  =   \int^{t^\eps_+(z^{*})}_0 \{I,H\}_{\Phi_t^{\eps,sm}\mathcal{R}_2z^{im}}dt $, so , $I^{im}(\theta,I)=I+\eps \tilde f_1^-(\theta,I;\eps) =\bar I -\eps \tilde f_1^+(\bar \theta,\bar I;;\eps)$ and  $\tilde f_1=\tilde f_1^- +\tilde f_1^+ $.

Let  $  G_1^-(\theta,I;\eps)  =   \int^0_{-t^\eps_-(z^{im})} \{\theta,H\}_{\Phi_t^{\eps,sm}z^{im}}dt -  \omega_2(I^{im}) t^\eps_-(z^{im}) $  and  $G_1^+(\theta,  I;\eps)  =   \int^{t^\eps_+(z^{*})}_0 \{\theta,H\}_{\Phi_t^{\eps,sm}\mathcal R_2z^{im}}dt-  \omega_2(I^{im}) t^\eps_+(z^{im}) $. 
Since $ \{\theta,H_{int}\}=\omega_2(I)$  and   $t^\eps_\pm(z^{im})$  are bounded, we conclude that  $G_1^\pm(\theta,I;\eps)=O(\eps) $ and are $C^r$ smooth (these functions are smooth also at the corner singularity curves, where the reflection does not correspond to a physical orbit). 

Since at impact the angle jumps from $\tw_2(I^{im})$ to   $2\pi -\tw_2(I^{im})$, namely, gains a  $2\pi -2\tw_2(I^{im})$  jump, we conclude that\begin{equation}\label{eq:bartheta1}
    \bar \theta-\theta=G_1^-(\theta,I;\eps)+G_1^+(\theta,I;\eps)+ \omega_2(I^{im}) (t^\eps_-(z^{im})+t^\eps_+(z^{im})) + 2(\pi-\tw(I^{im})).
\end{equation}
Let $G_2(\theta,I;\eps)=\omega_2(I^{im}) (t^\eps_-(z^{im})+t^\eps_+(z^{im})) + 2(\pi-\tw(I^{im}))-\Omega_{J_1}(I)$. 
Since  $I^{im}(\theta,I)=I+O(\eps) $, and since the return time of the impact flow to $\Sigma_h$ is $\eps$-close to the unperturbed return time:  $ (t^\eps_-(z^{im})+t^\eps_+(z^{im}))=T_1(I^{im})+O(\eps)=T_1(I)+O(\eps)$, and since, for  $h>h^w_\eps$,   $\tw(I)$ depends smoothly on $I$ near $I^{im}$, we obtain that $G_2(\theta,I;\eps)=O(\eps)$,  so indeed   $\bar \theta-\theta=\Omega_{J_1}(I)+\eps \tilde g_1 $ where 
$\eps\tilde g_1=G_1^-(\theta,I;\eps)+G_1^+(\theta,I;\eps)+ G_2(\theta,I;\eps)$ is the small, $C^r$ smooth remainder term.  

 So \eqref{eq:mapj1itheta} is established. 

Importantly, since the impact with the upper wall is transverse for all $(\theta,I)\in J_1^\eps$, within this region all the above defined  functions: $\tilde f_1, \tilde f_1^\pm,  G_1^\pm,G_2,\tilde g_1 , I^{im},z^{im}$ and  $t^\eps_\pm(z^{im})$  are regular in $\eps$ and depend $C^r$ smoothly on $(\theta,I)$.

\textbf{Time reversing orbit in $J_1^\eps$:} 

We now calculate the map in $J_1^\eps$ by fixing $K<0$ and varying $\phi \in \mathcal{J}_1^\eps(K)$. 

Notice that for all $K<0$, there exists  $\phi^{*}(K)\in \mathcal{J}_1^\eps(K)$ such that $p_1^{im}=0$ at this angle: this follows from the continuity of  $p_1^{im}$ and the fact that it is negative for $\phi \nearrow \phi_{1R}(K)$  and positive for  $\phi \searrow \phi_{01}(K)$, see Eq. \eqref{eq:p1epsUpsilon} and Figure \ref{fig:wallsection}. Let $(\theta^*,I^*)=(\phi^*(K),K+\It^\eps(\phi^*(K)))$ denote this parametric curve of initial conditions in $J_1^\eps$.

Initial conditions belonging to this curve impact the step at $z^{*,im}=z^{im}(\phi^{*}(K),K)=z^{im}(\theta^*,I^*)$ at a right angle, so the reflection $\mathcal{R}_2 z^{*,im} = \mathcal{R}_1  \mathcal{R}_2  z^{*,im}  $ exactly reverses the dynamics. Hence,
the forward image of $(\theta^*,I^*)$ at $\Sigma_h$ is exactly its reflection in $\theta$: $\mathcal{F}_\eps (\theta^*,I^*)=(-\theta^*,I^*)$.  Namely,  \begin{equation}\label{eq:istarthetastar}
    \bar \theta^{*}(K) =  -\theta^{*} (K), \quad  \bar I^* = I^*.
\end{equation}
Hence, $\tilde f_1(\theta^*,I^*;\eps)=\tilde f_1(\theta^*,K+\It^\eps(\theta^*);\eps)=0$ and
$\eps \tilde g_1 (\theta^*,I^*;\eps)=-2\theta^*-\Omega_{J_1}(I^*)$.

 Since $\tilde f_1,\tilde g_1$ are smooth in $\theta,I$ for $\theta \in \mathcal{J}_1^\eps(K)$, and  $|\mathcal{J}_1^\eps(K)|$ is  small for small $K$,  it follows, e.g. by the mean value theorem that  $\eps \tilde f_1(\theta,K+\It^\eps(\theta);\eps) =\eps \tilde f_1(\theta^*,K+\It^\eps(\theta^*);\eps)+  \eps (\theta-\theta^*)\tilde f_2(\theta,I;\eps)= \eps (\theta-\theta^*)\tilde f_2(\theta,I;\theta^*,\eps)$ where $\tilde f_2(\theta,I;\eps)$ is a $C^{r-1}$ function.
 So, for all  $\theta \in \mathcal{J}_1^\eps(K)$: \begin{equation}\label{eq:ibareqi}
     \bar I =I +\mathcal{O}_{C^{r-1}}(\eps(\theta-\theta^*))=I+ \mathcal{O}_{C^{r-1}}(\eps \sqrt {-K}).
 \end{equation} where the last equality follows from the fact that   $|\mathcal{J}_1^\eps(K)|=O(\sqrt{-K})$.
 
 In this interval $\bar \theta$  is also close to $\bar \theta^*$; indeed, since $ \Omega_{J_1}(I)$ is smooth and $I=K+\It^\eps(\theta)$ is also smooth, for all   $\theta \in \mathcal{J}_1^\eps(K)$ we have that:
 $\bar \theta -\bar \theta^{*}=\theta-\theta^*  + \Omega_{J_1}(I)-\Omega_{J_1}(I^*)+ \eps \tilde g_1 (\theta,I;\eps)-\eps \tilde g_1 (\theta^*,I^*;\eps) =  O(|\theta-\theta^*|)$. It thus follows from \eqref{eq:istarthetastar} and the $\eps$-closeness of $\It^\eps(\theta)$  to  $\It(h)$ (of   \eqref{eq:Itan})  that \begin{equation}\label{eq:itanthetastar}
     \It^\eps(\theta)=\It^\eps(\theta^{*} + O(|\theta-\theta^*|))=\It^\eps(-\bar \theta^* )+ O(\eps |\theta-\theta^*|)=\It^\eps(-\bar \theta )+ O(\eps |\theta-\theta^*|).
 \end{equation}
 
Since $K(\theta,I)=I-\It^\eps(\theta)$,  and $\bar K=K(\bar \theta, \bar I) = \bar I -\It^\eps(\bar \theta)$ by  Eq.  \eqref{eq:ibareqi} , we obtain that  $\bar K= I -\It^\eps(\bar \theta)=I -\It^\eps(\theta)+ \It^\eps(\theta)-\It^\eps(\bar \theta)=K+ \It^\eps(\theta)-\It^\eps(\bar \theta)$. By \eqref{eq:itanthetastar} we conclude that
 $\bar K= K -\It^\eps(\bar \theta) + \It^\eps(-\bar \theta) + O(\eps |\theta-\theta^*|) $, hence, $\bar K = K + \eps f(\bar \theta) + O(\eps |\theta-\theta^*|)$,  and using again $ |\mathcal{J}_1^\eps(K)|=O(\sqrt{-K})$,  the first line of   \eqref{eq:mapj1} follows. 

Finally, since $\Omega_{J_1}(I)=\Omega_{J_1}(\It^\eps(\theta)+K)=\Omega_{J_1}(\It(h)+K) + O(\eps) $ the second line of  \eqref{eq:mapj1} follows from the second line of  \eqref{eq:mapj1itheta}.

Since the perturbation terms in \eqref{eq:mapj1itheta} are $C^r$ functions and the computations around $\theta^*(K)$ involve one derivative the error terms in  \eqref{eq:mapj1}
 are $C^{r-1}$ functions of $\theta$.
\end{proof}

To complete the proof  we combine \eqref{eq:mapj0}, \eqref{eq:mapjr} and \eqref{eq:mapj1} and noticing that $\Omega_{J_1}(\It(h)+K)=\Omega_{J_1}(\It(h))+ (\Omega_0'(\It(h))-2 \frac{d \tw(I)}{dI}|_{\It(h)})K + O(K^2)$,  we have
established \eqref{eq:returnmapfull}.

The behavior of the correction terms is singular at small negative $K$ values. The requirement that the piecewise smooth map \eqref{eq:returnmapfull} corresponds to the return map of the HIS, namely that it preserves the symmetries and that it is piecewise symplectic imposes some restrictions on the form of the correction terms. We leave it to future studies to fully characterize the  correction terms and the dependence of the results on their form.

\section{Parameters for specific potentials \label{sec:calculationofparam}}

The resulting parameters  $\{\Omega_0,\tau_0, ,\tau_1, \lambda,\cos\tw\}$ of the truncated return map \eqref{eq:truncatedmap} for various combinations of the Harmonic potential: $V(q)=\frac{1}{2}\omega^2q^2$ and the Tan potential $V(q)=\frac{\omega^2}{2\alpha^2}\tan^2(\alpha q)=\frac{\omega^2}{2\alpha^2\cos^2(\alpha q)}-\frac{\omega^2}{2\alpha^2}$ may be explicitly calculated (since for these potentials the transformation to action-angle coordinates is explicit).

Since, for small $(\alpha, q)$ the Tan potential limits to the quadratic potential (with the above form), it suffices to compute the parameters for the Tan-Tan case and then notice the various limits.  Using
\begin{align*}
\theta_i\left(q_i,h_i\right) &  =\cos^{-1}\left(\sqrt{\alpha_i^{2}+\frac{\omega_i^{2}}{2h_i}}\frac{\sin\left(\alpha_i q_i\right)}{\alpha_i}\right) \\
h_i(I_i)&=\omega_i I_i +\alpha_i^2 I_i^2/2,\\\frac{d}{dI_i}h_i(I_i)&=\omega_i  +\alpha_i^2 I_i\\
\frac{d \theta_i^w}{dI_i}&=\frac{-1}{\sin \theta }\frac{d \cos \theta^w}{dh}\frac{d h}{dI} \\
&= \frac{1}{\sqrt{1-(\alpha^{2}+\frac{\omega^{2}}{2h})\left(\frac{\sin\left(\alpha q\right)}{\alpha}\right)^2}}\frac{\sin\left(\alpha q\right)}{\alpha}\frac{\omega^2(\omega +\alpha^2I)}{4h^2 \sqrt{\alpha^{2}+\frac{\omega^{2}}{2h}}}
\end{align*}
and
$\omega_2(\It(h))=\omega_2 +\alpha_2^2I_2=\sqrt{\omega_2^2+2 h_2\alpha_2^2}=\sqrt{\omega_2^2+2 (h-h_1^w)\alpha_2^2}$ with  $h_1^w=\frac{\omega_1^2}{2\alpha_1^2}\tan^2(\alpha_1 \qw_1)$, as $h=\hw_1+H_2(\It(h))$  and $\Omega_0(I)|_{\It(h)}=\frac{2 \pi \omega_2(I)}{\omega_1(I_1(I))}=\frac{2\pi (\omega_2+\alpha_2^2 I)}{\omega_1+\alpha_1^2 I_1(h-H_2(I))}|_{\It(h)}=\frac{2 \pi \sqrt{\omega_2^2+2 (h-h_1^w)\alpha_2^2}}{\sqrt{\omega_1^2+2 \alpha_1^2 h_1^w}}$, we get:
\begin{equation}
    \begin{array}{cc}
     \Omega_0 &=\frac{2 \pi \sqrt{\omega_2^2+2 (h-h_1^w)\alpha_2^2}}{\sqrt{\omega_1^2+2 \alpha_1^2 h_1^w}}      \\
    \tau_0     &=\frac{ 2\pi \alpha_2^2 (\omega_1^2+2 \alpha_1^2 h_1^w)^{3/2}+2\pi \alpha_1^2 (\omega_2^2+2 (h-h_1^w)\alpha_2^2)}{(\omega_1^2+2 \alpha_1^2 h_1^w)^2}  \\
    \lambda &= \frac{\sqrt{2}(\omega_2^2+2 (h-h_1^w)\alpha_2^2)^{3/4}}{\omega_1^2}\frac{|\alpha_1 \cos^3(\alpha_1 \qw_1)|}{|\sin(\alpha_1 \qw_1)|} \\
    \tw  &= \arccos\left( -\sqrt{\alpha_2^{2}+\frac{\omega_2^{2}}{2(h-\hw_1)}}\frac{\sin\left(\alpha_2 \qw_2\right)}{\alpha_2}\right) \\
    \tau_1 &=
-\frac{\sin\left(\alpha_2 \qw_2\right)}{\alpha_2}\frac{\omega_2^2}{(h-h_{1}^w)}\frac{1}{\sqrt{2(h-h_{1}^w)\cos^2\left(\alpha_2 \qw_2\right)-\omega_2^{2}\left(\frac{\sin\left(\alpha_2 \qw_2\right)}{\alpha_2}\right)^2}}.
    \end{array}
\end{equation}
So, for non-zero $(\alpha_1,\alpha_2)$, the parameters $\Omega_0,\tau_0, \lambda$ are positive and increasing with $h$ whereas $|\cos\tw|,|\tau_1|$ are monotone decreasing with $h$.  Note that $\mathrm{sign}\tau_1= -\mathrm{sign}\qw_2$. 

When one or both of the $\alpha_i$'s limit to zero (i.e. the potentials approach the quadratic potential) the above parameters attain  finite limits:
\begin{itemize}
    \item For a fixed $\alpha_1 \ne 0$, the limit $\alpha_2 \rightarrow 0$  makes $\Omega_0,\tau_0, \lambda$ positive and independent of $h$.
    \item For a fixed $\alpha_2\ne0$, the limit $\alpha_1 \rightarrow 0$  makes $\tau_0$ positive and independent of $h$.
    \item The limit $\alpha_1,\alpha_2 \rightarrow 0$  makes, naturally,  $\tau_0\rightarrow 0$ (no twist for the case of smooth potential which is the sum of two quadratic potentials), $\Omega_0 \rightarrow \frac{2 \pi \omega_2}{\omega_1}, \lambda\rightarrow \frac{\sqrt{2}\omega_2^{3/2}}{\omega_1^2 |\qw_1|} $ have  finite limits independent of $h$, and $|\cos\tw|,|\tau_1|$ are non-zero and monotone decreasing with $h$. In this limit the results regarding the  critical circle and the hovering orbits  need to be re examined as we assumed, for applying KAM theory, that $\tau_0\ne 0$.
\end{itemize} 

 The asymmetry between the first two cases is a result of concentrating on the tangential torus which is tangent to the right side of the step, for which the horizontal motion is tangential. 
\end{document}